\documentclass[10pt,twoside,twocolumn,german,english,prl,aps,amssymb,preprint,showpacs,superscriptaddress]{revtex4}
\usepackage{babel}
\usepackage[T1]{fontenc}
\usepackage[latin1]{inputenc}
\usepackage[tight]{subfigure}
\usepackage{float}
\usepackage{amsmath}
\usepackage{amssymb}
\usepackage{ae}
\usepackage{graphicx}
\usepackage{exscale}
\usepackage{mathrsfs}
\newlength{\formulalength}
\newcommand{\setformulalength}{
\setlength{\formulalength}{ \linewidth }
\addtolength{\formulalength}{- \labelwidth}
\addtolength{\formulalength}{- \fboxrule}
\addtolength{\formulalength}{- \fboxsep}
}
\newcommand{\mymathfont}[1]{\textrm{#1}}
\newcommand{\hs}[1]{\hspace*{#1 mm}}
\newcommand{\ds}{\displaystyle}
\newcommand{\mb}{\mathbb}

\newcommand{\ms}{\mathscr}   
\makeatletter

\newcommand{\lts}{}
\newcommand{\tr}{\triangle}
\newcommand{\dint}[2]{\int\!\!\!\!\int\lts_{#1}}
\newcommand{\Euler}{C_{\mathcal{E}}}
\newcommand{\surften}{\gamma}                     
\newcommand{\Subsec}[1]{Subsec.\,\ref{sub:#1}}
\newcommand{\Eq}[2][]{Eq#1.\,(\ref{#2})}
\newcommand{\Corr}{C}                                  
\newcommand{\zlimit}{L}

\newcommand{\fcstar}{f^{\circledast}}
\newcommand{\conc}{X}                             
\newcommand{\md}{\rho^{*}}			

\newcommand{\mnd}{{\rho}^{*}}
\newcommand{\ip}{\textrm{p}}		

\newcommand{\ibfu}{R_{\mathit{min}}}    
\newcommand{\pztr}{\mathcal{R}}		
\newcommand{\clrd}{d}					
\newcommand{\clr}[1]{\frac{\xi_{#1}}{\xi_1 + \xi_2}}	
\newcommand{\Clrd}{\mathbb{D}}				
\newcommand{\ep}{\Lambda}
\newcommand{\epu}{\ep ^{+}}				
\newcommand{\era}{\lambda}				
\newcommand{\erb}{\Gamma}				
\newcommand{\emi}{\ep ^{-}}				
\newcommand{\gammasym}{\gamma_\mathsf{asym}^{-}}
\newcommand{\gdcDEF}{\varphi}
\newcommand{\gdc}[2]{\gdcDEF_{#1}^{(#2)}}		
\newcommand{\gdcb}[2]{\bar{\gdcDEF}_{#1}^{(#2)}}   %
\newcommand{\dco}[1]{\gdc{#1}{1} [ \delta \mnd ]}			%
\newcommand{\dcob}[1]{\gdcb{#1}{1} [ \delta \mnd ]}			%
\newcommand{\dcop}[1]{\gdc{#1}{1} [ \partial_{u} \mnd_{c} ]}		%
\newcommand{\dcopb}[1]{\gdcb{#1}{1} [ \partial_{u} \mnd_{c} ]}		%
\newcommand{\dct}[1]{\gdc{#1}{2} \big{[} [ \delta \mnd ]^{2} \big{]}}	%
\newcommand{\dcws}{\omega}				
\newcommand{\dcw}[1]{\dcws^{(#1)}}			
\newcommand{\dcwh}[1]{\hat{\dcws}^{(#1)}}		
\newcommand{\gamw}{\gamma^{\wedge}}

\newcommand{\gamv}{\gamma^{\vee}}

\makeatother
\begin{document}
\selectlanguage{english}
\renewcommand{\figurename}{Fig.}
\renewcommand{\thesubfigure}{\,(\alph{subfigure})}
\title{Microscopic theory for interface fluctuations in binary liquid mixtures}

\author{Thorsten Hiester}

\affiliation{Institut f\"ur Theoretische Physik, Universit\"at Erlangen-N\"urnberg,
Staudtstr. 7, D-91058 Erlangen, Germany}

\author{S. Dietrich}

\affiliation{Max-Planck-Institut f\"{u}r Metallforschung, Heisenbergstr. 3,
D-70569 Stuttgart, Germany}

\affiliation{Institut f\"{u}r Theoretische und Angewandte Physik, Universit\"{a}t
Stuttgart, Pfaffenwaldring 57, D-70569 Stuttgart, Germany}

\author{Klaus Mecke}

\affiliation{Institut f\"ur Theoretische Physik, Universit\"at Erlangen-N\"urnberg,
Staudtstr. 7, D-91058 Erlangen, Germany}

\begin{abstract}
Thermally excited capillary waves at fluid interfaces in binary liquid mixtures exhibit simultaneously both
density and composition fluctuations.
Based on a density functional theory for inhomogeneous binary liquid mixtures
we derive an effective wavelength dependent Hamiltonian for fluid interfaces in these systems 
beyond the standard capillary-wave model.
Explicit expressions are obtained for the surface tension, the bending rigidities,
and the coupling constants of compositional capillary waves
in terms of the profiles of the two number densities characterizing the mixture. These results lead to
predictions for grazing-incidence x-ray scattering experiments at such interfaces. \\
\\
\end{abstract}

\keywords{Interface fluctuation, binary Liquids, capillary waves}
\pacs{68.05.-n, 68.03.-g, 82.65.+r, 05.70.Np, 64.75.+g}
\maketitle

\section{Introduction\label{sec:intro}}

\begin{figure}[t]

\centering\foreignlanguage{english}{\includegraphics[%
  scale=0.30]{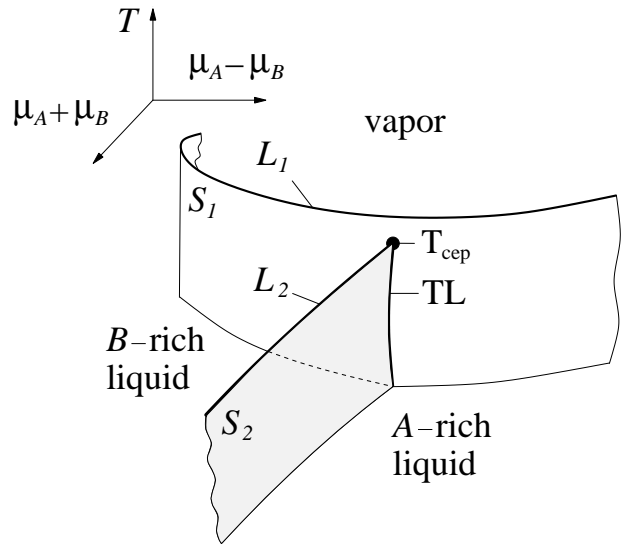}}

\caption{\label{cap:Phasdiagram} Schematic bulk phase diagram of a binary liquid mixture
in the thermodynamic parameter space $(\mu_{A}+\mu_{B},\mu_{A}-\mu_{B},T)$ spanned by 
the chemical potentials $\mu_{A}$ and $\mu_{B}$ of the two species and temperature $T$.
As mentioned in the introduction, the phase diagram exhibits a vapor phase, a mixed liquid phase, an $A$-rich liquid phase,
and a $B$-rich liquid phase separated by sheets $S_{1}$  and $S_{2}$ of first order-phase transitions which
are bounded by lines $L_{1}$ and $L_{2}$, respectively, of second-order phase transitions.
The line of intersection between $S_{1}$ and $S_{2}$ represents
the triple line TL where three phases coexist. $\mathrm{T}_{\mathrm{cep}}$ denotes the critical end point.
In the present context of fluid interfaces the solid phases of a binary liquid mixture, which 
occur at high pressures $(\sim \mu_{A}+\mu_{B})$ or low temperatures, are omitted for reasons of simplicity.
Further details can be found in Refs.\,\cite{Dietrich:9204(1989),Getta:1856(1993),Dietrich:178(1997)}.
}
\end{figure}

If two thermodynamically coexisting fluid phases are brought into
spatial contact via suitable boundary conditions, an interface forms
which interpolates smoothly between the bulk properties of the coexisting
phases. For more than a hundred years substantial theoretical and experimental
efforts have been devoted to resolve the structural  properties of this transition
region (see, e.g., Refs.\,\cite{Rowlinson:1982,Jasnow:1059(1984)}). \\
The reason for the persistence of these challenge resides in the difficulty to describe the
simultaneous occurrence of bulk fluctuations reaching the interface and of capillary wave-like
fluctuations of the local interface position \cite{Mecke:6766(1999)}. 
For the simplest example, i.e., the  liquid-vapor
interface of a one-component fluid, the concept of 
an effective interface Hamiltonian leads to quantitative predictions for a wavelength-dependent surface tension 
\cite{Helfrich:693(1973),Blokhuis:6986(1991),
Napiorkowski:1836(1993),
Blokhuis:397(1999),Mecke:6766(1999)}, which
has been confirmed experimentally for the interface structure factor down to microscopic
length scales for a wide variety of one-component fluids
\cite{Fradin:871(2000),Daillant:223(2001),Mora:216101(2003),Li:136102(2004),Lin:106(2005)}.
For these systems the wavelength dependent surface tension $\surften(q=2\pi/\lambda)$
is a function of temperature $T$ and a functional of the interaction pair potential of the fluids particles.
The macroscopic surface tension $\surften=\surften(0)$  of the liquid-vapor interface is obtained for $q\rightarrow0$ 
whereas $\surften(q)$ decreases for increasing values of $q$, reaches a substantial minimum,
and increases again for large $q$. This decrease of $\surften(q)$ is in a accordance with simulation data
\cite{Stecki:5002(1998),Milchev:81(2002),Vink:134905(2005)}
which, however, have not yet confirmed the predicted and experimentally observed re-increase of $\surften(q)$
at large $q$.\\
The present work aims at  extending this analysis to the case of binary liquid mixtures composed of species
$A$  and  $B$. This is motivated by the following reasons:

\renewcommand{\theenumi}{\roman{enumi}}
\renewcommand{\labelenumi}{(\theenumi)}
\begin{enumerate}
\item Binary liquid mixtures are governed by three pair potentials $w_{ij}(\mathbf{r})$
for the $A$-$A$ and $B$-$B$ interaction between the like species and the $A$-$B$ interaction
between unlike species. Provided a wavelength dependent surface tension can be introduced analogous
to the one for one-component fluids, it will therefore be a functional of three pair potentials. By exchanging
systematically one of the two components by a sequence of molecules with a quasi-continuously changing
architecture, this might open the possibility to tune the shape of the function $\surften(q)$ and thus to create
new interfacial phenomena.
\item Whereas for one-component fluids two-phase coexistence is confined to a liquid-vapor coexistence {\em{line}}
described by the chemical potential $\mu_{o}(T)$, in binary liquid mixtures two fluid phases can coexist on a
{\em{two}}-dimensional sheet in their thermodynamic parameter space $(\mu_{A},\mu_{B},T)$ spanned by the
chemical potentials $\mu_{A}$ and $\mu_{B}$ of the two species and temperature (see Fig.\,\ref{cap:Phasdiagram}).
This allows one to vary the
thermodynamic state of the system over a considerably larger parameter space without loosing two-phase
coexistence, which in turn increases the possibilities to vary $\surften(q)$ by changing thermodynamic variables 
such as the composition.
\item Generically, for one-component systems liquid and vapor are the only fluid phases and thus liquid-vapor
interfaces are the only possible fluid interfaces in such systems.
Binary liquid mixtures exhibit various fluid phases: a mixed vapor phase, a mixed liquid phase, an $A$-rich liquid
phase, and a $B$-rich liquid phase, separated from each other by sheets of first-order phase transitions which intersect
along a triple line of three-phase coexistence and which are delimited each by lines of critical points (see, e.g.,
Refs.\,\cite{Dietrich:9204(1989),Getta:1856(1993),Dietrich:178(1997)} and Fig.\,\ref{cap:Phasdiagram}).
Depending on the relative integrated strength of the attractive parts of the aforementioned three pair potentials, 
there is a wide range of rather different topologies of the bulk phase diagrams of binary liquid mixtures
\cite{vanKonynenburg:495(1980)}.
These topologies of the bulk phase diagrams allow for four distinct types of fluid interfaces: vapor|mixed fluid,
vapor|$A$-rich fluid, vapor|$B$-rich fluid, $A$-rich liquid|$B$-rich liquid. In contrast to one-component systems
this offers the possibility to vary significant features of fluid interfaces without changing the underlying interaction
potentials but only the thermodynamic state.
\item The description of inhomogeneous binary liquid mixtures requires two number density profiles, $\rho_{A}(z)$ and
$\rho_{B}(z)$, where $z$ denotes the  distance  from the  mean interface position along the $z$ axis. In many cases
it is suitable to introduce instead the total number density $\rho(z)=\rho_{A}(z)+\rho_{B}(z)$ and the concentration
$X(z)=\rho_{A}(z)-\rho_{B}(z)$ as linear combinations. Whereas for one-component systems it is straightforward 
(as in Ref.\,\cite{Mecke:6766(1999)}) to assign 
a local liquid-vapor interface position $f(x,y)$ to a given density configuration as the position of an isodensity surface (e.g., points
$\mathbf{s}$ where $\rho(\mathbf{s})=(\rho_{liq}+\rho_{vap})/2$), such a construction is not clear from the outset in the presence
of two fluctuating densities. Thus, the study of fluid interfaces in binary liquid mixtures raises the challenging conceptual issue
how and to which extent they can be described microscopically in terms of an effective Hamiltonian $\mathcal{H}[f]$ 
and a wavelength dependent surface tension $\surften(q)$.
\item It requires  special care to prepare a bona fide one-component fluid. Naturally, systems come as multicomponent samples.
Generally, segregation phenomena occur  at their interfaces, which might influence significantly  the interface fluctuations and vice
versa. By choosing suitable series of molecules of related architecture and appropriate concentrations, binary fluid mixtures offer the
possibility to interpolate systematically between the material properties of the corresponding limiting one-component systems,
which generates substantial application perspectives. Finally, binary liquid mixtures can serve as rudimentary polydisperse systems
as they occur in colloid suspensions. The study of interfacial properties in such systems has become very rewarding because they
can be analyzed in great detail by direct optical techniques
\cite{Aarts:1973(2004),Aarts:847(2004)},
allowing for quantitative comparisons with theoretical predictions on the scale of the particles.
\end{enumerate}
As mentioned above, two types of fluctuations occur simultaneously at interfaces: (a) fluctuations of the density 
as they occur in the bulk on length scales up to the bulk correlation length $\xi$; 
(b) in the absence of gravity and for large system sizes the mean position of the interface can be shifted without cost 
of free energy. This gives rise to thermally excited Goldstone modes leading to lateral fluctuations of the local interface 
position, with wavelengths reaching macroscopic values. Depending on which type of fluctuation is emphasized,
originally two different approaches for the theoretical understanding have emerged.

As put forward by van der Waals
\cite{vanderWaals:657(1894)},
the first approach leads to a so-called intrinsic density profile which interpolates smoothly between the constant densities in the
coexisting bulk phases. The interface is laterally flat and is kept in place by boundary conditions or a small gravitational field
acting along the interface normal. The width of the intrinsic profile
\cite{Jasnow:1059(1984),Weeks:3106(1977),Bedeaux:972(1985)}
is given by the bulk correlation length, which diverges upon approaching the critical point $T_{c}$ of the corresponding two-phase
coexistence, reflecting the disappearance of the interface at $T_{c}$. Accordingly, the van der Waals picture is expected to capture
the interfacial properties at  elevated temperatures close to $T_{c}$.

The second approach, conceived by Buff, Lovett, and Stillinger
\cite{Buff:621(1965)},
describes the width of an interface as a result of capillary-wave like fluctuations of a step-like intrinsic density profile.
Here only the local interface positions are the statistical variables.
The resulting mean density profile attains   the bulk values like a Gaussian whereas the van der Waals approach yields an
exponential decay for short-ranged forces between the fluid particles or inverse power laws in the presence of dispersion
forces \cite{Dietrich:1861(1991)}.
Within the capillary-wave model the width of the mean interface diverges upon switching off gravity or increasing the lateral system size.
This  roughening effect is missed by all available van der Waals approaches.
On the other hand the capillary wave model misses the fact
that the interfacial width diverges upon approaching $T_{c}$ on the scale of the bulk correlation length $\xi$.

Accordingly one can state that the van der Waals approach captures fluctuations on the length scale of $\xi$ and below and is suitable
at high temperatures whereas the capillary wave approach is valid at low temperatures and captures the fluctuations with wavelengths
larger than $\xi$. 
In Ref.\,\cite{Mecke:6766(1999)}
these two approaches have been reconciled by considering intrinsic density profiles, as obtained from density functional theory, which undergo
fluctuations of their lateral positions. Density functional theory provides expressions for the cost in free energy of such density configurations
relative to the free energy of a flat interface. This yields an effective interface Hamiltonian $\mathcal{H}[f]$ 
and thus provides the statistical weight of interfacial
fluctuations $f$. This statistical weight  can also be used to calculate correlations of the local interface normals
\cite{Mecke:204723(2005),Aarts:15(2005)}.

Inspired by the motivation described above, the present work extends the concept of Ref.\,\cite{Mecke:6766(1999)} 
to the description of fluid interfaces of binary liquid mixtures.
After a brief discussion of the bulk phase diagrams of binary liquid mixtures (Subsec.\,\ref{sub:phasedia})
we introduce the density functional theory which we use as the starting point for the description of spatially 
inhomogeneous fluids (\Subsec{DFT}).
We define the effective interface Hamiltonian $\mathcal{H}$ for mixtures in \Subsec{Eff_Interf_Hamiltonian}. 
After discussing the crude approximation of steplike density profiles (\Subsec{SKAP_1}), in \Subsec{CUREX_1} we 
introduce the central approximation which we actually use for further calculations. 
It involves the influence of the curvatures of the iso-density contours on the density profiles which has turned out 
to be crucial in order to describe the  fluctuations of a liquid-vapor interface (see Ref.\,\cite{Mecke:6766(1999)}). 
Since this approach cannot simply be transferred to binary 
liquid mixtures requiring two iso-density contours, Sec.\,\ref{sec:Effective-Hamiltonian}
is closed by remarks about how to overcome these additional problems.
In \Subsec{MeanSurfaceApprox} and Appendices\,\ref{app:Explicit-Form-of-H} and \ref{app:Second-Order-approximation}
we present the explicit expressions for the effective interface Hamiltonian $\mathcal{H}$ based on the above-mentioned 
approximations.
In order to be able to make
predictions for scattering experiments from such interfaces, in Sec.\,\ref{sec:Gaussian-Form-of_H} we present a 
Gaussian approximation of the
effective interface Hamiltonian $\mathcal{H}$ using various representations and we discuss the resulting contributions
(Subsecs.\,\ref{sub:General-Expression_H}-C). 
In Sec.\,\ref{Correlations} we analyze the temperature 
and the composition dependence of structural properties of interfaces in binary liquid mixtures as 
inferred from the correlation functions. 
We summarize our results in Sec.\,\ref{sec:Summary}.


\section{Effective Interface Hamiltonian\label{sec:Effective-Hamiltonian}}

\begin{figure}[t]

\centering\foreignlanguage{english}{\includegraphics[%
  scale=0.25]{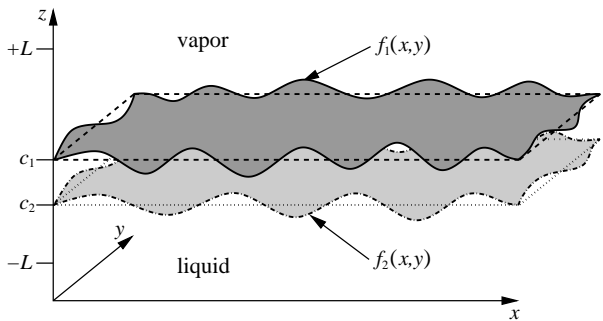}}

\caption{\label{cap:SketchSystem}Sketch of the liquid-vapor interface region
of a binary liquid mixture with its normal oriented along the $z$-axis; for each
of the two fluctuating densities of the mixture an interface position can be defined (dark and light grey,
see the main text, and Eqs. (\ref{eq:rho_c_prop}) and (\ref{eq:rho_f_prop})). In order to derive
the interface Hamiltonian, two interface configurations are considered:
the reference surface configuration is given by constant iso-density
contours for the two species $A$ and $B$ at $z=c_{1}$ and $z=c_{2}$, respectively, while a non-flat
configuration with iso-density contours $z=f_{1}(x,y)$ and $z=f_{2}(x,y)$
varies laterally around the position $c_{1}$ and $c_{2}$, respectively, so that
$\langle f_{1}\rangle = c_1$ and $ \langle f_{2}\rangle = c_2$.
Additionally, the non-flat surfaces are assumed to fluctuate mildly without
overhangs or bubbles, so that a Monge representation
with $f_{i}(x,y)$, $i\in\{1,2\}$, can be used. Vapor and liquid should be understood as
representations of two fluid phases including a mixed  liquid, an $A$-rich liquid, and a $B$-rich liquid 
(see Subsec.\,\ref{sub:phasedia}).}
\end{figure}

In this section we derive an effective interface Hamiltonian $\mathcal{H}$ for the
interface between two fluid phases of a simple binary liquid mixture consisting of spherical particles
with radially symmetric interaction potentials. The system with its interface is described
microscopically in terms of a simple, but for the present purpose appropriate
grand canonical density functional.
For each of the two equilibrium particle density distributions we specify implicitly
an iso-density contour as its interface surface assuming that this captures the
interface structure of the mixture as a whole.
The interface effective Hamiltonian, which counts the cost in free energy to deform the interfaces
from a given reference configuration, is  defined as the difference between
two grand canonical potentials corresponding to two different
surface configurations. Further simplifications are made to express
this Hamiltonian explicitly, rather in terms of the surfaces, in
terms of the yet unknown, inhomogeneous densities. Thus, by construction,
the microscopic interactions between the particles are taken into account transparently, which
finally lead to  effective interactions between the surfaces. To a large extent the functional
dependence on the interaction potentials is kept general. Ultimately, for numerical evaluations, we
assume long-ranged attractive  dispersion forces. \\
\\
The normal of the mean interface of the binary liquid mixture is taken to be
oriented along the $z$-axis such that, for instance, the liquid phase
and the vapor phase of the mixture are approached for $z\rightarrow-L$
and $z\rightarrow+L$, respectively (see Fig. \ref{cap:SketchSystem}).
Density functional theory assigns a free energy to each density configuration such that the
equilibrium configuration minimizes the functional and yields the corresponding grand canonical
potential. As the natural reference configuration we choose what we call the flat
state, in which the iso-density contours are laterally
constant surfaces and do not vary with $\mathbf{R}=(x,y)$ (see
Fig.\,\ref{cap:SketchSystem}). If present, gravity points into the negative $z$-direction.

\subsection{Bulk phase diagram of binary liquid mixtures\label{sub:phasedia}}

As stated in the introduction, binary liquid mixtures are composed of two species, 
called $A$ and $B$ particles. At high temperatures these particles mix in a gaseous phase.
Upon lowering the temperature the mixture exhibits a phase separation into a gas phase of
low density and a liquid phase of high density. In Fig.\,\ref{cap:Phasdiagram} this phase separation is indicated
by the sheet $S_{1}$ with $\mu_{A}+\mu_{B}$ as a measure of the total pressure of the system.
At sufficiently high temperatures in both these phases the two species remain mixed.
A further decrease of the temperature leads to an additional phase separation of the fluid phase 
into an $A$-rich liquid phase and a $B$-rich liquid phase (see sheet $S_{2}$ in Fig.\,\ref{cap:Phasdiagram}).
In the following any pair of the mixed gas, mixed fluid, $A$-rich liquid, and $B$-rich liquid are denoted
as liquid and vapor. 
Their coexistence corresponds to a point
on $S_{1}$ or $S_{2}$ and, for instance,  an increase of temperature at coexistence delineates a path on $S_{1}$
or $S_{2}$ approaching the line of critical points $L_{1}$ or $L_{2}$, respectively.
On the other hand, changing the composition of the mixture at a fixed temperature at coexistence corresponds
to a path on $S_{1}$ or $S_{2}$ intersecting a horizontal $(\mu_{A}+\mu_{B},\mu_{A}-\mu_{B})$-plane 
in Fig.\,\ref{cap:Phasdiagram}.
In Sec.\,\ref{Correlations} we shall discuss our results in two respects:
first, we study the influence of temperature and, second, we shall keep the temperature fixed and consider composition variations.

\subsection{Density Functional Theory\label{sub:DFT}}

\begin{figure}[t]
\centering\foreignlanguage{english}{\includegraphics[%
  scale=0.25]{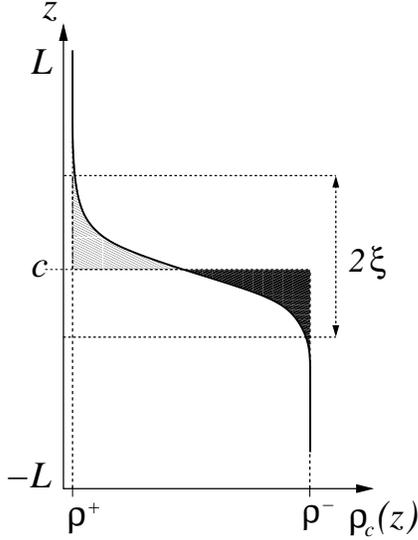}}

\caption{\label{cap:Sketch_rho_c}Sketch of the planar number density profile $\rho_{c}(z)$;
the width $2\xi=\xi^{+}+\xi^{-}$ of the transition region is roughly
the sum of the bulk correlation lengths $\xi^{+}$ and $\xi^{-}$ of the two coexisting phases
which, in general,
differ from each other (see also the remarks related to, c.f., Eq.\,(\ref{eq:corr_lengths_explicit})).
The length $\xi$ is also called the interfacial width of the (planar)
surface located at $z=c$ (see Eq.\,(\ref{eq:rho_c_prop})). The marked
regions are frequently used to define the interface position via
Gibbs' zero-adsorption criterion: the number of particles within the
dark and the light domain has to be the same for the interface at
$z=c$ \cite{Rowlinson:1982}. Within the sharp kink approximation one has $\xi=0$ so that a step-like 
profile results (see \Eq{eq:sk_profile}).}
\end{figure}

We consider a grand canonical density functional for a two-component
fluid which consists of particles $A$ and $B$ with a spherically symmetric interaction
potentials $W_{ij}(|\mathbf{r}|)=W_{ji}(|\mathbf{r}|)$, where the
indices $i,j\in\{1,\,2\}$ refer to the species $A$ and $B$. Following  standard procedure
\cite{Evans:143(1979)}
the interaction potential is split into a short-ranged repulsive
part $s_{ij}(r)$ and an attractive long-ranged part $w_{ij}(r)$.
For a system of volume $V=2L\, A$, where $A$ is the (flat) interfacial
area and $2L$ is the macroscopically large extension in $z$ direction, a simple version of the grand
canonical density functional $\Omega$  reads:\begin{eqnarray}
\Omega[\rho_{1}(\mathbf{r}),\rho_{2}(\mathbf{r})] & = &\;\mathcal{F}^{\mathsf{hs}}[\rho_{1}(\mathbf{r}),\rho_{2}(\mathbf{r})]\nonumber \\
 &  & +\,\sum_{i=1}^{2}\int\lts_{V}\! d^{3}r\;\big{(}\mu_{i}+V_{i}^{\mathsf{ex}}(\mathbf{r})\big{)}\rho_{i}(\mathbf{r})\nonumber \\
 &  & \hs{-27}+\,\frac{1}{2}\sum_{i,j=1}^{2}\,\int\lts_{V}\! d^{3}r\!\int\lts_{V}\! d^{3}r'\;\; w_{ij}(|\mathbf{r}-\mathbf{r}|)\rho_{i}(\mathbf{r})\rho_{j}(\mathbf{r}')\,.\label{eq:GranCanDF}\end{eqnarray}
 Here, $\rho_{i}(\mathbf{r})$ is the number density of the particles
of species $i\in\{1,2\}$ at $\mathbf{r}=(x,y,z)=(\mathbf{R},z)$, 
and $\mathcal{F}^{\mathsf{hs}}[\rho_{1}(\mathbf{r}),\rho_{2}(\mathbf{r})]$
is the reference free energy functional of a system governed by
the short-ranged contribution $s_{ij}(|\mathbf{r}|)$, expressed suitably in terms of a hard-sphere system.
In the following, we use $\mathcal{F}^{\mathsf{hs}}[\rho_{1}(\mathbf{r}),\rho_{2}(\mathbf{r})]$
within a local density approximation:\begin{equation}
\mathcal{F}^{\mathsf{hs}}[\rho_{1}(\mathbf{r}),\rho_{2}(\mathbf{r})]=\int\lts_{V}\! d^{3}r\; h\big{(}\rho_{1}(\mathbf{r}),\rho_{2}(\mathbf{r})\big{)}\,.\label{eq:hs_functional}\end{equation}
 In Eq.\,(\ref{eq:GranCanDF}) the chemical potential of species $i$
is denoted by $\mu_{i}$, while $V_{i}^{\mathsf{ex}}(\mathbf{r})$
represents its external potential, which in our case will be gravity acting along
the negative $z$-axis. The attractive part of the pair
interactions is given by $w_{ij}(|\mathbf{r}|)\equiv w_{ij}(r)$.
To a large extend our reasoning will not depend on specific choices for
$h\big{(}\rho_{1}(\mathbf{r}),\rho_{2}(\mathbf{r})\big{)}$,
$V_{i}^{\mathsf{ex}}$, and $w_{ij}(r)$. This will be required only for quantitative presentations.
Actually $w_{ij}(r)$ should be replaced by the direct correlation function $c_{ij}^{(2)}(r)$ which, however,
reduces to $w_{ij}(r)$ for large $r$. This replacement also does not alter our main results.
\\
With the notation $\mathbf{R}=(x,y)$ we introduce the bulk densities
\begin{equation}
\rho_{i}^{\pm}:=\rho_{i}(\mathbf{R},z\rightarrow\pm L)\,,\; i\in\{1,2\}\,\label{eq:Def_rho_pm}\end{equation}
characterizing the vapor ($\rho_{i}^{+}$) and the liquid phase ($\rho_{i}^{-}$) in the general sense described above.
In order to describe density configurations as shown in Fig.\,\ref{cap:SketchSystem}
we introduce $\rho_{c_{i}}(z)$ and $\rho_{f_{i}}(\mathbf{r})$ as the  density profiles of species $i$ which take a fixed
value $\bar{\rho}_{i}$ at the position $z=c_{i}$ for a flat configuration and at $z=f_{i}(\mathbf{R})$ for a non-flat configurations,
respectively. For the non-flat iso-density surfaces we assume a Monge
parameterization (see Fig.\,\ref{cap:SketchSystem}). Thus, the crossing
criterions are
\setformulalength
\begin{eqnarray}
\rho_{c_{i}}(z=c_{i}) & = & \bar{\rho}_{i}\label{eq:rho_c_prop}\\
\textrm{\parbox{0.6\formulalength}{and \hfill}}& &\textrm{\parbox{0.4\formulalength}{ \hfill}} \,\nonumber\\
\rho_{f_{i}}(\mathbf{R},z=f_{i}(\mathbf{R})) & = & \bar{\rho}_{i}\,.\label{eq:rho_f_prop}
\end{eqnarray}
The indices $c_{i}$ and $f_{i}$ indicate that these functions of $z$ only and of $\mathbf{r}=(\mathbf{R},z)$ take the constant 
value $\bar{\rho}_{i}$ at $z=c_{i}$ and at $z=f_{i}(\mathbf{R})$, respectively.
Reasonable choices for $\bar{\rho}_{i}$ would be $\bar{\rho}_{i}:=(\rho_{i}^{-}+\rho_{i}^{+})/2$
or the analogue of the Gibbs dividing surface concept in the
one-component fluids (see also Fig.\,\ref{cap:Sketch_rho_c}); however,
our results do not depend explicitly on the choices of $\bar{\rho}_{i}$.
Finally we introduce the density differences \begin{equation}\triangle\rho_{i}:=\rho_{i}^{-}-\rho_{i}^{+}\,\label{eq:density_diff}.
\end{equation}
\\
In the following we choose  $\rho_{c_{i}}(z)$
and $\rho_{f_{i}}(\mathbf{r})$ such that they minimize Eq.\,(\ref{eq:GranCanDF})
under the constraint given by \Eq{eq:rho_c_prop} and \Eq{eq:rho_f_prop}, respectively:
\begin{equation}
\Bigg{(}\frac{\delta\Omega[\rho_{1}(\mathbf{r}),\rho_{2}(\mathbf{r})]}{\delta\rho_{i}(\mathbf{r})}\Bigg{)}_{c,f}=0\,,\;i\in\{1,2\}\,,\label{eq:eq_condition_1}\end{equation}
where $\Big{(}\delta\Omega[\rho_{1},\rho_{2}]/\delta\rho_{i}(\mathbf{r})\Big{)}_{c,f}$
denotes the functional derivative of $\Omega$ w.r.t the density $\rho_{i}$,
under the constraint $c$ (see \Eq{eq:rho_c_prop}) or $f$ (see \Eq{eq:rho_f_prop}), respectively.\\
Within density functional theory, $\rho_{c_{i}}(z)$ and $\rho_{f_{i}}(\mathbf{r})$ are equilibrium density profiles
in the sense described before. Inter alia, \Eq{eq:eq_condition_1} will allow us to eliminate the explicit dependences on the
chemical potentials $\mu_{i}$ in our analytic expressions; for this purpose it is sufficient to use \Eq{eq:eq_condition_1} only
for the profiles $\rho_{c_{i}}(z)$.
Without constraint Eqs. (\ref{eq:GranCanDF}) and (\ref{eq:hs_functional}) lead to
\begin{eqnarray}
\frac{\delta\Omega[\rho_{1}(\mathbf{r}),\rho_{2}(\mathbf{r})]}{\delta\rho_{i}(\mathbf{r})} & = & \partial_{\rho_{i}}h\big{(}\rho_{1}(\mathbf{r}),\rho_{2}(\mathbf{r})\big{)}+\mu_{i}+V_{i}^{\mathsf{ex}}(\mathbf{r})\nonumber \\
 &  & \hs{-3}+\,\sum_{j=1}^{2}\int\lts_{V}\! d^{3}\! s\;\; w_{ji}\big{(}|\mathbf{r}-\mathbf{s}|\big{)}\rho_{j}(\mathbf{s})\,.\label{eq:eq_condition_2}
 \end{eqnarray}
Up to here there is no construction scheme provided  for determining $\rho_{c_{1}}$ and $\rho_{c_{2}}$. 
One can take solutions of Eq.\,(\ref{eq:eq_condition_2}) for $V_{i}^{\mathsf{ex}}\equiv 0$ and  shift the {\em{pair}} such that, e.g., 
the condition $\rho_{c_{1}}(z=c_{1})=\bar{\rho}_{1}$ 
is fulfilled (see \Eq{eq:rho_c_prop}), but in general $\rho_{c_2}(z)$ will not have the property
$\rho_{c_{2}}(z=c_{2})=\bar{\rho}_{2}$. This shows that there is only one degree of freedom in shifting, 
i.e., $c_{1}-c_{2}$. Thus, $c_{2}$ is not a free parameter but depends on $c_{1}$, which means that $\Omega$ cannot
be minimized for arbitrary pairs $(c_{1},c_{2})$.
As a consequence the effective interface Hamiltonian $\mathcal{H}$ depends only on the difference $c_{1}-c_{2}$, 
but we shall treat $c_{2}$ formally as a free parameter, 
which indicates the position of the planar interface of $\rho_{c_{2}}(z)$.

The free energy density 
$h\big{(}\rho_{1},\rho_{2}\big{)}=h^{\mathsf{id}}\big{(}\rho_{1},\rho_{2}\big{)}+h^{\mathsf{ex}}\big{(}\rho_{1},\rho_{2}\big{)}$
of the hard sphere part consists of an ideal gas contribution
$h^{\mathsf{id}}\big{(}\rho_{1},\rho_{2}\big{)}$
and an excess part $h^{\mathsf{ex}}\big{(}\rho_{1},\rho_{2}\big{)}$.
Our analytical formulae derived below will not depend on its functional form; for numerical calculations, however,
the Carnahan-Starling expression $h^{\mathsf{CS}}\big{(}\rho_{1},\rho_{2}\big{)}$
for the excess contribution is used \cite{Carnahan:635(1969)}. With ($\beta=(k_{B}T)^{-1}$ and the thermal de Broglie wavelength  $\lambda_{\mathsf{th}}$,
this means
\begin{equation}
h^{\mathsf{id}}(\rho_{1},\rho_{2})=\beta^{-1}\,\sum_{i=1}^{2}\rho_{i}\,\big{(}\ln(\lambda_{\mathsf{th},i}^{3}\,\rho_{i})-1\big{)}\label{eq:h_ideales_Gas}\end{equation}
and\begin{eqnarray}
h^{\mathsf{CS}}(\rho_{1},\rho_{2}) & = & -\, n_{o}\,\ln\big{(}1-n_{3}\big{)}+\frac{n_{1}\, n_{2}}{1-n_{3}}\nonumber \\
 &  & \hs{-3}+\, n_{2}^{3}\;\frac{n_{3}+(1-n_{3})^{2}\,\ln\big{(}1-n_{3}\big{)}}{36\pi n_{3}^{2}(1-n_{3})^{2}}\,,\label{eq:Carnahan_Starling}\end{eqnarray}
where\begin{eqnarray}
\begin{array}{ccc}
n_{o} & = & \ds{\sum_{k=1}^{2}\rho_{k}}\\
\\n_{2} & = & \ds{4\pi\sum_{k=1}^{2}\big{(}r_{o}^{(k)}\big)^{2}\rho_{k}}\end{array} &  & \begin{array}{ccc}
n_{1} & = & \ds{\sum_{k=1}^{2}r_{o}^{(k)}\rho_{k}}\\
\\n_{3} & = & \ds{\frac{4\pi}{3}\sum_{k=1}^{2}\big{(}r_{o}^{(k)}\big)^{3}\rho_{k}}\,.\end{array}\label{eq:weighted_desities}\end{eqnarray}
The weighted densities $n_{\alpha}$, $\alpha\in\{0,1,2,3\}$,
are composed of the densities $\rho_{i}$ and the particle
radii $r_{o}^{(i)}$ of species $i\in\{1,2\}$.\\
In order to model the van der Waals forces of simple fluids we take
for the attractive part of the interactions
\begin{equation}
w_{ij}(R,z)=-\frac{w_{o}^{(ij)}\,(r_{o}^{(ij)})^{6}}{\big{(}(r_{o}^{(ij)})^{2}+R^{2}+z^{2}\big)^{3}}\,,\; i,j\in\{1,2\}\;,\label{eq:w_ij}\end{equation}
which gives the correct large distance behavior $w_{ij}(r\gg r_{o}^{(ij)})=-\, w_{o}^{(ij)}r^{-6}$. 
The quantity $w_{o}^{(ij)}$ represents the depth of the potential,
while $r_{o}^{(ij)}=r_{o}^{(i)}+r_{o}^{(j)}$ is defined as the sum
of the particle radii. The functional form of $w_{ij}(r)$ for small $r$ is chosen for analytic convenience; most of our
results do not depend on this choice.\\

Independent of the explicit form of the potentials $w_{ij}$ we introduce their integrals
\setformulalength
\begin{eqnarray}
w_{ij}^{(1)}(|\mathbf{R}|,z) & :=&\int\lts_{\delta c_{ij}}^{z}\! d\bar{z}\,\, w_{ij}(|\mathbf{R}|,|\bar{z}|)\hs{10}\label{eq:Def_w^(1)}\\
\textrm{\parbox{0.3\formulalength}{and \hfill}}& &\textrm{\parbox{0.7\formulalength}{ \hfill}} \,\nonumber\\
w_{ij}^{(2)}(|\mathbf{R}|,z) & :=&\int\lts_{\delta c_{ij}}^{z}\! d\bar{z}\,\, w_{ij}^{(1)}(|\mathbf{R}|,|\bar{z}|)\label{eq:Def_w^(2)}
\end{eqnarray}
where $\delta c_{ij}:=c_{i}-c_{j}$.
They fulfill the symmetry relations\begin{equation}
w_{ij}^{(k)}(|\mathbf{R}|,z)=(-1)^{k}w_{ji}^{(k)}(|\mathbf{R}|,-z)\,,\; k\in\{1,2\}\,.\label{eq:w^k_sym_prop}\end{equation}
In the present context the binary liquid mixture is exposed to a gravitational
field acting along the $z$ direction:\begin{eqnarray}
V_{i}^{\mathsf{ex}}(z) & = & G\, m_{i}\,(z-c_{i})\,,\label{eq:Ext_Pot}\end{eqnarray}
where $G$ is the acceleration of gravity, $m_{i}$ is the particle
mass and $c_{i}$ the equilibrium flat interface position of species
$i$ (see Eq.\,(\ref{eq:rho_c_prop})). In the following
the first integral of $V_{i}^{\mathsf{ex}}(z)$ is frequently used:
\begin{eqnarray}
V_{i}^{(1)}(z) & := & \frac{1}{2}\, G\, m_{i}\,(z-c_{i})^{2}\,.\label{eq:Ext_Pot_(1)}\end{eqnarray}

In the following subsection the effective interface
Hamiltonian will be defined on the basis of the density
functional $\Omega$ introduced above.

\subsection{Effective Interface Hamiltonian\label{sub:Eff_Interf_Hamiltonian}}

The envisaged effective interface Hamiltonian $\mathcal{H}$ for a binary liquid mixture
provides the cost in free energy to maintain interface configurations described by the
iso-density contours $f_{i}(\mathbf{R})$, $i\in{1,2}$, relative to flat configurations $\rho_{c_{i}}(z)$.
We expect that $\mathcal{H}$ depends on the differences $f_{i}^{c}(\mathbf{R}):=f_{i}(\mathbf{R})-c_{i}$.
Therefore, we introduce the abbreviations\begin{eqnarray}
\mathbf{f}(\mathbf{R},\mathbf{R}') & := & \left(\begin{array}{c}
f_{1}^{c}(\mathbf{R})\\
f_{2}^{c}(\mathbf{R}')\end{array}\right)\,,\label{eq:Def_f_vector}\end{eqnarray}
$\mathbf{f}(\mathbf{R}):=\mathbf{f}(\mathbf{R},\mathbf{R})$,
$\delta f_{ij}^{c}(\mathbf{R},\mathbf{R}'):=f_{i}^{c}(\mathbf{R})-f_{j}^{c}(\mathbf{R}')$,
$\delta f_{ij}(\mathbf{R},\mathbf{R}'):=f_{i}(\mathbf{R})-f_{j}(\mathbf{R}')$, and
$\delta f_{ij}(\mathbf{R}):=f_{i}(\mathbf{R})-f_{j}(\mathbf{R})$.
In terms of these quantities, the effective interface Hamiltonian $\mathcal{H}$
is defined as the difference of the corresponding grand canonical
potentials:\begin{equation}
\mathcal{H}[\mathbf{f}(\mathbf{R},\mathbf{R}')]:=\Omega[\rho_{f_{1}}(\mathbf{r}),\rho_{f_{2}}(\mathbf{r}')]-\Omega[\rho_{c_{1}}(z),\rho_{c_{2}}(z')]\,.\label{eq:Def_Hamiltonian}\end{equation}
Our main goal is to derive an explicit expression of
$\mathcal{H}$ in terms of $f_{i}^{c}(\mathbf{R})$. \\
We rewrite $\mathcal{H}$ by carrying out partial integrations such that $\mathcal{H}$ is expressed mostly in terms
of derivatives of profiles which are mainly confined to the interfacial region and vanish for $z\rightarrow\pm L$.
Due to Eq.\,(\ref{eq:Def_rho_pm}) one has $|f_{i}^{c}(\mathbf{R})|\ll L$ for all $\mathbf{R}$, i.e.,
the interface deviations are much smaller than the sample size.
According to the structure of $\Omega$ in \Eq{eq:GranCanDF}, $\mathcal{H}$ is the sum of four terms.
The first term, which we shall treat later in  Subsec.\,\ref{sub:MeanSurfaceApprox} is given directly as
\begin{equation}
\mathcal{H}_{h}\big{(}\mathbf{f}(\mathbf{R})\big{)}:=\int\lts_{-\zlimit}^{\zlimit}\!\! dz\,\Big{[}h\big{(}\rho_{f_{1}}(\mathbf{r}),\rho_{f_{2}}(\mathbf{r})\big{)}-h\big{(}\rho_{c_{1}}(z),\rho_{c_{2}}(z)\big{)}\Big{]}\,.\label{eq:Def_H_h}\end{equation}
The second expression stems from the external potential $V_{i}^{\mathsf{ex}}(z)$
and has the form\begin{equation}
\mathcal{H}_{V}\big{(}\mathbf{f}(\mathbf{R})\big{)}:=-\sum_{i=1}^{2}\int\lts_{-\zlimit}^{\zlimit}\!\! dz\, V_{i}^{(1)}(z)\,\big{[}\partial_{z}\rho_{f_{i}}(\mathbf{r})-\partial_{z}\rho_{c_{i}}(z)\big{]}\,.\label{eq:Def_H_V}\end{equation}
The third contribution involves the chemical potentials $\mu_{i}$
and additional boundary contributions which arise from the interaction potentials.
With the constants\begin{eqnarray}
K_{k} & := & \mu_{k}+4\pi\sum_{j=1}^{2}\,\,\int\lts_{A}\!\! d^{2}R\,\big{[}\bar{\rho}_{j}\, w_{jk}^{(1)}(|\mathbf{R}|,z=\zlimit)\nonumber \\
 &  & \hs{23}-\rho_{j}^{-}\, w_{jk}^{(1)}(|\mathbf{R}|,z=0)\big{]}\label{eq:Def_K_Konstanten}\end{eqnarray}
it reads\begin{equation}
\mathcal{H}_{b}\big{(}\mathbf{f}(\mathbf{R})\big{)}:=-\sum_{i=1}^{2}\,K_{i}\,\int\lts_{-\zlimit}^{\zlimit}\!\! dz\, (z-c_{i})\,\big{[}\partial_{z}\rho_{f_{i}}(\mathbf{r})-\partial_{z}\rho_{c_{i}}(z)\big{]}\,.\label{eq:Def_H_b}\end{equation}
Finally, the contribution to $\mathcal{H}$ due to the attractive part of the interactions can be expressed as
\begin{eqnarray}
\mathcal{H}_{w}\big{(}\mathbf{f}(\mathbf{R},\mathbf{R}')\big{)} & := & -\frac{1}{2}\sum_{i,\, j=1}^{2}\int\lts_{-\zlimit}^{\zlimit}\!\! dz\!\int\lts_{-\zlimit}^{\zlimit}\!\! dz'\; w_{ij}^{(2)}(|\delta\mathbf{R}|,\delta z)\times\nonumber \\
 &  & \hs{-15}\times\Big{[}\partial_{z}\rho_{f_{i}}(\mathbf{r})\,\partial_{z'}\rho_{f_{j}}(\mathbf{r}')-\partial_{z}\rho_{c_{i}}(z)\,\partial_{z'}\rho_{c_{j}}(z')\Big{]}\,\label{eq:Def_H_w}\end{eqnarray}
where $\delta\mathbf{R}:=\mathbf{R}-\mathbf{R}'$ and $\delta z:=z-z'$.
Thus, Eqs.\,(\ref{eq:Def_Hamiltonian})-(\ref{eq:Def_H_w}) lead to
\begin{eqnarray}
\mathcal{H}[\mathbf{f}(\mathbf{R})] & = & \int\lts_{A}\!\! d^{2}R\,\Big{[}\mathcal{H}_{h}\big{(}\mathbf{f}(\mathbf{R})\big{)}+\mathcal{H}_{b}\big{(}\mathbf{f}(\mathbf{R})\big{)}+\mathcal{H}_{V}\big{(}\mathbf{f}(\mathbf{R})\big{)}\Big{]}\nonumber \\
 &  & +\,\dint{A}\!\! d^{2}R\, d^{2}R'\;\;\mathcal{H}_{w}\big{(}\mathbf{f}(\mathbf{R},\mathbf{R}')\big{)}\,.\label{eq:H_part_int}\end{eqnarray}
In the following two subsections  we analyze
two different models for the profiles $\rho_{c_{i}}(z)$ and $\rho_{f_{i}}(\mathbf{r})$  in order to obtain analytic results
for $\mathcal{H}$. The first approach assumes that at the interface position the densities vary discontinuously
between the corresponding bulk values. This so-called  sharp kink approximation
will be discussed in Subsec.\,\ref{sub:SKAP_1}. The second approach (Subsec.\,\ref{sub:CUREX_1})  is based
on continuous density profiles and takes the influence of the curvature of the iso-density contours on the densities
into account.\\
The validity of our approach is also based on the assumption that in the thermodynamic limit, i.e., $A\rightarrow \infty$, 
all lateral boundary contributions to $\mathcal{H}$  vanish in \Eq{eq:H_part_int}.

\subsection{Sharp Kink Approximation\label{sub:SKAP_1}}

The sharp kink approximation replaces the actual smooth variations of the density profiles (see Fig.\,\ref{cap:Sketch_rho_c}) 
on the scale of the bulk correlation length by step functions:
\begin{eqnarray}
\rho_{f_{i}}(\mathbf{r}) & := & -\triangle\rho_{i}\,\Theta(z-f_{i}(\mathbf{R}))+\rho_{i}^{-}\label{eq:sk_profile}\end{eqnarray}
where $\Theta(x)$ is the Heaviside function so that
\begin{eqnarray}
\partial_{z}\rho_{f_{i}}(\mathbf{r}) & = & -\triangle\rho_{i}\,\delta(z-f_{i}(\mathbf{R}))\,.\label{eq:sk_delta}\end{eqnarray}
Similar expressions hold for $\rho_{c_{i}}(z)$ with
$f_{i}(\mathbf{R})$ replaced by $c_{i}$.
For one-component fluids this approximation has turned out to be surprisingly successful in describing liquid-vapor
interfaces
\cite{Napiorkowski:1836(1993)} 
 and wetting phenomena \cite{Dietrich:1861(1991)}.
From Eqs.\,(\ref{eq:H_part_int}) and Eq.\,(\ref{eq:sk_delta}) together with the expansion
(see Eqs.\,(\ref{eq:Def_w^(2)}), (\ref{eq:w^k_sym_prop}), and (\ref{eq:Def_f_vector}))\begin{equation}
w_{ij}^{(2)}(|\delta\mathbf{R}|,\delta f_{ij}(\mathbf{R},\mathbf{R}'))\approx\frac{1}{2}\, w_{ij}(|\delta\mathbf{R}|,\delta c_{ij})\,\left[\delta f_{ij}^{c}(\mathbf{R},\mathbf{R}')\right]^{2}\label{eq:w^(2)-expansion_sk}\end{equation}
we find\begin{eqnarray}
 &  & \mathcal{H}^{\mathsf{sk}}[\mathbf{f}(\mathbf{R},\mathbf{R}')]\nonumber \\
 & \approx & \int\lts_{A}\!\! d^{2}R\,\,\sum_{i=1}^{2}\tr\rho_{i}\,\bigg{[}\frac{G}{2}\, m_{i}\,\left[f_{i}^{c}(\mathbf{R})\right]^{2}+\, f_{i}(\mathbf{R})\,\times\nonumber \\
 &  & \times\Big{(}\partial_{\rho_{i}}h\big{(}\rho_{f_{1}},\rho_{f_{2}}\big{)}\Big|_{z=f_{i}(\mathbf{R})}-\partial_{\rho_{i}}h\big{(}\rho_{c_{1}},\rho_{c_{2}}\big{)}\Big|_{z=c_{i}}\Big{)}\,\bigg{]}\nonumber \\
 &  & -\,\dint{A}\!\! d^{2}R\, d^{2}R'\,\sum_{i,j=1}^{2}\frac{\tr\rho_{i}\,\tr\rho_{j}}{4}\,\times\nonumber \\
 &  & \hs{20}\times w_{ij}(|\delta\mathbf{R}|,\delta c_{ij})\left[\delta f_{ij}^{c}(\mathbf{R},\mathbf{R}')\right]^{2}\,.\label{eq:H_skap}\end{eqnarray}
In Eq.\,(\ref{eq:H_skap}) the expressions 
$\partial_{\rho_{j}}h|_{z=f_{i}}-\partial_{\rho_{j}}h|_{z=c_{i}}$
are basically not determinable because at least one density has to be
evaluated at the interface position of the second which is unknown.
For instance, in order to evaluate $\rho_{f_{2}}(\mathbf{R},f_{1}(\mathbf{R}))$ 
in the case $c_{1}=c_{2}=0$
one would need the information whether $f_{1}(\mathbf{R})>f_{2}(\mathbf{R})$
or $f_{1}(\mathbf{R})<f_{2}(\mathbf{R})$: the first case
yields $\rho_{f_{2}}(\mathbf{R},f_{1}(\mathbf{R}))=\rho_{2}^{+}$,
the second gives $\rho_{f_{2}}(\mathbf{R},f_{1}(\mathbf{R}))=\rho_{2}^{-}$.
As a consequence, the expressions depend on the differences $f_{2}(\mathbf{R})-f_{1}(\mathbf{R})$ and $c_{1}-c_{2}$
and even vanish for $f_{2}(\mathbf{R})-f_{1}(\mathbf{R})=0$ and $c_{1}-c_{2}=0$, because each density is evaluated at its isodensity
surface resulting in the same value (see Eqs. (\ref{eq:rho_c_prop}), (\ref{eq:rho_f_prop})).
Using the expansion\begin{eqnarray}
 &  & \partial_{\rho_{i}}h\big{(}\rho_{f_{1}},\rho_{f_{2}}\big{)}\Big|_{z=f_{i}(\mathbf{R})}-\partial_{\rho_{i}}h\big{(}\bar{\rho}_{1},\bar{\rho}_{2}\big{)}\nonumber \\
 & \approx & \sum_{k=1}^{2}\partial_{\rho_{k}}\partial_{\rho_{i}}h\big{(}\bar{\rho}_{1},\bar{\rho}_{2}\big{)}\,\Big{[}\rho_{f_{k}}\big|_{z=f_{i}(\mathbf{R})}-\bar{\rho}_{k}\Big{]}\label{eq:h_sk_approx}\end{eqnarray}
and similarly for $\partial_{\rho_{i}}h\big{(}\rho_{c_{1}},\rho_{c_{2}}\big{)}\Big|_{z=c_{i}}$
leads to\begin{eqnarray}
 &  & \mathcal{H}^{\textrm{sk}}[\mathbf{f}(\mathbf{R},\mathbf{R}')]\nonumber \\
 & \approx & \,\sum_{i=1}^{2}\tr\rho_{i}\,\frac{G}{2}\, m_{i}\,\int\lts_{A}\!\! d^{2}R\,\left[f_{i}^{c}(\mathbf{R})\right]^{2}\label{eq:H_skap_final}\\
 &  & +\,\sum_{i,j=1}^{2}\frac{\tr\rho_{i}\tr\rho_{j}}{4}\,\,\partial_{\rho_{j}\rho_{i}}^{2}h(\bar{\rho}_{1},\bar{\rho}_{2})\,\times\nonumber \\
 &  & \times\bigg{[}\,\int\lts_{A}\!\! d^{2}R\,\,|\delta f_{ij}(\mathbf{R})|\,\,2\,\Theta\big{(}-\delta c_{ij}\delta f_{ij}(\mathbf{R})\big{)}\nonumber \\
 &  & -\,\dint{A}\!\! d^{2}R\, d^{2}R'\, w_{ij}(|\delta\mathbf{R}|,\delta c_{ij})\left[\delta f_{ij}^{c}(\mathbf{R},\mathbf{R}')\right]^{2}\bigg{]}\,.\nonumber \end{eqnarray}
Note, that the Heaviside function vanishes if $\delta c_{ij}$
and $\delta f_{ij}$ have the same signs. Its prefactor
$|f_{1}(\mathbf{R})-f_{2}(\mathbf{R})|$ prevents an appropriate Fourier analysis
because the resulting expressions cannot be ordered in products
of $\hat{f}_{i}\hat{f}_{j}$, where $\hat{f}$ denotes the Fourier
transformed function of $f$ (see Eq.\,(\ref{eq:Def_Fourier_1})). Therefore,
within this sharp kink approximation, the cost in free energy for deforming the
interface can be studied only  for the case $f_{1}\equiv f_{2}$
but not for the more general situation $f_{1}\neq f_{2}$.\\
For $f_{1}(\mathbf{R})=f_{2}(\mathbf{R})\equiv f(\mathbf{R})$
and $c_{1}=c_{2}=0$ the aforementioned problematic
expressions in Eq.\,(\ref{eq:H_skap}) drops out. With the Fourier transformation
\begin{eqnarray}
\hat{f}(\mathbf{q}) & := & \frac{1}{(2\pi)^{n/2}}\,\int\lts_{\mathbb{R}^{n}}\! d^{n}R\, f(\mathbf{R})\, e^{-\, i\mathbf{qR}}\,,\label{eq:Def_Fourier_1}\\
f(\mathbf{R}) & = & \frac{1}{(2\pi)^{n/2}}\,\int\lts_{\mathbb{R}^{n}}\! d^{n}q\,\hat{f}(\mathbf{q})\, e^{+\, i\mathbf{qR}}\,,\label{eq:Def_Fourier_2}\end{eqnarray}
and\begin{eqnarray}
\hat{w}_{ij}(\mathbf{q},z) & = & \frac{1}{2\pi}\int\lts_{\mathbb{R}^{2}}\!\! d^{2}R\; w_{ij}(|\mathbf{R}|,u)\, e^{-i\mathbf{qR}}\nonumber \\
 & = & \int\lts_{0}^{\infty}\!\! dR\; R\, J_{o}(qR)\, w_{ij}(R,z)\label{eq:w_Fourier_general_form}\end{eqnarray}
where $J_{o}(x):=\frac{1}{\pi}\int_{0}^{\pi}\, e^{-ix\,\cos{t}}dt$
is the zeroth order Bessel function, $\mathcal{H}^{\mathsf{sk}}$ can be expressed as
\begin{equation}
\mathcal{H}^{\textrm{sk}}[\hat{f}(\mathbf{q})]=\frac{1}{4\pi}\int\lts_{\mb R^{2}}\!\! d^{2}q\,\,|\hat{f}(\mathbf{q})|^{2}\,\,\Big{[}G\,\mathcal{G}^{\mathsf{sk}}+q^{2}\,\gamma^{\,\mathsf{sk}}(q)\Big{]}\,,\label{eq:H_skap_fourier}\end{equation}
with $\mathcal{G}^{\mathsf{sk}}:=\sum_{j=1}^{2}\tr\rho_{j}\, m_{j}$
and a wavelength-dependent surface tension\begin{equation}
\gamma^{\,\mathsf{sk}}(q):=\frac{1}{q^{2}}\,\sum_{i,j=1}^{2}\tr\rho_{i}\,\tr\rho_{j}\,\big{[}\hat{w}_{ij}(|\mathbf{q}|,0)-\hat{w}_{ij}(0,0)\big{]}\,.\label{eq:Def_gamma_skap}\end{equation}
Equation\,(\ref{eq:Def_gamma_skap}) is the generalization of the
corresponding result for a one-component fluid
\cite{Napiorkowski:1836(1993),Mecke:6766(1999)} 
assuming a \emph{single steplike} interface in the binary case.
For fluids governed by dispersion forces  (Eq.\,(\ref{eq:w_ij}))
one obtains in the limit of long wavelengths $1/q$
\begin{eqnarray}
\gamma^{\,\mathsf{sk}}(q\rightarrow0) & = & \frac{1}{16}\sum_{i,j=1}^{2}\tr\rho_{i}\,\tr\rho_{j}\, w_{o}^{(ij)}\,(r_{o}^{(ij)})^{4}\times\label{eq:gamma_sk_lim_0}\\
 &  &\!\!\!\!\!\!\!\!\!\!\!\!\!\!\!\!\!\! \Big{(}1+\,\frac{1}{4}q^{2}\,(r_{o}^{(ij)})^{2}\,\Big{[}\log(\frac{q\, r_{o}^{(ij)}}{2})+C\Big{]}\Big{)}+\mathcal{O}(q^{4})\nonumber \end{eqnarray}
with Euler's constant $\Euler=0.5772\ldots$ and $C=\Euler-\frac{3}{4}$;
\begin{equation}
\gamma_{o}^{\,\mathsf{sk}}\,:=\,\gamma^{\,\mathsf{sk}}(q=0)=\frac{1}{16}\sum_{i,j=1}^{2}\tr\rho_{i}\,\tr\rho_{j}\, w_{o}^{(ij)}\,(r_{o}^{(ij)})^{4}\label{eq:gamma_sk_0}\end{equation}
is the macroscopic surface tension within the sharp kink approximation.
At short wavelengths, i.e., $q\rightarrow\infty$, one finds
$\gamma^{\,\mathsf{sk}}(q\rightarrow\infty)\rightarrow0$,
which means that distortions with short  wavelengths are
insufficiently suppressed.

While the previous calculations are based on intrinsic steplike density profiles,
in the following subsection we consider the more realistic case of smoothly varying
intrinsic profiles, including changes of their shape due to local curvatures of their interfaces.

\subsection{Curvature Expansion\label{sub:CUREX_1}}

In this subsection we consider continuous density profiles $\rho_{c_{i}}(z)$
(see Fig.\,\ref{cap:Sketch_rho_c}) and $\rho_{f_{i}}(\mathbf{r})$.
The thickness of the transition region or  the width of the
interface is of the order of the bulk correlation length $\xi$.

In order to take the influence of local curvatures on the density
profile $\rho_{f_{i}}$ into account, first we introduce normal
coordinates for each surface $f_{i}(\mathbf{R})$ followed by a transformation
of the density $\rho_{f_{i}}(\mathbf{r})$ to  its normal coordinate
system. Second, the transformed density is expanded in powers of
local curvatures. %
\begin{figure}[t]
\centering\foreignlanguage{english}{\includegraphics[%
  scale=0.32]{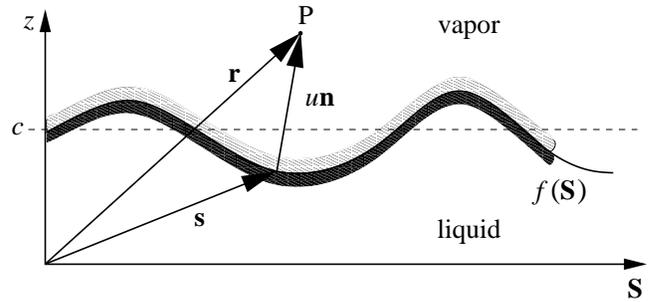}}

\caption{\label{cap:Sketch_normal_coord}Sketch of an interface $f(\mathbf{S})$, $\mathbf{S}\in \mathbb{R}^{2}$,
and its normal coordinate system (NCS) consisting of a point $\mathbf{s}\in \mathbb{R}^{3}$ on $f$, 
its normal vector $\mathbf{n}$, and the normal distance
$u$. Thus each point $\textrm{P}$ has two representations, either
as a vector $\mathbf{r}$ or as $\mathbf{s}+u\,\mathbf{n}$ within the NCS
(see Eq.\,(\ref{eq:Def_Normal_Trafo_general})).
Additional assumptions are required to assure the unique
equivalence of both (see main text). In the present picture the condition
$|u|<\ibfu$ for all $u$ is violated because the  normal distance of the point $P$ is larger than
the radius of curvature of the right bump.}
\end{figure}
 (For the following general remarks we omit the index $i$.)

To this end we consider the points $\mathbf{s}(\mathbf{S})=\big{(}\mathbf{S},f(\mathbf{S})\big)^{t}$
of the Monge parameterized surface $f(\mathbf{S})$, the normal vector
$\mathbf{n}(\mathbf{S})$,  and the map $\mathcal{T}:\mathbb{R}^{2}\times\mathbb{R}\rightarrow\mathbb{R}^{3}$
(see Fig.\,\ref{cap:Sketch_normal_coord}) so that
\begin{equation}
\mathcal{T}(\mathbf{S},u):=\mathbf{s}(\mathbf{S})+u\,\mathbf{n}(\mathbf{S})\,.\label{eq:Def_Normal_Trafo_general}\end{equation}
Thus, each spatial point $\mathbf{r}$ can be expressed in terms
of a point on the surface $\mathbf{s}$ and its normal distance $u$ from the surface.
However, finding $\mathbf{S}$ and $u$ for a given point $\mathbf{r}$,
i.e., finding the solution of the equation $\mathbf{r}=\mathcal{T}\big{(}\mathbf{S}(\mathbf{r}),u(\mathbf{r})\big{)}$,
is generally not a trivial task. However, in order to obtain  a  unique map $\mathcal{T}^{-1}$
we have to restrict the range of values of $u$ to $(-\ibfu,\ibfu)$
where $\ibfu>0$ denotes the absolute value of the minimal
radius of curvature of the manifold $f(\mathbf{S})$. Therefore, the constraint
$|u|<\ibfu$ guarantees, that the Jacobian $J_{f}$ of the transformation
$\mathcal{T}$,\begin{eqnarray}
J_{f}(\mathbf{S},u) & = & \sqrt{g(\mathbf{S})}\,(1-2H(\mathbf{S})\, u+K(\mathbf{S})\, u^{2})\,,\label{eq:Jacbobian}\end{eqnarray}
does not vanish in the  domain $\mathbb{R}^{2}\times(-\ibfu,\ibfu)$.
Here, $H\equiv H(\mathbf{S})$ is the local mean curvature, $K\equiv K(\mathbf{S})$
is the local Gaussian curvature, and $g\equiv g(\mathbf{S})$ is the
metric of the manifold $f(\mathbf{S})$ which in Monge representation takes
the form $g(\mathbf{S})=1+\big{(}\nabla f(\mathbf{S})\big)^{2}$.
\\
\\
If $f(\mathbf{S})$ is an iso-density contour of $\rho_{f}(\mathbf{r})$,
we may write for points $(\mathbf{S},u)\in\mathbb{R}^{2}\times(-\ibfu,\ibfu)$
\begin{equation}
\rho_{f}(\mathbf{r})=\tilde{\rho}_{f}(\mathbf{S},u)\,.\label{eq:Transformed_desity}\end{equation}
These expressions hold also for a flat surface which results in $\rho_{c}(z)=\tilde{\rho}_{c}(u)$
with $u=z-c$. \\
\\
Similar to Eq.\,(\ref{eq:Def_rho_pm}) we assume \begin{equation}
\tilde{\rho}_{f_{i}}(\mathbf{S},u\rightarrow\pm\ibfu)=\rho_{i}^{\pm}\,,\; i\in\{1,2\}\,.\label{eq:rho_u_limit_pm}\end{equation}
Since \Eq{eq:rho_u_limit_pm} can be strictly valid only for macroscopicly large values of $u$,
we assume that $\ibfu$ is sufficiently large so that \Eq{eq:rho_u_limit_pm}  is fulfilled for all
practical purposes. Now, we propose an expansion of the transformed density profile
$\tilde{\rho}_{f_{i}}(\mathbf{S},u)$
into powers of the local curvatures $H_{i}\equiv H_{i}(\mathbf{S})$
and $K_{i}\equiv K_{i}(\mathbf{S})$, $i\in\{1,2\}$:
\begin{subequations}
\begin{eqnarray}
\tilde{\rho}_{f_{i}}(\mathbf{S},u) & = & \tilde{\rho}_{c_{i}}(u)+\delta\rho_{f_{i}}(\mathbf{S},u)\label{eq:rho_pertubation_general}\\
\delta\rho_{f_{i}}(\mathbf{S},u) & \approx &\sum_{{{\alpha,\beta=0\atop \alpha+\beta\geq1}}}(2H_{i})^{\alpha}\, K_{i}^{\beta}\,\rho_{H_{i}^{\alpha}K_{i}^{\beta}}(u)\;.\label{eq:Def_curex}
 \end{eqnarray}
 \end{subequations}
For each term this implies a factorization of the dependencies on
the lateral coordinates $\mathbf{S}$
and the normal distance $u$, reflecting the condition that the width
of the interface $\xi$ should be small compared with the minimal radius of curvature,
i.e., $\xi\ll\ibfu$. For the following calculations, it is not necessary
to specify the  functions $\rho_{\lambda}(u)$, $\lambda\in\{ H,K,H^{2},HK,\ldots\}$,
which depend only on the normal distance $u$ but are so far unknown explicitly.
However, for quantitative predictions one has to use a model for  $\rho_{\lambda}(u)$ (see, c.f.,  
Subsec.\,\ref{sub:General-Expression_H}).

\subsection{Mean Surface Approximation\label{sub:MeanSurfaceApprox}}

\begin{figure}[t]
\centering\foreignlanguage{english}{\includegraphics[%
  scale=0.32]{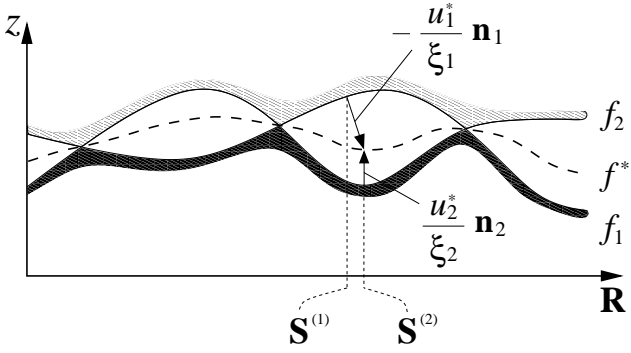}}

\caption{\label{cap:Sketch_Mean_Surface}Sketch of the construction scheme for
the mean surface points $\mathbf{r}^{*}$. Since the points $\mathbf{r}^{*}$
represent an iso-density manifold $f^{*}$ of the mean density $\md(\mathbf{r})$
they are determined in terms of the distances $u_{i}^{*}$ between
the surfaces $f_{i}$ and $\mathbf{r}^{*}$ (see Eq.\,(\ref{eq:implicit_mean_surface}))
assuming that, normal to each surface $f_{i}$, the density profile
has the form given by Eq.\,(\ref{eq:form_of_planar_rho}). }
\end{figure}

Except for $\mathcal{H}_{h}$ (see Eq.\,(\ref{eq:Def_H_h})) the formulas derived above
can be used to transform and to expand the various contributions  of the Hamiltonian $\mathcal{H}$.
For $\mathcal{H}_{h}$ both densities have to be evaluated
at the same spatial point, but there is no rule telling  which normal
coordinate set should be used for the transformation, i.e.,
which local curvatures have to be used. In order to resolve this issue
we construct a mean density distribution $\md(\mathbf{r})$ with a corresponding
iso-density contour $f^{*}(\mathbf{R})$, such that
$h\big{(}\rho_{1}(\mathbf{r}),\rho_{2}(\mathbf{r})\big{)}=h^{*}(\md(\mathbf{r}))$
where $h^{*}$ is a function to be determined, and use the normal coordinate
system associated with $f^{*}(\mathbf{R})$ in order to transform $\mathcal{H}_{h}$ and
to expand $h^{*}(\md(\mathbf{r}))$. Since we know the relation between
$f^{*}(\mathbf{R})$ and $f_{i}(\mathbf{R})$ explicitly, we are able
to express the results in terms of $f_{i}$.
We stress that this problem would equally arise
for more sophisticated density functionals beyond the local
density approximation used in Eq.\,(\ref{eq:hs_functional}). \\
We start our approach with an implicit definition of the  points $\mathbf{r}^{*}$
fulfilling  the constraint $\md(\mathbf{r}^{*})=\mathit{const}$.
First, we assume that the flat configuration can be written as (see Eqs.\,(\ref{eq:rho_c_prop})-(\ref{eq:density_diff}))
\begin{equation}
\tilde{\rho}_{c_{i}}(u)=-\,\frac{\triangle\rho_{i}}{2}\,\ip(u/\xi_{i})+\bar{\rho}_{i}\label{eq:form_of_planar_rho}\end{equation}
with an odd function $\ip(-\, x)=-\,\ip(x)$ so that $\ip(0)=0$. Second, we define, with
a not yet specified prefactor $\triangle\md$, the mean density
\begin{equation}
\md(\mathbf{r}):=\triangle\md\,\sum_{i=1}^{2}\,\frac{\tilde{\rho}_{f_{i}}\big{(}\mathbf{S}^{(i)}(\mathbf{r}),u_{i}(\mathbf{r})\big{)}}{\triangle\rho_{i}}\,\,.\label{eq:Def_mean_density}\end{equation}
The equation for the corresponding iso-density contour reads\begin{equation}
\md(\mathbf{r}^{*})=\overline{\md}\,:=\,\triangle\md\,\sum_{i=1}^{2}\frac{\bar{\rho}_{i}}{\triangle\rho_{i}}.\label{eq:mean_isodens_cond}\end{equation}
Using the expansion introduced in Eq.\,(\ref{eq:Def_curex}),  in lowest
order this leads to
\begin{equation}
\sum_{i=1}^{2}\,\frac{\tilde{\rho}_{c_{i}}\big{(}u_{i}(\mathbf{r}^{*})\big{)}}{\triangle\rho_{i}}=0\label{eq:Mean_surface_construction}\end{equation}
and hence to the condition (using Eq.\,(\ref{eq:form_of_planar_rho})) with $u_{i}^{*}\equiv u_{i}(\mathbf{r}^{*})$
\begin{equation}
\frac{u_{1}^{*}}{\xi_{1}}+\frac{u_{2}^{*}}{\xi_{2}}=0\,.\label{eq:implicit_mean_surface}\end{equation}
Thus, we postulate that the normal distances $u_{i}^{*}$ between
$f_{i}$ and a point $\mathbf{r}^{*}$ on the  iso-density manifold $f^{*}$,
measured in units of the width $\xi_i$ of the corresponding interface, are equal (see Fig.\,\ref{cap:Sketch_Mean_Surface});
this is a construction scheme for $f^{*}$.
In general, the lateral coordinates $\mathbf{S}^{(1)}$ and
$\mathbf{S}^{(2)}$ which belong to the same point $\mathbf{r}^{*}$
are different. We write $\mathbf{s}_{i}(\mathbf{r}^{*}):=\big{(}\mathbf{S}^{(i)}(\mathbf{r}^{*}),f_{i}(\mathbf{S}^{(i)}(\mathbf{r}^{*}))\big{)}\equiv\mathbf{s}_{i}^{*}$
for the corresponding point on the surface $f_{i}$ while $\mathbf{n}_{i}(\mathbf{r}^{*})\equiv\mathbf{n}_{i}^{*}$
denotes its normal vector there. Expressing $\mathbf{r}^{*}$ as
\begin{equation}
\mathbf{r}^{*}=a_{1}\,\mathbf{s}_{1}^{*}+a_{2}\,\mathbf{s}_{2}^{*}+\alpha_{1}\,\mathbf{n}_{1}^{*}+\alpha_{2}\,\mathbf{n}_{2}^{*}\label{eq:r_star_Ansatz}\end{equation}
with the coefficients $a_{i},\alpha_{i}\in\mathbb{R}$ to be determined.
In combination with Eq.\,(\ref{eq:implicit_mean_surface}) one obtains
\begin{equation}
0=\sum_{{{i,j=1\atop i\neq j}}}^{2}\,\mathbf{s}_{i}^{*}\,\Big{(}\frac{a_{i}-1}{\xi_{i}}\,\mathbf{n}_{i}^{*}+\frac{a_{i}}{\xi_{j}}\,\mathbf{n}_{j}^{*}\Big{)}+\frac{\zeta_{i}}{\xi_{1}\xi_{2}}\,,\label{eq:r_star_in_DefGl}\end{equation}
with $\alpha_{1}=\zeta_{1}/(\xi_{2}+\xi_{1}\,\mathbf{n}_{1}^{*}\mathbf{n}_{2}^{*})$,
$\alpha_{2}=\zeta_{2}/(\xi_{1}+\xi_{2}\,\mathbf{n}_{1}^{*}\mathbf{n}_{2}^{*})$,
and coefficients $\zeta_{i}\in\mathbb{R}$.\\
For the special case $\mathbf{s}_{1}^{*}=\mathbf{s}_{2}^{*}$ one has
$\mathbf{r}^{*}=\mathbf{s}_{1}^{*}=\mathbf{s}_{2}^{*}$ which leads to the relation
$a_{1}+a_{2}=1$ and $\alpha_{i}\sim|\mathbf{s}_{1}^{*}-\mathbf{s}_{2}^{*}|$.
For symmetry reasons we set $a_{1}=a_{2}=\frac{1}{2}$ in order to
treat the surfaces equally. We now consider the case $f_{2}(\mathbf{S})=-f_{1}(\mathbf{S})$
and $\xi_{1}=\xi_{2}$, which implies $\mathbf{S}^{(1)}=\mathbf{S}^{(2)}\equiv\mathbf{S}$
and $\mathbf{r}^{*}\,\mathbf{e}_{z}=0$, where $\mathbf{e}_{z}$ is
the unit vector in the $z$ direction (see Fig.\,\ref{cap:Sketch_Mean_Surface}).
With $\zeta_{1}=\frac{\xi_{1}}{2}\,(\mathbf{s}_{2}^{*}-\mathbf{s}_{1}^{*})\,\mathbf{n}_{2}^{*}$
and $\zeta_{2}=\frac{\xi_{2}}{2}\,(\mathbf{s}_{1}^{*}-\mathbf{s}_{2}^{*})\,\mathbf{n}_{1}^{*}$
this leads to\begin{equation}
\mathbf{r}^{*}=\left(\begin{array}{c}
\mathbf{S}+f_{1}(\mathbf{S})\nabla f_{1}(\mathbf{S})\\
0\end{array}\right)\;\textrm{ and }\; u_{1}^{*}=-\sqrt{g_{1}(\mathbf{S})}\, f_{1}(\mathbf{S})\,,\label{eq:r_star_special_case}\end{equation}
so that Pythagoras'  theorem, $u_{1}^{*2}=(\mathbf{R}^{*}-\mathbf{S})^{2}+f_{1}^{2}(\mathbf{S})$,
is fulfilled. (For different choices of $\zeta_{i}$ this is generally
not the case.)
This means that for $f_{2}(\mathbf{S})=-\,f_{1}(\mathbf{S})$ the manifold $f^{*}$ is the plane $z=0$.
Using the same choice for $\zeta_{i}$ in the case $\xi_{1}\neq \xi_{2}$, Eq.\,(\ref{eq:r_star_Ansatz}) yields
 \begin{equation}
\alpha_{1}=\frac{\xi_{1}\,\big{[}\mathbf{s}_{2}^{*}-\mathbf{s}_{1}^{*}\big{]}\,\mathbf{n}_{2}^{*}}{2\,(\xi_{2}+\xi_{1}\,\mathbf{n}_{1}^{*}\mathbf{n}_{2}^{*})}\quad,\quad\alpha_{2}=\frac{\xi_{2}\,\big{[}\mathbf{s}_{1}^{*}-\mathbf{s}_{2}^{*}\big{]}\,\mathbf{n}_{1}^{*}}{2\,(\xi_{1}+\xi_{2}\,\mathbf{n}_{1}^{*}\mathbf{n}_{2}^{*})}\label{eq:Def_alpha_factors}\end{equation}
so that  the distances $u_{i}^{*}$ fulfill
\begin{equation}
\frac{u_{1}^{*}}{\xi_{1}}=\frac{\alpha_{1}}{\xi_{1}}-\frac{\alpha_{2}}{\xi_{2}}=-\frac{u_{2}^{*}}{\xi_{2}}\,.\label{eq:u_star_distances}\end{equation}
Nonetheless, Eq.\,(\ref{eq:r_star_Ansatz}) still is an implicit expression
for $\mathbf{r}^{*}$ which represents approximately points on the 
iso-density manifold $f^{*}$ of $\md(\mathbf{r})$. \\
Since $f_{1}$ and $f_{2}$ are assumed to not exhibit  strong variations on short
scales, this translates to $f^{*}$ so that $\mathbf{r}^{*}$ allows for a Monge parametrisation
$\mathbf{r}^{*}=(\mathbf{R}^{*},f^{*}(\mathbf{R}^{*}))$, too. Hence in Eq.\,(\ref{eq:r_star_Ansatz})
we can use a Taylor expansion $f_{i}(\mathbf{S}^{(i)})\approx f_{i}(\mathbf{R}^{*})+\nabla f_{i}(\mathbf{R}^{*})\,(\mathbf{S}^{(i)}-\mathbf{R}^{*})$
which leads in lowest order to (Eq.\,(\ref{eq:Def_alpha_factors}))
\begin{equation}
\frac{f^{*}(\mathbf{R}^{*})}{\xi^{*}}=\frac{1}{2}\,\sum_{i=1}^{2}\frac{f_{i}(\mathbf{R}^{*})}{\xi_{i}}\quad,\;\xi^{*}:=\frac{2\xi_{1}\xi_{2}}{\xi_{1}+\xi_{2}}\,.\label{eq:f_star_approx}\end{equation}
A more sophisticated calculation, which takes additional curvature
corrections in Eq.\,(\ref{eq:Mean_surface_construction}) into account,
i.e., using the next higher order terms in Eq.\,(\ref{eq:Def_curex}),
shows that corrections to Eq.\,(\ref{eq:f_star_approx}) are of the
order $\mathcal{O}(f_{i}^{n}f_{j}^{m},n+m\geq3)$. Since  in
Sec.\,\ref{sec:Gaussian-Form-of_H} we shall consider
the Hamiltonian $\mathcal{H}$ within a Gaussian approximation
the expression in Eq.\,(\ref{eq:f_star_approx}) is sufficient.
Furthermore, from Eq.\,(\ref{eq:f_star_approx}) it
follows that the surface with a smaller interfacial width $\xi$ contributes stronger
to $f^{*}$. \\
\\
To summarize Sec.\,\ref{sec:Effective-Hamiltonian}, from
the density functional $\Omega$ in Eq.\,(\ref{eq:GranCanDF}) for a binary liquid mixture,
in Eq.\,(\ref{eq:Def_Hamiltonian}) we have introduced  an effective interface
Hamiltonian $\mathcal{H}$  by specifying
an iso-density contour $f_{i}$ for each density profile $\rho_{i}$
as its corresponding interface, which compose the interface of the mixture as a whole, and
by comparing them with the corresponding flat reference
configurations. In order to express $\mathcal{H}$
in terms of the manifolds $f_{i}$ we used an expansion of the
densities in powers of curvatures of $f_{i}$ (see Eq.\,(\ref{eq:Def_curex})).
Since the hard sphere contribution  $\mathcal{H}_{h}$ cannot
be treated within  this approximation, we have constructed an effective
mean surface $f^{*}$ (see Eq.\,(\ref{eq:f_star_approx})), which
itself is an iso-density contour of a composed density $\md$, so
that the curvature expansion can be performed regarding $f^{*}$ and
$\md$. The results of the curvature expansion of $\mathcal{H}$ up
to second order are presented in the next chapter. Since all expressions
would become rather clumsy without using  short notations, we shall
introduce additional abbreviations in order to obtain a clear presentation of
the structure of the formulas.


\section{Gaussian Approximation\label{sec:Gaussian-Form-of_H}}

In the previous chapter we have illustrated the basic ideas and have
derived the general expressions which
arise upon introducing the effective interface Hamiltonian. In this
section we carry out the curvature expansion in \Eq{eq:Def_curex}
up to second order. Higher order terms are given explicitly in
Appendix\,\ref{app:Explicit-Form-of-H}.
Since in the following the profiles $\rho_{c_{i}}(z)$ do no longer occur we drop the tilde
in $\tilde{\rho}_{c}(u)$ and write  $\rho_{c}(u)$ instead
(see Eqs.\,(\ref{eq:Transformed_desity})-(\ref{eq:Def_curex})).\\
First, we provide some numerical aspects which enter into the graphical
presentation given below. Within the Gaussian approximation
$\mathcal{H}$ is determined by the profiles  $\rho_{c}$ and $\rho_{H}$ and the interaction potentials
$w_{ij}$ given in \Eq{eq:w_ij}.  For $\rho_{c_{i}}$ we use Eq.\,(\ref{eq:form_of_planar_rho}) with an
intrinsic profile $\ip(x)=\tanh(x/2)$ and, guided by Ref.\,\cite{Mecke:6766(1999)},
for $\rho_{H_{i}}$ we choose
\begin{equation}
\frac{\rho_{H_{i}}(u)}{\tr\rho_{i}}=C_{N}\, r_{o}^{(i)}\,\frac{x\,\ip(2\, x)}{2\pi\,\cosh(x)}\;,\; x\equiv\frac{u}{2\,\xi_{i}}\,,\label{eq:rho_H_explicit}\end{equation}
with a dimensionless positive number $C_{N}$. Comparing this expression with the analogous one,
$\rho_{H}$, for the one-component fluid introduced in Eq.\,(3.27) in Ref.\,\cite{Mecke:6766(1999)}
with a prefactor $C_H$, one obtains $C_{N}\, r_{o}^{(i)}=C_{H}\,\xi_{i}$.
Thus, different from Ref.\,\cite{Mecke:6766(1999)} here we assume,
that the prefactor $C_{N}\, r_{o}^{(i)}$ does not vary with temperature.
This choice here translates into that in Ref.\,\cite{Mecke:6766(1999)} if there one
takes $C_{H}\sim\xi^{-1}\rightarrow 0$ for $T\rightarrow T_{c}$.
Therefore, $\rho_{H_{i}}/\tr\rho_{i}$ remains bounded for all temperatures,
so that the curvature influence characterized  by $\rho_{H_{i}}$ vanishes
$\sim\tr\rho_i$ for $T\rightarrow T_{c}$.
In Ref.\,\cite{Mecke:6766(1999)}, $\rho_{H}/\tr\rho_{i}$
diverges $\sim\xi$ for $T\rightarrow T_{c}$, so that
in that temperature range the influence of the curvature may even dominate.
Therefore we prefer the choice given in \Eq{eq:rho_H_explicit}
A more detailed discussion of a possible temperature
dependence of $C_{H}$ can be found in Ref.\,\cite{Hiester:2005}.
For reasons of simplicity, in the following we consider only the case $C_{N}=1.0$.
We emphasize that the structural properties found in Ref.\,\cite{Mecke:6766(1999)}
do not change if $C_H\sim\xi^{-1}$ instead of being constant.
\\
While the ratio  $r_{o}^{(1)}/r_{o}^{(2)}$ of the radii of the particles
is a free parameter, the temperature dependent correlation lengths
$\xi_{i}$ are determined by the bulk correlation functions.
In terms of the total bulk density $\rho$ the concentrations
$\rho_{i}=\conc_{i}\,\rho$ fulfill $\conc_{1}+\conc_{2}=1$.
From the Ornstein-Zernike theory for mixtures one has\begin{eqnarray}
\xi_{i}^{2} & = & -\frac{\rho^{2}}{2}\,\chi_{T}\,\sum_{i=1}^{2}\int_{V}\, d^{3}r\, r^{2}w_{ij}(r)\label{eq:corr_lengths_explicit}\end{eqnarray}
with the isothermal compressibility $\chi_{T}=\frac{1}{\rho}\big{(}\frac{\partial\rho}{\partial p}\big)_{T,V}$,
which can be expressed as\begin{equation}
\chi_{T}=\frac{1}{\rho^{2}}\,\bigg{[}\,\frac{d^{2}h(\rho \conc_{1},\rho \conc_{2})}{d\rho^{2}}+\sum_{i,j=1}^{2}\int_{V}\, d^{3}r\,\conc_{i}\conc_{j}\, w_{ij}(r)\,\bigg]^{-1}\,.\label{eq:isotherm_compress_explicit}\end{equation}
For the two coexisting phases liquid and vapor the two corresponding total densities $\rho^{\pm}$
lead to different values $\chi_{T}^{\pm}$ and thus $\xi_{i}^{\pm}\equiv\xi_{i}(\rho^{\pm})$.
In the subsequent numerical calculations we use $\xi_{i}=(\xi_{i}^{+}+\xi_{i}^{-})/2$
(see also the caption of Fig.\,\ref{cap:Sketch_rho_c}).

\subsection{General Expression\label{sub:General-Expression_H}}

As stated at the beginning of this section we consider only contributions to $\mathcal{H}$ up
to 2nd order in the deviations $f_{i}^{c}$ of the local interface height from the flat configuration.
 With $\hat{\mathbf{f}}(\mathbf{q})$ as the Fourier transform of the
vector $\mathbf{f}(\mathbf{R})$ (see Eqs.\,(\ref{eq:Def_f_vector})
and (\ref{eq:Def_Fourier_1})), one has
\begin{equation}
\mathcal{H}^{G}[\hat{\mathbf{f}}(\mathbf{q})] = \frac{1}{4\pi}\int_{A}\! d^{2}q\;\;\hat{\mathbf{f}}^{\dag}(\mathbf{q})\;\mathbb{E}(q)\;\hat{\mathbf{f}}(\mathbf{q})\,\label{eq:H_Gauss_general}\
\end{equation}
with
\begin{equation}
\mathbb{E}(q)  :=  \mathbb{G}(q)+\mathbb{W}(q)+q^{4}\,\mathbb{K}\,.\label{eq:Def_E_matrix}\end{equation}
Here, the matrix $\mathbb{G}(q)$ represents the contributions stemming
from gravity (see Eq.\,(\ref{eq:Def_G_Matrix_f})), the matrix $\mathbb{W}(q)$
captures the influence of the attractive interaction potentials
(see Eq.\,(\ref{eq:Def_W_Matrix_f})), and the constant matrix $\mathbb{K}$
involves hard sphere contributions (see Eq.\,(\ref{eq:Def_kappa_ij})).
The explicit expressions for $\mathbb{G}$, $\mathbb{W}$, and $\mathbb{K}$
are derived in Appendix\:\ref{app:Second-Order-approximation}, where
the equilibrium condition for the planar densities $\rho_{c_{i}}(u)$ (Eq.\,(\ref{eq:eq_condition_2}))
is frequently used to obtain
the final form of $\mathbb{E}(q)$. \\
In order to be able to present our results in a compact form we introduce the
following abbreviations. For an integer $n\geq0$ and arbitrary
expressions $\mathcal{A}_{i}\equiv\mathcal{A}_{i}(u,\ldots)$ we define
the moments
(similar as in Ref.\,\cite{Mecke:6766(1999)})\begin{eqnarray}
\delta_{n}[\mathcal{A}_{i}] & = & \frac{1}{\tr\rho_{i}}\,\int\lts_{-\ibfu}^{\ibfu}\!\! du\, u^{n}\,\mathcal{A}_{i}(u,\ldots)\,.\label{eq:Def_delta_MAINTEXT}\end{eqnarray}
With this notation, the matrix elements of $\mathbb{G}(q)$ can be expressed as
($\delta_{\alpha\beta}$ is the Kronecker symbol)
\begin{eqnarray}
\mathbb{G}_{11}&=&G\, m_{1}\tr\rho_{1}\,\Big{[}1-2q^{2}\,\Big{(}\delta_{o}[\rho_{H_{1}}]\nonumber\\
&&\!\!\!\!\!\!\!+\,\sum_{k=1}^{2}(\delta_{k1}-\clrd_{2}^{2})\,\big{(}\delta_{2}[\partial_{u}\rho_{c_{k}}]+q^{2}\,\delta_{2}[\rho_{H_{k}}]\big{)}\Big{)}\Big{]}\label{eq:G_ii_explicit}
\end{eqnarray}
with $\clrd_{i}\,=\,\xi_{i}/(\xi_{1}+\xi_{2})$ and similarly for $\mathbb{G}_{22}$ by
interchanging the indices $1\leftrightarrow2$ and ($\mathbb{G}_{12}=\mathbb{G}_{21}$)
\begin{equation}
\mathbb{G}_{12}=2q^{2}\,\clrd_{1}\clrd_{2}\sum_{i=1}^{2}G\, m_{i}\tr\rho_{i}\,\Big{(}\delta_{2}[\partial_{u}\rho_{c_{i}}]+q^{2}\,\delta_{2}[\rho_{H_{i}}]\Big{)}\,.\label{eq:G_12_explicit}\end{equation}
The first part of Eq.\,(\ref{eq:G_ii_explicit}) up to $\delta_{o}[\rho_{H_{1}}]$  is identical with
the corresponding expression in Ref.\,\cite{Mecke:6766(1999)}, and is recovered
by setting $\tr\rho_{2}=0$ and  $\xi_{2}\rightarrow \infty$ (which results in $\clrd_{1}=0$ and $\clrd_{2}=1$), 
which consequently implies $f^{*}\equiv f_{1}$ from Eq.\,(\ref{eq:f_star_approx}). 
All further parts in Eq.\,(\ref{eq:G_ii_explicit}) arise due to the
presence of a second interface. This is somewhat surprising, because
the gravity terms of the density functional $\Omega$ are diagonal
in the densities and thus, one expects $\mathbb{G}(q)$ to be diagonal
w.r.t. the surfaces, too. Actually, the additional terms in Eq.\,(\ref{eq:G_ii_explicit})
and $\mathbb{G}_{12}(q)\neq0$ emerge by applying the equilibrium
condition in Eq.\,(\ref{eq:eq_condition_2}) in order to get rid of certain hard-sphere
contributions from $\mathcal{H}_{h}$ (Eq.\,(\ref{eq:Def_H_h})). All remaining
hard-sphere contributions are captured by the matrix $\mathbb{K}=\big{(}\mathbb{K}_{ij}\big{)}_{i,j\in\{1,2\}}$
with\begin{equation}
\mathbb{K}_{ij}=\int\lts_{-\ibfu}^{+\ibfu}\! du\;\partial_{\rho_{i}\rho_{j}}^{2}h\big{(}\rho_{c_{1}}(u),\rho_{c_{2}}(u)\big{)}\,\,\rho_{H_{i}}(u)\rho_{H_{j}}(u)\,.\label{eq:kappa_ij_explicit}\end{equation}
$\mathbb{K}$ has the form expected as the generalization to two components of
the analogous term $\kappa$ in Ref.\,\cite{Mecke:6766(1999)}.\\
The matrix $\mathbb{W}(q)$ depends on the pair potentials. Hence,
for $k\in\{0,1,2\}$ it is convenient to use  the short notation $\hat{w}_{ij}^{(k)}[q,\delta c_{ij}]:=\hat{w}_{ij}^{(k)}(q,u'-u''+\delta c_{ij})$
for the Fourier transformed interaction potential (see Eq.\,(\ref{eq:w_Fourier_general_form}))
or the integrals of it (see Eqs.\,(\ref{eq:Def_w^(1)}) and  (\ref{eq:Def_w^(2)})), respectively. Moreover,
similar to Eq.\,(\ref{eq:Def_delta_MAINTEXT}) for an expression
$\mathcal{A}_{ij}\equiv\mathcal{A}_{ij}(u',u'',\ldots)$ we define the moments
\begin{equation}
\dcwh{k}(q,\mathcal{A}_{ij}):=\iint\lts_{-\ibfu}^{\ibfu}\!\! du'du''\,\,\hat{w}_{ij}^{(k)}[q,\delta c_{ij}]\,\mathcal{A}_{ij}(u',u'',\ldots)\label{eq:Def_dcwh_MAINTEXT}\end{equation}
and the differences
\begin{equation}
\delta\dcwh{k}(q,\cdots):=\dcwh{k}(q,\cdots)-\dcwh{k}(0,\cdots)\,,\label{eq:Def_delta_dcwh_MAINTEXT}\end{equation}
but on the lhs we suppress the indices $c$ and the square brackets around $\mathcal{A}_{ij}$
indicating the functional dependence on $\mathcal{A}_{ij}$.
In addition we use $\dcwh{0}(\ldots)\equiv\hat{\dcws}(\ldots)$ due to $w_{ij}^{(0)}(\ldots)\equiv w_{ij}(\ldots)$
and similarly $\delta\dcwh{0}(\ldots)\equiv\delta\hat{\dcws}(\ldots)$.
Then, the entries of the
matrix $\mathbb{W}(q)$ can be written as ($\partial_{'}\equiv\partial_{u'}$,
$\partial_{''}\equiv\partial_{u''}$)
\begin{multline}
\mathbb{W}_{11}(q)=-\,\hat{\dcws}\big{(}0,\partial_{'}\rho_{c_{1}}\partial_{''}\rho_{c_{2}}\big{)}+\delta\hat{\dcws}\big{(}q,\partial_{'}\rho_{c_{1}}\partial_{''}\rho_{c_{1}}\big{)}\\
+\,2\, q^{2}\,\Big{[}\delta\hat{\dcws}\big{(}q,\rho_{H_{1}}(u')\partial_{''}\rho_{c_{1}}\big{)}-\hat{\dcws}\big{(}0,\rho_{H_{1}}(u')\partial_{''}\rho_{c_{2}}\big{)}\\
+\,\sum_{i,j=1}^{2}(\clrd_{2}^{2}-\delta_{i1})\,\dcwh{1}\big{(}0,u'\partial_{'}\rho_{c_{i}}\partial_{''}\rho_{c_{j}}\big{)}\Big{]}\\
+\, q^{4}\,\Big{[}\,2\sum_{i,j=1}^{2}(\clrd_{2}^{2}-\delta_{i1})\,\dcwh{1}\big{(}0,u'\rho_{H_{i}}(u')\partial_{''}\rho_{c_{j}}\big{)}\\
+\hat{\dcws}\big{(}q,\rho_{H_{1}}(u')\rho_{H_{1}}(u'')\big{)}+2\sum_{j=1}^{2}\dcwh{2}(0,\rho_{H_{1}}(u')\partial_{''}\rho_{c_{j}})\Big{]}\label{eq:W_11_explicit}\end{multline}
and ($\mathbb{W}_{12}=\mathbb{W}_{21}$)\begin{eqnarray}
\mathbb{W}_{12}(q) & = & \hat{\dcws}\big{(}q,\partial_{'}\rho_{c_{1}}\partial_{''}\rho_{c_{2}}\big{)}\label{eq:W_12_explicit}\\
 &  & +\, q^{2}\sum_{i,j=1}^{2}\bigg{\{}\,(1-\delta_{ij})\,\hat{\dcws}\big{(}q,\rho_{H_{i}}(u')\partial_{''}\rho_{c_{j}}\big{)}\nonumber \\
 &  & \qquad\qquad\;+\,2\,\clrd_{1}\clrd_{2}\,\dcwh{1}\big{(}0,u'\partial_{'}\rho_{c_{i}}\partial_{''}\rho_{c_{j}}\big{)}\nonumber \\
 &  & \qquad+\, q^{2}\,\Big{[}2\,\clrd_{1}\clrd_{2}\,\dcwh{1}\big{(}0,u'\rho_{H_{i}}(u')\partial_{''}\rho_{c_{j}}\big{)}\nonumber \\
 &  & \qquad+\,\frac{(1-\delta_{ij})}{2}\,\hat{\dcws}\big{(}q,\rho_{H_{i}}(u')\rho_{H_{j}}(u'')\big{)}\Big{]}\bigg{\}}\,.\nonumber \end{eqnarray}
 $\mathbb{W}_{22}(q)$ is obtained by interchanging the labels $1\leftrightarrow2$
in $\mathbb{W}_{11}(q)$. Again, the result for a single interface
is included as a limiting case by setting $\tr\rho_{2}=0$, $\clrd_{1}=0$,
and $\clrd_{2}=1$ (or equivalently $\xi_{2}\rightarrow\infty$) in Eqs.\,(\ref{eq:W_11_explicit})
and (\ref{eq:W_12_explicit}), which gives $\mathbb{W}_{12}=\mathbb{W}_{22}=0$ and $f^{*}\equiv f_{1}$ from
Eq.\,(\ref{eq:f_star_approx}).
All additional terms are generated by applying the equilibrium condition in \Eq{eq:eq_condition_2}.
Although these expressions are useful to determine numerically the matrix elements
$\mathbb{E}_{ij}(q)$, the formal structure of
$\mathbb{E}(q)$ might be more transparent in the presentation given in Eq.\,(\ref{eq:E_ij_explicit}).\\
In order to obtain further insight into the nature of $\mathcal{H}$ it is useful
to transform $\mathbb{E}$ into a diagonal matrix. To this end, we define
\setformulalength
\begin{eqnarray}
\era_{i}(q) & := & \mathbb{E}_{ii}(q)+\mathbb{E}_{12}(q)\,,\label{eq:Def_era_i}\\
\epu(q) & := & \era_{1}(q)+\era_{2}(q)\label{eq:Def_epu}\,,\\
\textrm{\parbox{0.35\formulalength}{and \hfill}}& &\textrm{\parbox{0.65\formulalength}{ \hfill}} \,\nonumber\\
\emi(q) & := & \frac{\det\mathbb{E}(q)}{\epu(q)}\,.\label{eq:Def_emi}
\end{eqnarray}
Furthermore, we define a {}``mean'' surface $f^{+}$, and a {}``relative''
surface $f^{-}$ via
\setformulalength
\begin{eqnarray}
\hat{f}^{+}(\mathbf{q}) & := & \sum_{i=1}^{2}\frac{\era_{i}(q)}{\era_{1}(q)+\era_{2}(q)}\,\,\hat{f}_{i}^{c}(\mathbf{q})\label{eq:Def_f+}\\
\textrm{\parbox{0.35\formulalength}{and \hfill}}& &\textrm{\parbox{0.65\formulalength}{ \hfill}} \,\nonumber\\
\hat{f}^{-}(\mathbf{q}) & := &
\hat{f}_{2}^{c}(\mathbf{q})-\hat{f}_{1}^{c}(\mathbf{q})\,,\label{eq:Def_f-}\end{eqnarray}
the fluctuations of which are decoupled within the Gaussian approximation, in contrast to
$f_{1}^{c}$ and $f_{2}^{c}$.
This leads to
\begin{equation}
\mathcal{H}^{G}[\hat{\mathbf{f}}(\mathbf{q})]=\frac{1}{4\pi}\int_{A}\! d^{2}q\,\sum_{\alpha\in\{+,-\}}\,\ep^{\alpha}(q)\,|\hat{f}^{\alpha}(\mathbf{q})|^{2}\,.\label{eq:H_Gauss_diagonal}\end{equation}
This resembles some similarity to the decomposition of the two-body problem in classical
mechanics. It is important to note
that $\ep^{\pm}(q)$ are not the eigenvalues of $\mathbb{E}(q)$ since
the  coordinate transformation used and defined by Eqs.\,(\ref{eq:Def_f+})
and (\ref{eq:Def_f-}) is not orthonormal. Moreover, this transformation
makes sense only for $\era_{j}\neq0$. Therefore, the limiting case of a single
component is better discussed in terms of
Eqs.\,(\ref{eq:H_Gauss_general}) and (\ref{eq:Def_E_matrix}) 
as mentioned above.

\subsection{Energy Density $\epu(q)$\label{sub:energy_dens_plus}}
\begin{figure}[t]
\centering\foreignlanguage{english}{\includegraphics[%
  width=0.35\textwidth,
  angle=-90]{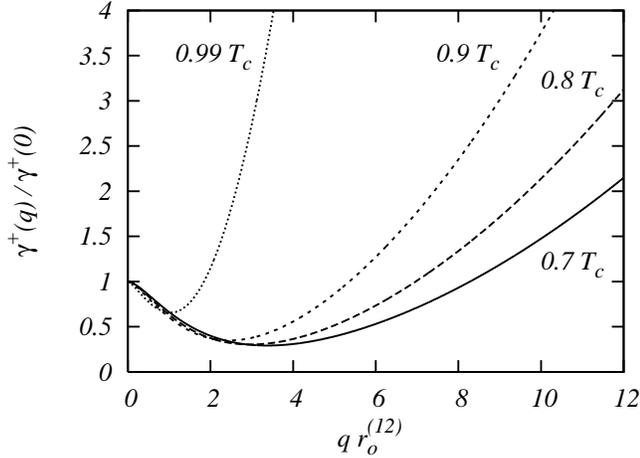}}

\caption{\label{cap:gamma_plus_norm}Normalized wave-vector dependent surface
energy $\gamma^{+}(q)/\gamma^{+}(0)$ as defined by Eq.\,(\ref{eq:Def_gamma_plus})
for different temperatures $T=0.7\ldots0.99\, T_{c}$, where $T_{c}$
is the bulk critical point of the binary liquid mixture in coexistence
with its vapor, computed from the grand canonical density functional
in Eq.\,(\ref{eq:GranCanDF}). The size 
ratio is $r_{o}^{(2)}/r_{o}^{(1)}=1.001$ with $r_{o}^{(12)}=r_{o}^{(1)}+r_{o}^{(2)}$
and $w_{o}^{(22)}/w_{o}^{(11)}=1.05$,
$w_{o}^{(12)}/w_{o}^{(11)}=0.5$ for the depths of the interaction potentials (see Eq.\,(\ref{eq:w_ij})).
$\delta c_{12}$ is set to  $0$, but 
$\gamma^{+}(q)/\gamma^{+}(0)$ depends only weakly on $\delta c_{12}$  for
$|\delta c_{12}|\leq5\, r_{o}^{(12)}$. Here, $C_{N}=1$ (see Eq.\,(\ref{eq:rho_H_explicit})); another choice $C_{N}>1$
modifies mainly the bending constants $\mathbb{K}_{ij}$ (see Eq.\,(\ref{eq:Def_kappa_ij}))
and thus the curves increase stronger for larger values of $q$ beyond the  minimum which  is
also shifted to larger wavelengths $1/q$. The opposite behavior is observed for $C_{N}<1$.
For low temperatures the minimum is rather deep and occurs at large $q$ values. Upon raising the temperature 
the minimum becomes more and more shallow and moves to smaller values of $q$, too. This is the 
same qualitative behavior as observed for the one component case \cite{Mecke:6766(1999)}.}
\end{figure}

The quantity $\epu(q)$ can be written as\renewcommand{\arraystretch}{1.35}
\begin{equation}
\epu(q)  =  G\,\mathcal{G}^{+}(q)  +  q^{2} \gamma^{+}(q)\label{eq:epu_Formel}
\end{equation}
where $\mathcal{G}^{+}(q)=\mathcal{G}_{1}(q)+\mathcal{G}_{2}$ with
$\mathcal{G}_{i}=m_{j}\tr\rho_{j}\,\big{(}1-2q^{2}\,\delta_{o}[\rho_{H_{j}}]\big{)}$ (Eq.\,(\ref{eq:Def_G_j_Fourier})), 
and $\gamma^{+}(q)$ takes the form
\begin{equation}
 \gamma^{+}(q) := \sum_{i,j=1}^{2}\,\gamma_{ij}(q)\label{eq:Def_gamma_plus}
\end{equation}
with 
\setformulalength
\begin{eqnarray}
\gamma_{ij}(q) & = &\frac{\delta\hat{\dcws}\big{(}q,\partial_{'}\rho_{c_{i}}\partial_{''}\rho_{c_{j}}\big{)}}{q^{2}}+\,2\,\delta\hat{\dcws}\big{(}q,\rho_{H_{i}}\partial_{''}\rho_{c_{j}}\big{)} \nonumber \\
\textrm{\parbox{0.1\formulalength}{\hfill}}& &\textrm{\parbox{.9\formulalength}{\hs{-12}$+\, q^{2}\,\Big{[}\,\hat{\dcws}\big{(}q,\rho_{H_{i}}\rho_{H_{j}}\big{)}+\mathbb{K}_{ij}+\,2\dcwh{2}(0,\rho_{H_{i}}\partial_{''}\rho_{c_{j}})\Big{]}$.}}\label{eq:Def_gamma_ij} 
\end{eqnarray}

This formula is the generalization of the corresponding result derived for a one-component
fluid \cite{Mecke:6766(1999)}, except that the term $\dcwh{2}(0,\rho_{H_{i}}\partial_{''}\rho_{c_{j}})$
shows up additionally.  $\gamma^{+}(q)$ is plays the role of a
wavevector dependent surface tension for $f^{+}$, which then 
behaves similarly  as for a single interface. Since according to Eq.\,(\ref{eq:Def_f+}) 
$\hat{f}^{+}(q)$ is a linear combination of the surfaces, $\hat{f}_{j}^{c}$
can be considered as the prime or mean surface of the binary fluid.
The functional form of $\gamma^{+}(q)$ is shown in Fig.\,\ref{cap:gamma_plus_norm} 
for various temperatures.

\subsection{Energy Density $\emi(q)$\label{sub:energy_dens_minus}}
\begin{figure}[t]
\centering\foreignlanguage{english}{\includegraphics[%
  width=0.35\textwidth,
  angle=-90]{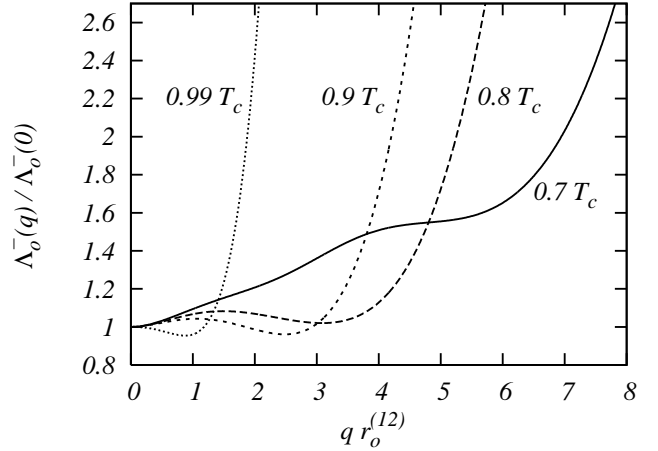}}

\caption{\label{cap:Lambda_o_minus_norm}Normalized wavevector dependent free energy
density $\emi_{o}(q)/\emi_{o}(0)$ as given by Eq.\,(\ref{eq:emi_canonical_form})
for various temperatures $T=0.7\ldots0.99\, T_{c}$. The choices for the interaction
potentials are  $r_{o}^{(2)}/r_{o}^{(1)}=1.001$ for the radii and $w_{o}^{(22)}/w_{o}^{(11)}=1.05$,
$w_{o}^{(12)}/w_{o}^{(11)}=0.5$ for the interaction strengths (see Eq.\,(\ref{eq:w_ij})).
$\delta c_{12}$ is taken  to be $0$. Distinct from the normalized
surface tension $\gamma^{+}(q)$ shown in Fig.\,\ref{cap:gamma_plus_norm},
$\emi_{o}(q)$ shows a monotonic increase at low  temperatures, but forms a minimum at a nonzero
value of $q$ if the temperature is increased. As for $\gamma^{+}(q)$, this minimum is shifted to larger
wavelengths upon further increasing the temperature. It is important to
note, that $\emi_{o}(q)$ reflects the free energy density of different
surface configurations of $f^{-}$ but it is not a surface tension because
no Goldstone modes, i.e., translational shifts without cost in free energy, exist for $f^{-}$ as they do for $f^{+}$. Moreover,
for $q>0$ one expects $0<\emi_{o}(q)\sim\det\mathbb{E}(q)$ for stability
reasons.}
\end{figure}
$\emi(q)$ is determined by Eq.\,(\ref{eq:Def_emi}). It exhibits 
a more complex structure than $\epu(q)$. The explicit
expression for $\det\mathbb{E}(q)$ is given by Appendix\:\ref{app:Explicit-correlations}.
The intrinsic behavior of
the surfaces is given by the pair potentials of the particles alone,
independent of the external field. Thus in the absence of gravity,
i.e., for $G=0$, one obtains the undisturbed energy density
$\emi_{o}(q)=\emi(q,G=0)$ of the different surface
configurations  $f^{-}$. Similar to $\epu(q)$ it can be decomposed
into a wave-vector dependent surface tension
$\gamma^{-}(q)$ and an additional contribution that does not vanish
for $q\rightarrow0$ and which depends parametrically only on the
interaction potential $w_{12}$ between the two species and the planar
density profiles $\partial_{u}\rho_{c_{i}}$ (see Eq.\,(\ref{eq:Def_dcwh_MAINTEXT}) 
for $\hat{\dcws}$):\begin{equation}
\emi_{o}(q)=-\,\hat{\dcws}\big{(}0,\partial_{'}\rho_{c_{1}}\partial_{''}\rho_{c_{2}}\big{)}+q^{2}\,\gamma^{-}(q)\,.\label{eq:emi_canonical_form}\end{equation}
$\emi_{o}(q)$ describes the free energy required to deform the relative surface 
into a corrugated one  with a wave-vector $q$ in the presence of
the microscopic interactions of the particles. For $q=0$ one has $\emi_{o}(0)>0$,
which corresponds to the free energy needed to separate the flat equilibrium
surfaces $c_{1}$ and $c_{2}$ from each other.
This is in accordance with the facts, that $\emi_{o}(0)$ depends on $w_{12}$ and significantly  weakens
for larger $\delta c_{12}$ for which  $f_{2}^{c}$ and $f_{1}^{c}$ decouple.
In addition, at low temperatures $\emi_{o}(q)$ varies sensitively upon changes
of $\delta c_{12}$, but it  hardly changes its
character at higher temperatures. This can be explained heuristically
by noting that for $T\lesssim T_{c}$ the dominant length scale is
set by the diverging bulk correlation length $\xi$ so that  the difference
$\delta c_{12}$ becomes irrelevant for the statistical  weights.

$\emi_{o}(q)$ differs from $\gamma^{+}(q)$ qualitatively (see Fig.\,\ref{cap:Lambda_o_minus_norm} and
note that $\epu(q,G=0)/\epu(q\rightarrow0,G=0)=\gamma^{+}(q)/\gamma^{+}(0)$ according to \Eq{eq:Def_gamma_plus}):
for temperatures close to the triple point, it shows a monotonic increase
implying that surface configurations $f^{-}$ with 
nonzero wavelengths are energetically suppressed. But
for higher temperatures a minimum at $q\neq0$ evolves. This minimum is also shifted
towards longer wavelengths for further increased temperatures but
does not change its depth. Thus, together with the behavior of
$\gamma^{+}(q)$, this means  that for low temperatures the mean
surface $f^{+}$ is more easily excited thermally than the relative
surface $f^{-}$. But for higher temperatures, the thermal fluctuations
have a stronger influence on $f^{-}$ while $f^{+}$ becomes more
rigid. This behavior is quantitatively controlled by the curvature
corrections characterized by $\rho_{H}$ (see Eq.\,(\ref{eq:rho_H_explicit}))
and thus  $C_{N}$. The influence of $C_{N}$ on $\emi_{o}(q)$
becomes mainly visible through a shift of the depth of the minimum, which increases
strongly for larger values of $C_{N}$. Thus, $\emi_{o}(q)$ may even
become negative for certain values of $C_{N}$. In the absence of
gravity, i.e. for $G=0$, one has $\emi_{o}(q)\sim\det\mathbb{E}(q)$ and thus
$\emi_{o}(q)<0$ means that the system becomes unstable. By switching
off all interactions between the two species, i.e., for $w_{12}=0$  the simple expression
\begin{equation}
\emi_{o}(q)=q^{2}\,\frac{\gamma_{11}(q)\,\gamma_{22}(q)}{\gamma_{11}(q)+\gamma_{22}(q)}\quad,\quad\,w_{12}\equiv0\,,\label{eq:emi_o_simple_form}\end{equation}
emerges, so that from Eqs.\,(\ref{eq:epu_Formel}) and (\ref{eq:emi_canonical_form}) the relation 
$\gamma^{-}(q)\,\gamma^{+}(q)=\gamma_{11}(q)\,\gamma_{22}(q)$ follows. \\

\begin{figure}[t]
\centering\foreignlanguage{english}{\includegraphics[%
  width=0.35\textwidth,
  angle=-90]{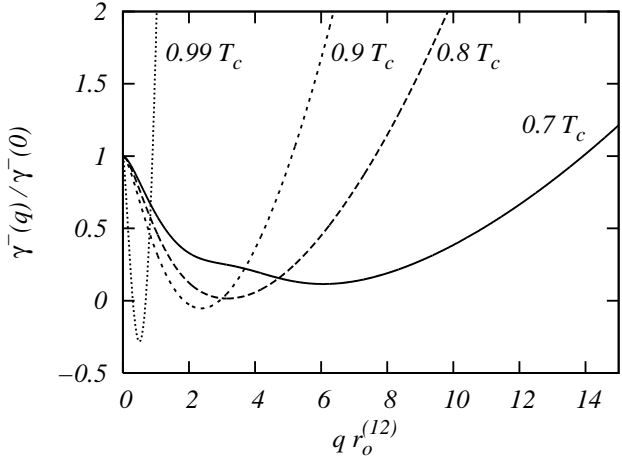}}

\caption{\label{cap:gamma_minus_norm}Normalized wavevector dependent surface
free energy $\gamma^{-}(q)/\gamma^{-}(0)$ as given by Eqs.\,(\ref{eq:emi_canonical_form}) and (\ref{eq:gamma_minus})
for various temperatures $T=0.7\ldots0.99\, T_{c}$. The parameters are chosen as in Fig.\,\ref{cap:Lambda_o_minus_norm}.
Similar to $\gamma^{+}(q)$,
$\gamma^{-}(q)$ shows a minimum which  shifts towards smaller
wavevectors but increases in depth upon increasing the temperature.
Thus, the thermally activated creation of additional surface area of the relative surface
becomes easier for longer wavelengths at higher temperatures.
For large values of $C_{N}$ (see \Eq{eq:rho_H_explicit}) $\gamma^{-}(q)$
becomes negative which probably indicates a breakdown of the Gaussian approximation or even of the
concept of the relative surface.}
\end{figure}

In the general case of nonzero $w_{12}$ the expressions for $\gamma_{ij}$
in Eq.\,(\ref{eq:Def_gamma_ij}), for $\gamw_{12}$ in Eq.\,(\ref{eq:Def_gamma_wedge}), 
and for $\gamv_{ij}$  in Eq.\,(\ref{eq:Def_gamma_vee}) together  with
\setformulalength
\begin{eqnarray}
\erb_{1}(q) & := & \gamma_{11}(q)+\frac{\gamw_{12}(q)+\gamv_{12}(q)}{q^{2}}\label{eq:Def_erb_1}\\
\textrm{\parbox{0.3\formulalength}{and \hfill}}& &\textrm{\parbox{0.7\formulalength}{ \hfill}} \,\nonumber\\
\erb_{2}(q) & := & \gamma_{22}(q)+\frac{\gamw_{12}(q)+\gamv_{21}(q)}{q^{2}}\label{eq:Def_erb_2}\end{eqnarray}
lead to the following expression for $\gamma^{-}(q)$ as defined in \Eq{eq:emi_canonical_form}  (see Eq.\,(\ref{eq:Def_clrd_entry}) for $\clrd_{i}$
and Eq.\,(\ref{eq:Def_msW_ij}) for $\ms W_{ij}$)
\begin{eqnarray}
\gamma^{-}(q) & = & \frac{\hat{\dcws}\big{(}0,\partial_{'}\rho_{c_{1}}\partial_{''}\rho_{c_{2}}\big{)}-\gamw_{12}(q)}{q^{2}}+\frac{\erb_{1}(q)\erb_{2}(q)}{\gamma^{+}(q)}\nonumber \\
 &  & -\,2\,\clrd_{1}\clrd_{2}\ds{\sum_{i,j=1}^{2}}\ms W_{ij}(q)+\gammasym(q)\,.\label{eq:gamma_minus}\end{eqnarray}
The last term $\gammasym(q)$ (see \Eq{eq:Def_gammasym} in Appendix\:\ref{app:Explicit-correlations})
turns out to be the smallest contribution and it is determined by the contrast between the two species.
The terms in Eq.\,(\ref{eq:gamma_minus}) are listed according to their quantitative
importance. The behavior of $\gamma^{-}(q)$ is determined mainly by the
first and the second term, while the third one is about one order of
magnitude smaller than the previous ones, and the last one may be smaller by even
two orders of magnitude. $\gamma^{-}(q)$ captures the wavelength dependence of $\emi_{o}(q)$ 
(see \Eq{eq:emi_canonical_form}).
A comparison between $\gamma^{+}$ and $\gamma^{-}$
(see Fig.\,\ref{cap:gamma_minus_norm}) shows, that $\gamma^{-}(q)/\gamma^{-}(0)$
also exhibits a minimum at a nonzero wavevector but its depth increases
with increasing temperature. Hence, for large values of $C_{N}$, $\gamma^{-}(q)$
and even $\gamma^{-}(0)$, which depends on $C_{N}$, too (see Eq.\,(\ref{eq:gamma_minus})),
may become negative which probably indicates a breakdown of the
Gaussian approximation or even of the concept of a relative surface.
Nevertheless, for increasing temperatures its minimum is shifted to smaller
values of $q$, analogous to the behavior of $\gamma^{+}(q)/\gamma^{+}(0)$.
Quantitatively, one finds $\gamma^{-}(q)/\gamma^{+}(q)\lesssim0.1$
for all values of $q$ and temperatures $T\lesssim0.9\, T_{c}$ ($\gamma^{-}(q)/\gamma^{+}(q)\lesssim0.2$
for $T\lesssim0.99\, T_{c}$), so that $\gamma^{-}(q)$ is about one
order of magnitude smaller than $\gamma^{+}(q)$. Therefore, one may
regard the mean surface $f^{+}$ to be more rigid than the relative
surface $f^{-}$.\\
Recently diffuse X-ray scattering data from the liquid-vapor interfaces of 
Bi:Ga, Tl:Ga, and Pb:Ga binary liquid alloys rich in Ga have been reported \cite{Li:235426(2006)}.
In order to interpret these data the authors put forward an expression similar to \Eq{eq:Def_gamma_plus} 
(see Eqs.\,(8)-(11) in Ref.\,\cite{Li:235426(2006)}), in which, however, different than in 
Eqs.\,(\ref{eq:Def_gamma_plus}) and (\ref{eq:Def_gamma_ij}) only the curvature correction profile $\rho_{H_{2}}(u)$ 
of the segregated component was used without taking into account the profile $\rho_{H_{1}}(u)$ of the
majority component, i.e., $\rho_{H_{1}}(u)\equiv0$. Thus it appears to be highly rewarding to reanalyze these experimental
data in a future contribution on the basis of the present full statistical description. This description might also
provide an understanding of recent synchrotron X-ray reflectivity data on the interfacial width, broadened by capillary 
waves, of the liquid-liquid interface of nitrobenzene and water \cite{Luo:4527(2006),Benjamin:407(1997)}. 
However, this would require to extent the present analysis to binary dipolar fluids \cite{Szalai:2873(2005)}.
Furthermore, a recent analysis of the fluctuation spectrum of lipid bilayer shows similarities to our 
description in terms of two interfaces  \cite{Brannigan:1501(2006)}.
These authors also define a mean and a relative surface (see Eqs.\,(5) and (6) in Ref.\,\cite{Brannigan:1501(2006)}) in order
to take into account conformations of the bilayer via 
fluctuation modes of the bilayer thickness and of the bending modes of the mean surface of the bilayer.
Their choice of boundary conditions leads to a decoupling of these modes in real space and an effective free 
energy for the bilayer deformations
on both short and long wavelengths is derived (Eq.\,(21) in Ref.\,\cite{Brannigan:1501(2006)}). 
The main difference to our approach consists in their specific choice of boundary conditions for lipid bilayers, 
which cannot be applied to fluid interfaces. Accordingly, within the Gaussian approximation, in our approach 
a mode decoupling between $f^{+}$ and $f^{-}$ is achieved only in Fourier space, where $\hat{f}^{+}$ is defined as a 
{\em{wavelength-dependent weighted sum}} of the Fourier modes of the surfaces $\hat{f}_1$ and $\hat{f}_2$ 
(see Eqs.\,(\ref{eq:Def_f+})-(\ref{eq:H_Gauss_diagonal})). Consequently, $f^{+}$ consists of a sum of corresponding 
convolutions, which, 
in general, cannot be written as a sum of $f_{1}$ and $f_{2}$ with constant weights as it is done in Ref.\,\cite{Brannigan:1501(2006)}.
Nevertheless, in Subsec.\,\ref{sub:MeanSurfaceApprox} we have also introduced a concept  similar to the one used
in Ref.\,\cite{Brannigan:1501(2006)}. In Subsec.\,\ref{sub:MeanSurfaceApprox} the mean surface $f^{*}$ 
(see Fig.\,\ref{cap:Sketch_Mean_Surface} and Eq.\,(\ref{eq:f_star_approx})) is defined in real space in order to analyze 
$\mathcal{H}_{h}$ using the curvature expansion (Eq.\,(\ref{eq:Def_curex})) and in order to express $\mathcal{H}$ in terms 
of $f_{1}$ and $f_{2}$ taking into account the full coupling between them. 

\section{Correlation functions\label{Correlations}}

\begin{figure}[H]
\centering{
\subfigure{\label{cap:Corr_T_f1f1}\includegraphics[%
  width=0.31\textwidth,
  angle=-90]{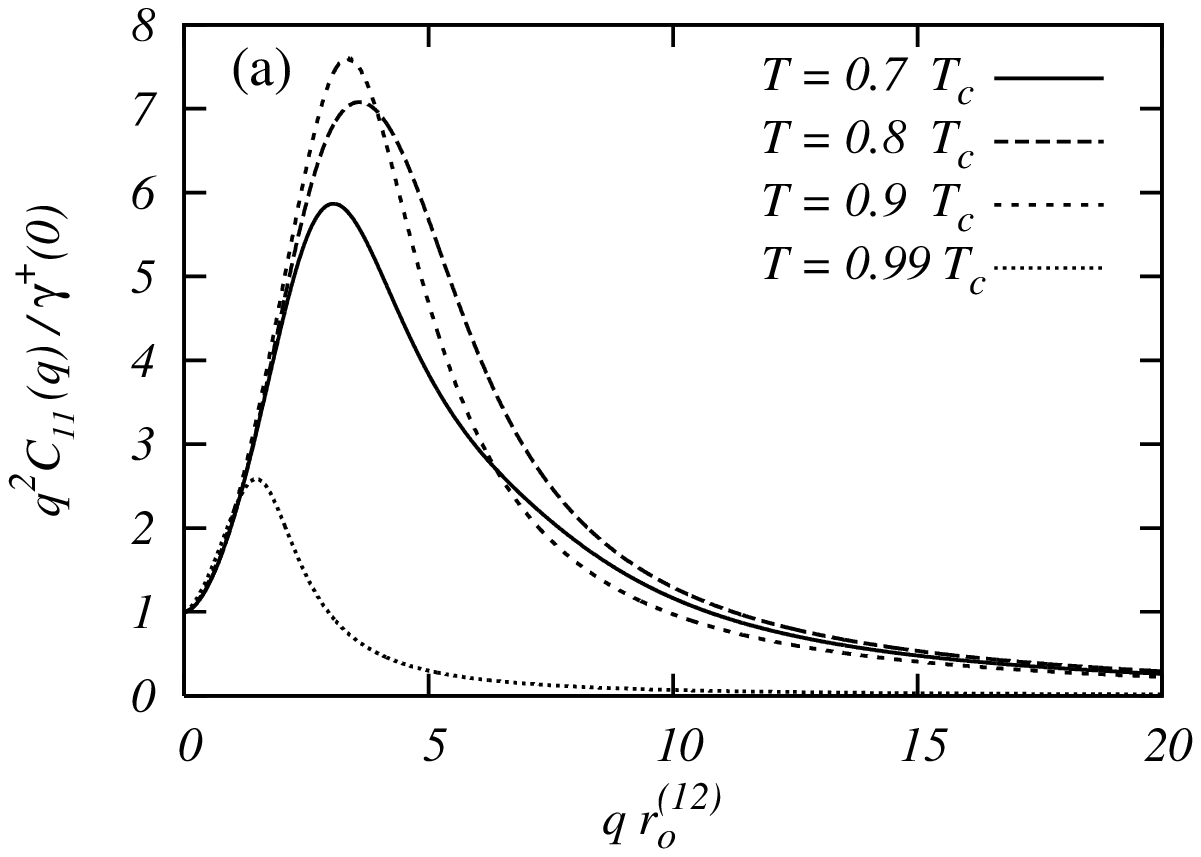}}
\subfigure{\label{cap:Corr_T_f2f2}\includegraphics[%
  width=0.31\textwidth,
  angle=-90]{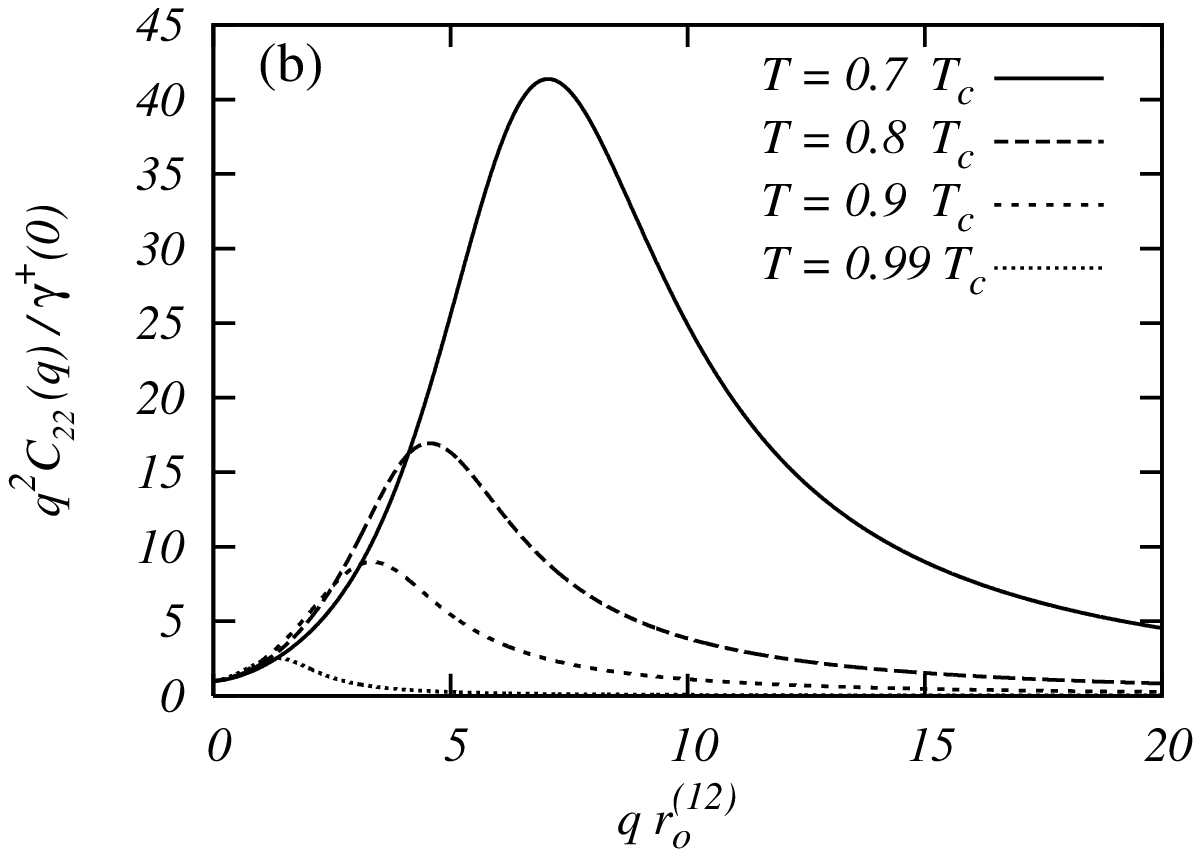}}
\subfigure{\label{cap:Corr_T_f2f1}\includegraphics[%
  width=0.31\textwidth,
  angle=-90]{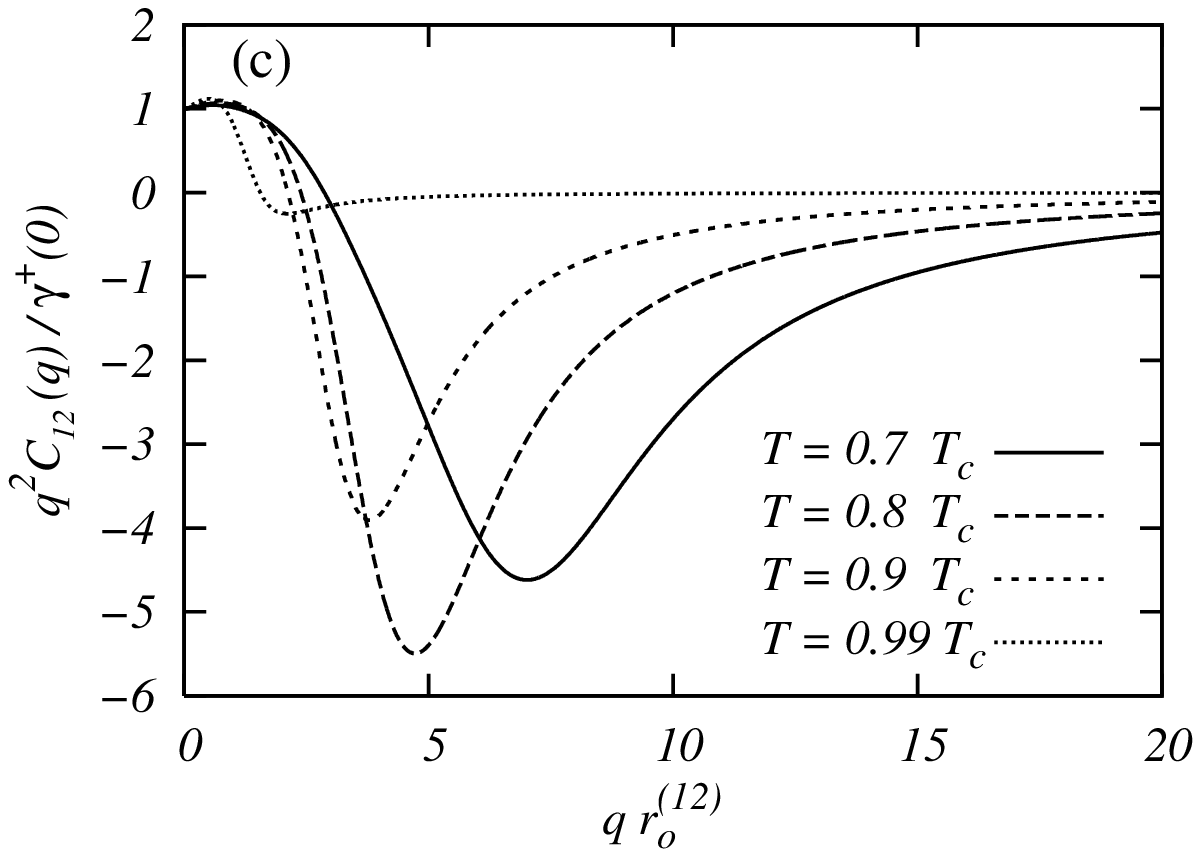}}
}
\caption{Normalized height-height correlation functions
$q^{2} \Corr_{ij}(q,G=0)$  (Eqs.\,(\ref{eq:correlations_general})-(\ref{eq:Corr_G_0_12})) 
for various temperatures $T=0.7\ldots0.99\, T_{c}$. 
The comparison between \subref{cap:Corr_T_f1f1} and
\subref{cap:Corr_T_f2f2} 
reveals that the correlation of the Fourier modes of $f_{1}^{c}$ and
$f_{2}^{c}$ are similar at elevated temperatures but differ on short length 
scales, i.e., large $q$ values, at temperatures close to the triple point due
to differences between $\gamma_{11}(q)$ and $\gamma_{22}(q)$ (see \Eq{eq:Def_gamma_ij}).
For $q>q_o(T)$ the correlation function $q^{2}\Corr_{12}(q,G=0)$  is negative 
(see \subref{cap:Corr_T_f2f1})
so that the Fourier modes of the two different surfaces are anti-correlated.
Parameters are given in the main text.
}
\end{figure}

\begin{figure}[H]
\centering{
\subfigure{\label{cap:Corr_conc_f1f1}\includegraphics[%
  width=0.31\textwidth,
  angle=-90]{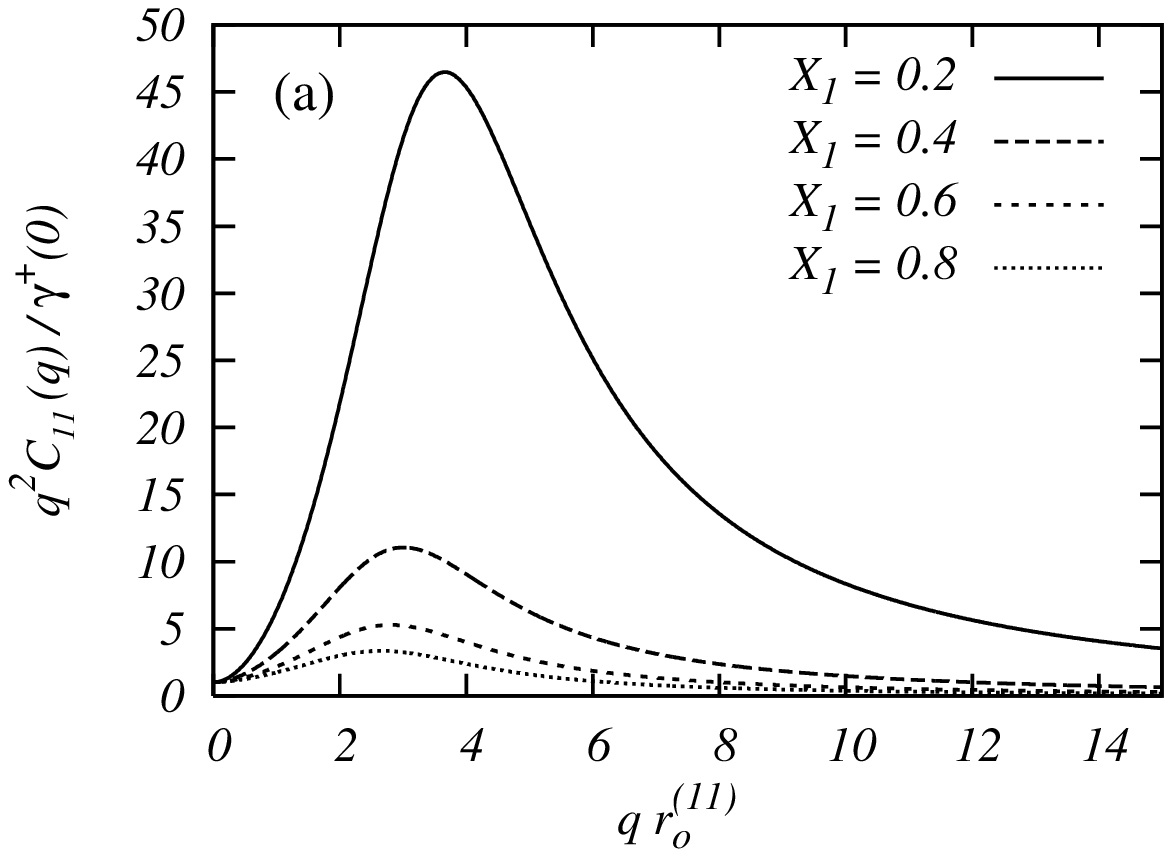}}
\subfigure{\label{cap:Corr_conc_f2f2}\includegraphics[%
  width=0.31\textwidth,
  angle=-90]{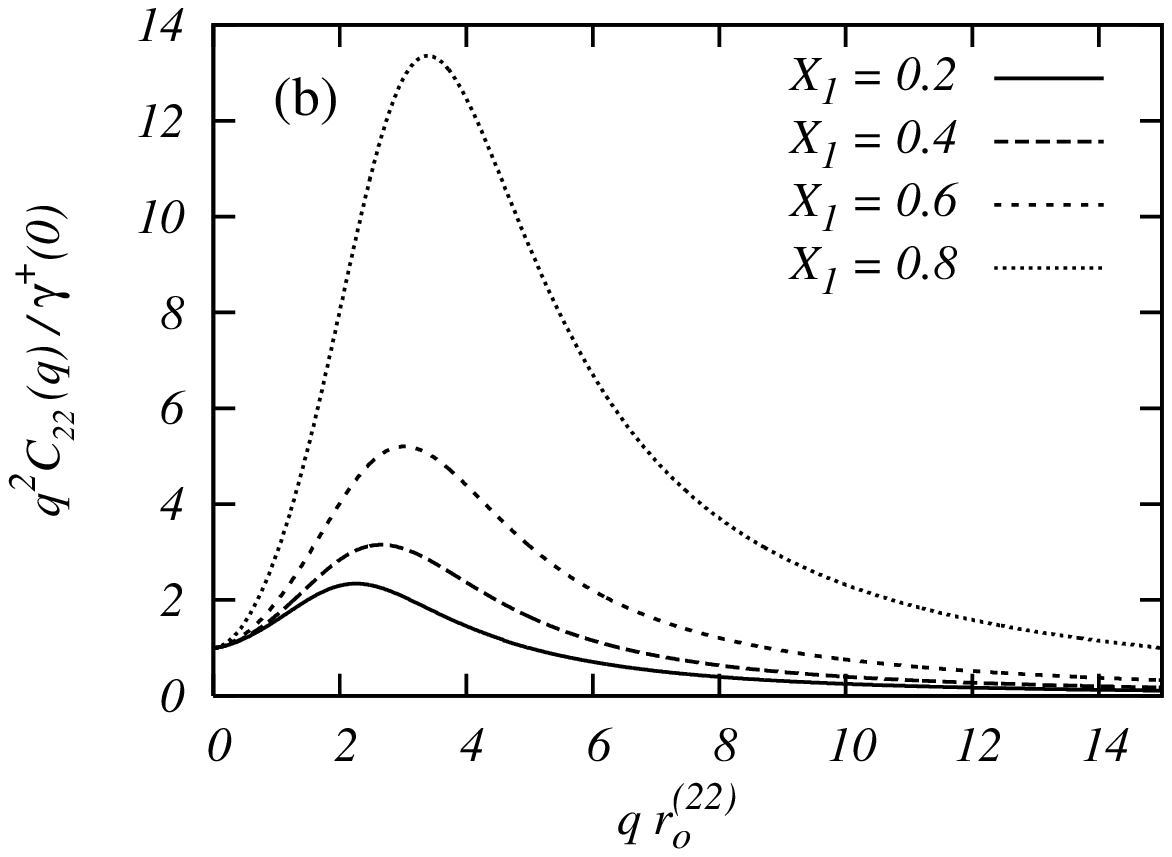}}
\subfigure{\label{cap:Corr_conc_f2f1}\includegraphics[%
  width=0.31\textwidth,
  angle=-90]{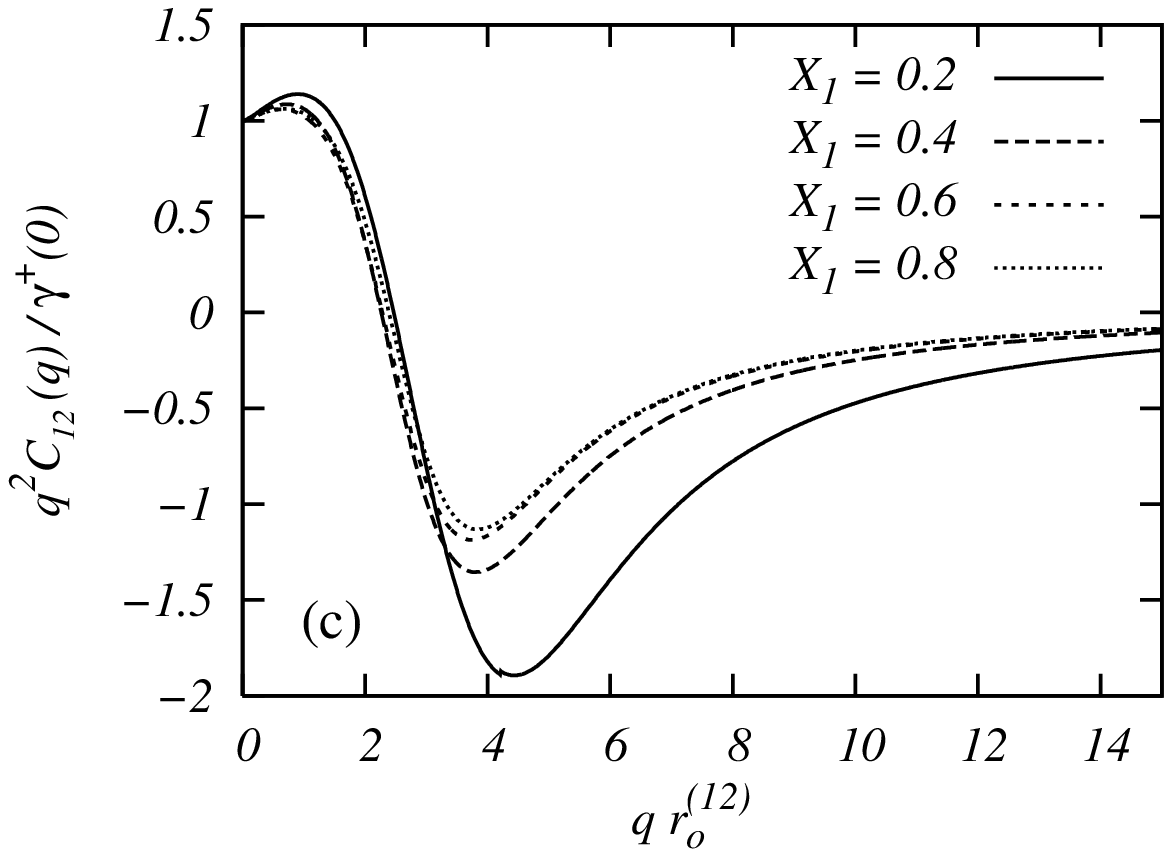}}
}
\caption{\label{cap:Corr_conc}Normalized height-height correlation functions $q^{2} \Corr_{ij}(q,G=0)$
of the Fourier modes $\hat{f}_{i}^{c}(\mathbf{q})$ for various concentrations
$X_{1}$ of component $1$ and for $T=0.7\,T_{c}$. The choices for the interaction parameters
are given in the main text. For high concentrations $X_{1}\geq 0.6$ one finds
for  $q^{2} \Corr_{11}(q,G=0)$ the well known
structure for a height-height correlation function as predicted by classical
capillary wave theory (see \subref{cap:Corr_conc_f1f1}). 
But for low concentrations
$X_{1}\leq 0.4$ this function exhibits  a  peak for $2\leq q\,r^{(11)} \leq 6$ indicating
the increasing influence of the second component on the interface $f_{1}$. This effect can be seen in reverse in
\subref{cap:Corr_conc_f2f2} for small $X_{2}=1-X_{1}$.
Since $ \Corr_{12}(q)=\Corr_{21}(q)$ it does not depend strongly on the
concentration (see \subref{cap:Corr_conc_f2f1}).
}
\end{figure}

From the diagonalization in Eq.\,(\ref{eq:H_Gauss_diagonal}) it is clear,
that the surfaces $f^{+}$ and $f^{-}$ are uncorrelated within the
Gaussian approximation and thus statistically independent. In order to
obtain  insight into the structure of the original interfaces, one has to consider
the correlation functions
\begin{equation}
\langle\hat{f}_{i}^{c}(\mathbf{q})\hat{f}_{j}^{c}(\mathbf{q}')\rangle=2\pi\delta(\mathbf{q}+\mathbf{q}')\,\Corr_{ij}(q)\,,\label{eq:correlations_general}
\end{equation}
where $\beta \Corr_{ij}(q)=\mathbb{E}_{ij}^{-1}(q)$ denote the matrix elements of the inverse matrix $\mathbb{E}^{-1}(q)$ (see \Eq{eq:Def_E_matrix}).
The corresponding detailed expressions are given in Appendix\:\ref{app:Explicit-correlations}.
In the limits $G\rightarrow0$ and $q\rightarrow0$, one has
$\beta\Corr_{ij}(q)=[G\,\mathcal{G}^{+}+q^{2}\,\gamma^{+}(q)]^{-1}$
for all pairs $i,j\in\{1,2\}$ as already predicted in Ref.\,\cite{Tarazona:1357(1985)}
(see also Eq.\,(\ref{eq:Appendix_fifj_corr_limit}) in Appendix\:\ref{app:Explicit-correlations}).\\
However, our approach allows us to go beyond  the limits  $G\rightarrow0$ and
$q\rightarrow0$ in order to obtain the interfacial structure of a binary liquid
mixture on smaller wavelengths. Although all expression derived above include the
influence of the external potential, i.e., gravity, we restrict our
considerations in this section to  $G=0$ in order to simplify  the following discussion.\\
One obtains from Eq.\,(\ref{eq:Appendix_f1f1_corr_explicit})
the  \emph{positive} function (see Eqs. (\ref{eq:Def_gamma_eff_11})-(\ref{eq:Def_gamma_eff_12})
for $\gamma_{ij}^{\,\mathsf{eff}}$)
\begin{equation}
\frac{1}{\beta q^{2}\Corr_{11}(q)}= \gamma_{11}^{\,\mathsf{eff}}(q)+\frac{\emi_{o}(0)}{q^{2}}-\frac{1}{q^{2}}\,\frac{\big{[}\emi_{o}(0)-q^{2}\gamma_{12}^{\,\mathsf{eff}}(q)\big]^{2}}{\emi_{o}(0)+q^{2}\gamma_{22}^{\,\mathsf{eff}}(q)}\,\label{eq:h_ii_for_G_0}
\end{equation}
and a similar expression for $\Corr_{22}(q)$ by interchanging the indices $1\leftrightarrow 2$.
As mentioned above, one has $\beta q^{2}\Corr_{ii}(q r^{(ii)}\ll1)=1/\gamma^{+}(q)$.
In the limiting case $w_{12}\equiv 0$ one obtains
$\beta q^{2}\Corr_{ii}(q)=1/\gamma_{ii}(q)$.
On the other hand, for
$\Corr_{12}(q)$ (\Eq{eq:correlations_general})
one has
\begin{eqnarray}
\beta q^{2}\Corr_{12}(q,G=0) & = & \frac{\emi_{o}(0)-q^{2}\gamma_{12}^{\,\mathsf{eff}}(q)\,(q)}{\gamma^{+}(q)\,\emi_{o}(q)}\,;\label{eq:Corr_G_0_12}
\end{eqnarray}
however, Eq.\,(\ref{eq:Corr_G_0_12}) does not exhibit the form of Eq.\,(\ref{eq:h_ii_for_G_0}), because
$\gamma_{12}^{\,\mathsf{eff}}(q)$, defined in Eq.\,(\ref{eq:Def_gamma_eff_12}),
is also a positive function for all values of $q$.
Thus, $\Corr_{12}(q)$
changes its sign at a certain value $q_{o}$, which depends crucially
on $\emi_{o}(0)$. This means that for $q>q_{o}$ the Fourier modes of the surfaces $f_{1}$
and $f_{2}$ are anti-correlated (see Figs.\,\ref{cap:Corr_T_f2f1} and \ref{cap:Corr_conc_f2f1}).

Figures\,\ref{cap:Corr_T_f1f1}\;-\subref{cap:Corr_T_f2f1} show the correlation
functions $q^{2} \Corr_{ij}(q,G=0)$ for different temperatures and the parameter
choices  $r_{o}^{(2)}/r_{o}^{(1)}=1.002$, $w_{o}^{(22)}/w_{o}^{(11)}=1.05$, and $w_{o}^{(12)}/w_{o}^{(11)}=0.5$.
Although the parameter differences of the two components is small the
correlation functions  $q^{2} \Corr_{11}(q)$ and $q^{2} \Corr_{22}(q)$
exhibit a different behavior for small wavelenths and temperatures close to
the triple point (see Figs.\,\ref{cap:Corr_T_f1f1} and \subref{cap:Corr_T_f2f2}). 
This result indicates a structural difference between the surfaces $f_{1}^{c}$ and
$f_{2}^{c}$ on short length scales which vanishes  for higher
temperatures. Similar the correlation function $q^{2} \Corr_{12}(q)$ indicates a
(anti-)correlation between the Fourier modes of the surfaces for low
temperatures which becomes weaker for high tempertures (see Fig.\,\ref{cap:Corr_T_f2f1}).

\begin{figure}[t]
\centering{
\subfigure{\label{cap:Corr_conc_f2f2_pro_f1f1}\includegraphics[%
  width=0.31\textwidth,
  angle=-90]{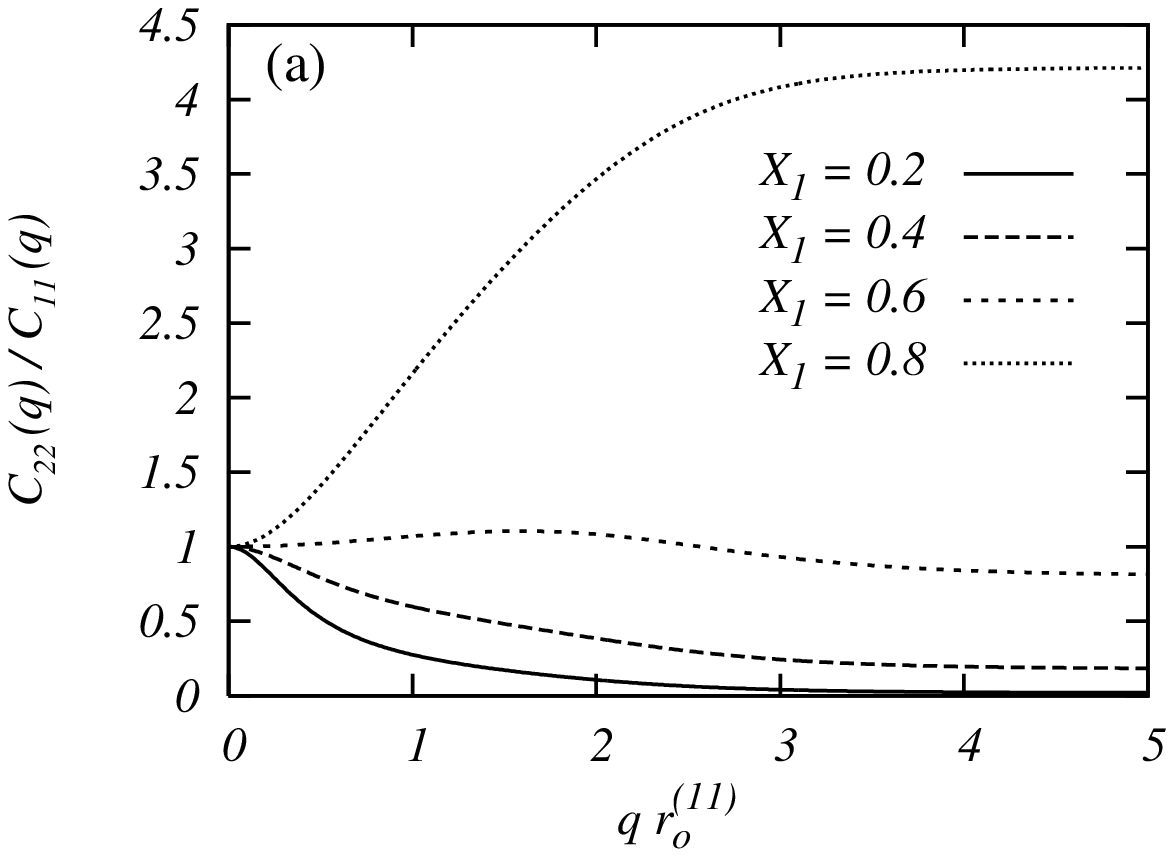}}
\subfigure{\label{cap:Corr_conc_f2f1_pro_f2f2}\includegraphics[%
  width=0.31\textwidth,
  angle=-90]{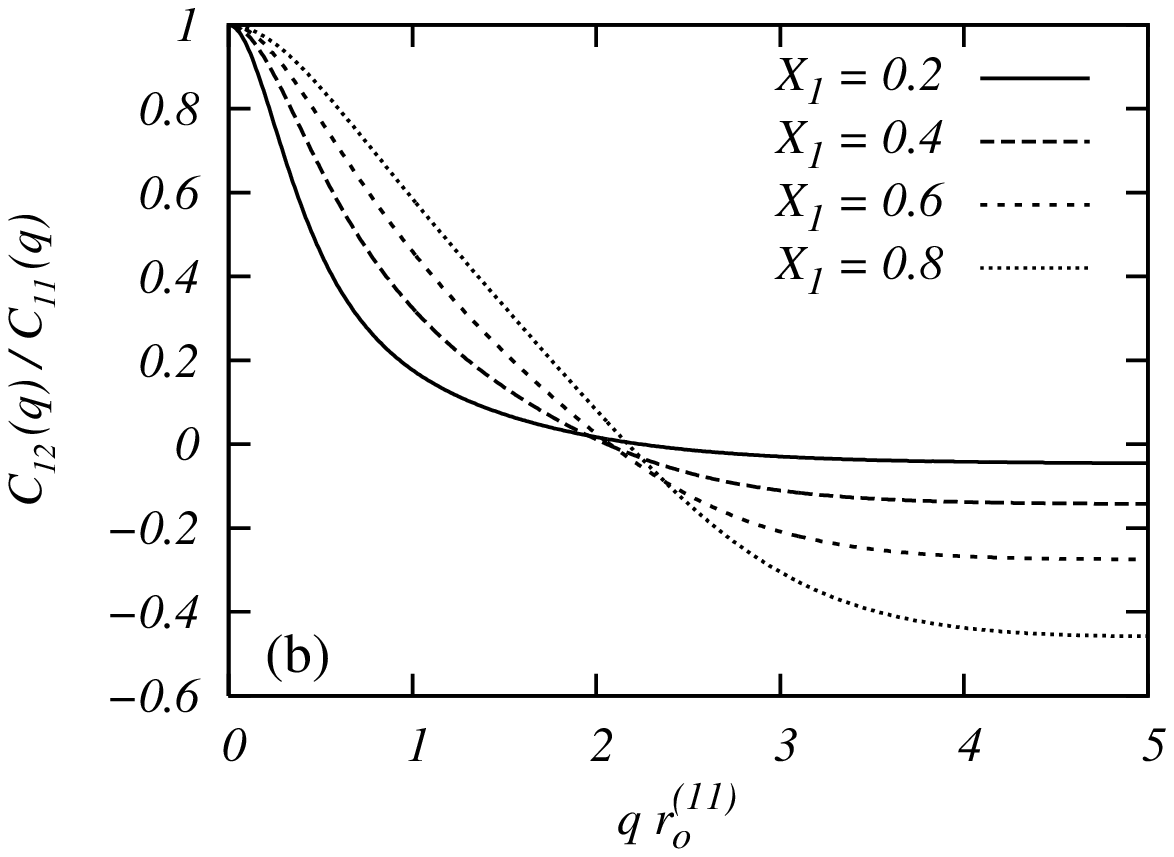}}
}
\caption{Ratio of the correlation
  functions  $\Corr_{22}(q,G=0)/\Corr_{11}(q,G=0)$
  and $\Corr_{12}(q,G=0)/\Corr_{11}(q,G=0)$
  for various  concentrations $X_{1}$ of component $1$ and for
  $T=0.7\,T_{c}$. Since all correlation functions attain the same
  value for $q=0$, one has  $\Corr_{ij}(q=0)/\Corr_{11}(q=0)=1$. One finds,
  that these ratios change significantly for different choices of the
  concentration: both, the slope for small $q$ values and the limit 
  $q\rightarrow\infty$ are influenced by $X_1$. 
Parameters are the same as in Fig.\,\ref{cap:Corr_conc}.
}
\end{figure}

Figures\,\ref{cap:Corr_conc_f1f1}\;-\subref{cap:Corr_conc_f2f1} show
the influence of the composition $\conc_{1}=1-\conc_{2}$ on the correlation functions
$q^{2} \Corr_{ij}(q)$ at a fixed temperature $T$ close to the triple point for the parameter choices
$r_{o}^{(2)}/r_{o}^{(1)}=1.2$, $w_{o}^{(22)}/w_{o}^{(11)}=1.4$,
and $w_{o}^{(12)}/w_{o}^{(11)}=0.586$. 
We infer from Fig.\,\ref{cap:Corr_conc_f2f2}  that for low concentrations of species $1$
the correlation function of the height
of the interface associated with the more attractive component $2$ does not show  particular features
at large $q$ values
whereas the interface of the less attractive component $1$ seems
to be more ordered by the  stronger species. In Fig.\,\ref{cap:Corr_conc_f1f1} this is indicated by a weaker
decay of the corresponding height-height correlation function $q^{2} \Corr_{11}(q)$ for $X_{1}\leq 	0.4$.
Since $q^{2} \Corr_{12}(q)$ is a symmetric w.r.t. the label exchange $1\leftrightarrow2$,
one would not expect a strong dependence of the corresponding height-height
cross correlation function on the concentration. This expectation is confirmed by Fig.\,\ref{cap:Corr_conc_f2f1}.
\section{Summary\label{sec:Summary}}

We have considered the liquid-vapor or liquid-liquid interface region
of a binary liquid mixture (Fig.\,\ref{cap:Phasdiagram}) which is characterized by two phase
separating surfaces $f_{i}(\mathbf{R})$ for the two species $i\in\{1,2\}$ (Fig.\,\ref{cap:SketchSystem}).
We have obtained the following main results:

(1) Based on a grand canonical density functional $\Omega[\rho_{1}(\mathbf{r}),\rho_{2}(\mathbf{r})]$ 
for binary liquid mixtures, we have
defined an effective interface Hamiltonian $\mathcal{H}[f_{1}(\mathbf{R}),f_{2}(\mathbf{R})]$
providing the statistical weight $\exp(-\,\beta\,\mathcal{H})$ for nonflat surface configurations (Eq.\,(\ref{eq:Def_Hamiltonian})).
This approach takes into account and keeps track of both 
the presence of long-ranged dispersion forces (Eq.\,(\ref{eq:w_ij})),
smoothly varying intrinsic density profiles (Fig.\,\ref{cap:Sketch_rho_c}),
and the thermodynamic state of the system
(Fig.\,\ref{cap:Phasdiagram}). In particular, it captures liquid-vapor as well as 
liquid-liquid interfaces (see the remarks (i)-(iii) in Sec.\,\ref{sec:intro}).

(2) Using a local normal coordinate system
(Fig.\,\ref{cap:Sketch_normal_coord}, Eq.\,(\ref{eq:Def_Normal_Trafo_general}))
we have  incorporated changes of the intrinsic density profiles
caused by the curvatures of the fluctuating interfaces (Eq.\,(\ref{eq:Def_curex})).
To this end we have introduced the concept of a mean surface $f^{*}$
(Subsec.\,\ref{sub:MeanSurfaceApprox}, Fig.\,\ref{cap:Sketch_Mean_Surface},
and Eq.\,(\ref{eq:f_star_approx})). Within this approximation the
two surfaces $f_{i}$ and their width $\xi_{i}$ are used to form a nominal
surface $f^{*}$ for which the above mentioned
coordinate change is applied without using additional parameters (see Sec.\,\ref{sec:intro}\,(iv)).

(3) This approach leads to an explicit expression of $\mathcal{H}[f_{1}(\mathbf{R}),f_{2}(\mathbf{R})]$
in terms of the two surfaces (\Eq{eq:H_part_int} and Subsecs.\,\ref{sub:CUREX_1} and \ref{sub:MeanSurfaceApprox}). 
In particular it contains the  coupling between the two surfaces based on the microscopic interactions
between the two species (Appendix\,\ref{app:Explicit-Form-of-H}).

(4) Within a Gaussian approximation the Hamiltonian $\mathcal{H}[f_{1}(\mathbf{R}),f_{2}(\mathbf{R})]$
takes a bilinear form (Eq.\,(\ref{eq:H_Gauss_general})).
In order to simplify the further discussion we define a mean surface $f^{+}$
and a relative surface $f^{-}$ (Eqs. (\ref{eq:Def_f+}) and (\ref{eq:Def_f-}))
which leads to a diagonalization of $\mathcal{H}$ (Eq.\,(\ref{eq:H_Gauss_diagonal})).
Thus, the wavevector dependent free energy density $\epu(q)$ (Eqs.\,(\ref{eq:epu_Formel})-(\ref{eq:Def_gamma_ij})) 
of the mean surface $f^{+}$ (\Eq{eq:Def_f+}) generalizes the corresponding
expression for the liquid-vapor interface of a one-component fluid. Therefore
$f^{+}$ plays the role of the overall interface of the binary
mixture which  remains even in the special case $f_{1}=f_{2}=f^{+}$ or $f^{-}\equiv0$,
respectively, which is equivalent to a two-component system modelled
by a single interface. $\epu(q)$ contains a gravity
part, $G\,\mathcal{G}^{+}(q)$, and a contribution  stemming
from the interactions among and between the species which can be considered
as a wavelength dependent surface tension $\gamma^{+}(q)$ (Eq.\,(\ref{eq:Def_gamma_plus}), 
Appendix\,\ref{app:Second-Order-approximation}).
$\gamma^{+}(q)$ decreases as  function of $q$, attains a minimum and increases again (Fig.\,\ref{cap:gamma_plus_norm}).
The minimum occurs at smaller values of $q$ and becomes less deep upon raising the temperature.
This resembles the behavior of the wavelength dependent surface tension of the corresponding one-component fluid.
The second $q$-dependent free energy density $\emi(q)$ is linked to the
relative surface $f^{-}$. Even in the absence of the gravity  ($G=0$)
and thus different from $\epu(q)$
it consists of  an effective surface tension component $\gamma^{-}(q)$,
and an additional constant contribution depending on the flat  intrinsic
density profiles and the interaction potential $w_{12}$ between the two
species only (Eq.\,(\ref{eq:emi_canonical_form})). This
constant describes the cost in free energy for a separation of the two surfaces
against the attraction between the two species.
For temperatures close to the triple point
$\emi_o(q)=\emi(q,G=0)$ increases monotonicly as a function of $q$, but
for higher temperatures it developes a minimum
that is gradually shifted to larger wavelengths (Fig.\,\ref{cap:Lambda_o_minus_norm}).
\\
The surface tension $\gamma^{-}(q)$ has a similar structure as
$\gamma^{+}(q)$, but it develops a minimum the depth of which increases upon
increasing temperature. Thus, depending on the strength $C_N$ of the influence of curvatures
on the intrinsic profiles (Eq.\,(\ref{eq:rho_H_explicit})), $\gamma^{-}(q)$
may even become negative signalling probably the breakdown of the Gaussian approximation
or even of the mean interface concept for large values of $C_N$
and certain temperatures (Fig.\,\ref{cap:gamma_minus_norm}).
However, the total energy density $\emi_{o}(q)$ remains
positive for all values of $q$.

(5) Finally, we have discussed the Fourier transforms of the height-height correlation
functions (Eqs. (\ref{eq:Appendix_f1f1_corr_explicit}) and (\ref{eq:Appendix_f1f2_corr_explicit})).
Figures\;\ref{cap:Corr_T_f1f1}\;-\subref{cap:Corr_T_f2f1} illustrate their temperature dependence
whereas Figs.\,\ref{cap:Corr_conc_f1f1}\;-\subref{cap:Corr_conc_f2f1} demonstrate the influence
of the concentration on the height-height correlation functions (see Sec.\,\ref{sec:intro}\,(v)). 
\renewcommand{\appendixname}{\textbf{APPENDIX}}
\appendix
\section{Explicit Form of the effective interface Hamiltonian\label{app:Explicit-Form-of-H}}

In this appendix we drop the tilde of $\delta\tilde{\rho}_{f_{\alpha}}$ (see Eq.\,(\ref{eq:Def_curex}))
and write $\delta\rho_{\alpha}$ instead and we frequently omit the
full list of arguments an expression depends on. Hence, one should  keep
in mind that $\delta\rho_{\alpha}(\mathbf{S},u)$ and $\rho_{c}(u)$
depend on the set of normal coordinates.
Here, we present our results for the effective interface Hamiltonian up to second order
in the height displacements. Higher order terms and further details
concerning the derivation can be found in Ref.\,\cite{Hiester:2005}.

\subsection{Gravity Part}

After a transformation into appropriate normal coordinates the general form of the gravity parts
can be expressed in terms of $\tr\varrho_{j}\equiv m_{j}\tr\rho_{j}$ as
\begin{equation}
\int_{A}\!\! d^{2}R\,\mathcal{H}_{V}\big{(}\mathbf{f}(\mathbf{R})\big{)}=\frac{G}{2}\int_{A}d^{2}S\,\,\sum_{j=1}^{2}\tr\varrho_{j}\,\big{[}\mymathfont{G}_{j}^{\partial}+\mymathfont{G}_{j}^{\delta}+\mymathfont{G}_{j}^{\pztr}\big{]}\,.\label{eq:G_Term_Trafo}\end{equation}
Similar as in Ref.\,\cite{Mecke:6766(1999)} in order to proceed and for later purposes we define moments
with $n\geq0$ and the metric $g_{i}(\mathbf{S})=1+\big{(}\nabla f_{i}(\mathbf{S})\big{)}^2$ (see \Eq{eq:Jacbobian})
of arbitrary expressions $\mathcal{A}_{i}\equiv\mathcal{A}_{i}(u,\ldots)$
 \setformulalength
\begin{eqnarray}
\bar{\delta}_{n}[\mathcal{A}_{i}] & := & \frac{1}{\tr\rho_{i}\,\sqrt{g_{i}^{n}}}\,\int\lts_{-\ibfu}^{\ibfu}\!\! du\, u^{n}\,\mathcal{A}_{i}(u,\ldots)\label{eq:Def_delta_bar}\\
\textrm{\parbox{0.25\formulalength}{and \hfill}}& &\textrm{\parbox{0.75\formulalength}{ \hfill}} \,\nonumber\\
\delta_{n}[\mathcal{A}_{i}] & := & \sqrt{g_{i}^{n}}\,\bar{\delta}_{n}[\mathcal{A}_{i}]\,.\label{eq:Def_delta}
\end{eqnarray}
Without carrying out the curvature expansion we find up to second order in $f_{j}$:
\begin{flalign}
\mymathfont{G}_{j}^{\partial} & =-\,2f_{j}^{c}\,\bar{\delta}_{1}[\partial_{u}\rho_{c_{j}}]+2H_{j}\sqrt{g_{j}}\,\bar{\delta}_{3}[\partial_{u}\rho_{c_{j}}]\label{eq:Def_G_partial}\\
 & \quad-\, K_{j}g_{j}\,\bar{\delta}_{4}[\partial_{u}\rho_{c_{j}}]+\,(f_{j}^{c})^{2}\nonumber \\
 & \quad+\,\big{(}(\nabla f_{j}^{c})^{2}+2f_{j}^{c}(2H_{j})\sqrt{g_{j}}\big{)}\,\bar{\delta}_{2}[\partial_{u}\rho_{c_{j}}]\,,\nonumber \end{flalign}
\begin{flalign}
\mymathfont{G}_{j}^{\delta} & =\frac{2\bar{\delta}_{1}[\delta\rho_{j}]}{\sqrt{g_{j}}}+f_{j}^{c}\,\frac{2\delta_{o}[\delta\rho_{j}]}{\sqrt{g_{j}}}\,-3\,(2H_{j})\,\bar{\delta}_{2}[\delta\rho_{j}]\label{eq:Def_G_delta}\\
 & \quad-\,4f_{j}^{c}(2H_{j})\,\bar{\delta}_{1}[\delta\rho_{j}]+4K_{j}\sqrt{g_{j}}\,\bar{\delta}_{3}[\delta\rho_{j}]\,,\nonumber \end{flalign}
and
\begin{equation}
\mymathfont{G}_{j}^{\pztr}=\frac{-\,1}{\tr\rho_{j}\sqrt{g_{j}}}\,\int\lts_{-\ibfu}^{\ibfu}\!\!\! du\,(f_{j}^{c}+\frac{u}{\sqrt{g_{j}}})^{2}\,\nabla f_{j}(\mathbf{S})\nabla\delta\rho_{j}(\mathbf{S},u)\,.\label{eq:Def_G_pztr}
\end{equation}
Using the curvature expansion in Eq.\,(\ref{eq:Def_curex}) for $\delta\rho$
up to quadratic order  one obtains\begin{flalign}
\mymathfont{G}_{j}^{\delta} & =\,(2H_{j})\,\frac{2\bar{\delta}_{1}[\delta\rho_{j}]}{\sqrt{g_{j}}}+(2H_{j})^{2}\Big{(}\frac{2\,\bar{\delta}_{1}[\rho_{H_{j}^{2}}]}{\sqrt{g_{j}}}-3\,\bar{\delta}_{2}[\rho_{H_{j}}]\Big{)}\nonumber \\
 & \quad+\frac{2}{\sqrt{g_{j}}}\Big{(}f_{j}^{c}(2H_{j})\,\delta_{o}[\rho_{H_{j}}]+K_{j}\,\bar{\delta}_{1}[\rho_{K_{j}}]\Big{)}+\mathcal{O}(f^{3})\,.\label{eq:G_delta_CUREX}
\end{flalign}
By carrying out integration by parts with respect to the lateral coordinates one obtains for $\mymathfont{G}_{j}^{\pztr}$ in Eq.\,(\ref{eq:G_Term_Trafo})
\begin{eqnarray}
\mymathfont{G}_{j}^{\pztr} & = & \frac{2H_{j}\,\tr f_{j}}{\sqrt{g_{j}}}\,\bar{\delta}_{2}[\rho_{H_{j}}]+\mathcal{O}(f^{3})\,.\label{eq:G_pztr_CUREX}
\end{eqnarray}
where $\tr f_{j}$ means the Laplacian of the surface $f_{j}(\mathbf{S})$.

\subsection{Interaction Part\label{sub:Interaction-Part_APPENDIX}}

In this subsection we use the following notation: expressions with
an index $i$ depend on $u'$ and $\mathbf{R}'$ whereas
terms with an index $j$ depend on $u''$ and $\mathbf{R}''$.
Moreover we use the following symbolic notation: $\partial_{'}\equiv\partial_{u'}$,
$\nabla_{'}\equiv(\partial_{s'_{x}},\partial_{s'_{y}})$, similarly
$\partial_{''}$, $\nabla_{''}$, and
\begin{equation}
\partial\delta[\rho_{f_{i}},\rho_{f_{j}}]:=\partial_{'}\rho_{c_{i}}\partial_{''}\delta\rho_{f_{j}}+\partial_{'}\delta\rho_{f_{i}}\partial_{''}\rho_{c_{j}}+\partial_{'}\delta\rho_{f_{i}}\partial_{''}\delta\rho_{f_{j}}\,.\label{eq:Def_partial_delta_rhorho}\end{equation}
In addition, we introduce the short notation $w_{ij}^{(k)}[f_{i}(\mathbf{S}'),f_{j}(\mathbf{S}'')]\equiv w_{ij}^{(k)}[f_{i},f_{j}]$
for $k\in\{0,1,2\}$ with (see Eq.\,(\ref{eq:Def_Normal_Trafo_general})
for $\mathcal{T}_{f}$ and Eqs. (\ref{eq:Def_w^(1)}) and (\ref{eq:Def_w^(2)})
for $w_{ij}^{(k)}$)
\begin{eqnarray}
w_{ij}^{(k)}[f_{i},f_{j}] & := & w_{ij}^{(k)}\big{(}|\mathcal{T}_{f_{i}}(\mathbf{S}',u')-\mathcal{T}_{f_{j}}(\mathbf{S}'',u'')|\big{)}\label{eq:Def_w_ff}\\
w_{ij}^{(k)}[\delta c_{ij}] & := & w_{ij}^{(k)}(|\mathbf{S}'-\mathbf{S}''|,|u'-u''+\delta c_{ij}|)\,,\qquad\label{eq:Def_w_cc}
\end{eqnarray}
where $w_{ij}^{(0)}\equiv w_{ij}$. Similar to Eq.\,(\ref{eq:Def_delta_bar}),
it is convenient to use the following abbreviation for an arbitrary expression
$\mathcal{A}_{ij}\equiv\mathcal{A}_{ij}(u',u'',\ldots)$:
\begin{flalign}
\dcw k_{f}[\mathcal{A}_{ij}] & :=\iint\lts_{-\ibfu}^{\ibfu}\! du'du''\,\, w_{ij}^{(k)}[f_{i},f_{j}]\,\mathcal{A}_{ij}(u',u'',\ldots)\label{eq:Def_dcw_f}
\end{flalign}
and similarly for $\dcw{k}_{c}[\mathcal{A}_{ij}]$ for using $w_{ij}^{(k)}[\delta c_{ij}]$ (Eq.\,(\ref{eq:Def_w_cc}))
instead of $w_{ij}^{(k)}[f_{i},f_{j}]$ in \Eq{eq:Def_w_ff}.
$\dcw k_{f}$ still depends on $\mathbf{S}'$ and $\mathbf{S}''$;
hence, terms like $\nabla_{\mathbf{S}'}\dcw k_{f}\equiv\nabla_{'}\dcw k_{f}$
(and similarly for $\nabla_{''}$) are defined. The symbol $\underleftrightarrow{(i,')\leftrightarrow(j,'')}$
is introduced to shorten the formulae below. It states that the
first part of the formula has to be repeated with interchanged labels,
i.e., in each expression $i$ is replaced by $j$ and  $'$ is replaced
by $''$, and vice versa. \\
\\
After the transformation to normal coordinates the total expression for the interaction parts  can be written as (\Eq{eq:Def_H_w})
\begin{flalign}
 & \qquad\iint_{A}d^{2}R'd^{2}R''\,\,\mathcal{H}_{w}\big{(}\mathbf{f}(\mathbf{R},\mathbf{R}')\big{)}\nonumber \\
= & -\frac{1}{2}\sum_{i,\, j=1}^{2}\iint_{A}d^{2}S'd^{2}S''\,\,\Big{[}\mymathfont{W}_{ij}^{\partial}+\mymathfont{W}_{ij}^{\partial\delta}+\mymathfont{W}_{ij}^{\partial\pztr}\Big{]}\label{eq:W_Term_Trafo}\end{flalign}
with\begin{flalign}
\mymathfont{W}_{ij}^{\partial} & =\frac{1}{2}\,\big{(}\,\dcw2_{f}[\partial_{'}\rho_{c_{i}}\partial_{''}\rho_{c_{j}}]-\dcw2_{c}[\partial_{'}\rho_{c_{i}}\partial_{''}\rho_{c_{j}}]\,\big{)}\label{eq:Def_W_partial}\\
 & \quad-\,2H_{i}\,\dcw2_{f}[u'\partial_{'}\rho_{c_{i}}\partial_{''}\rho_{c_{j}}]+\, K_{i}\,\dcw2_{f}[(u')^{2}\partial_{'}\rho_{c_{i}}\partial_{''}\rho_{c_{j}}]\nonumber \\
 & \quad+\,2H_{i}H_{j}\,\dcw2_{f}[u'u''\partial_{'}\rho_{c_{i}}\partial_{''}\rho_{c_{j}}]+\underleftrightarrow{(i,')\leftrightarrow(j,'')}\,,\nonumber \end{flalign}
\begin{flalign}
\mymathfont{W}_{ij}^{\partial\delta} & =\frac{1}{2}\,\dcw2_{f}\big{[}\partial\delta[\rho_{f_{i}},\rho_{f_{j}}]\big{]}\label{eq:Def_W_partial_delta}\\
 & -2H_{i}\,\dcw2_{f}\big{[}u'\,\partial\delta[\rho_{f_{i}},\rho_{f_{j}}]\big{]}+K_{i}\,\dcw2_{f}\big{[}(u')^{2}\,\partial\delta[\rho_{f_{i}},\rho_{f_{j}}]\big{]}\nonumber \\
 & \quad+\,2H_{i}H_{j}\,\dcw2_{f}\big{[}u'u''\,\partial\delta[\rho_{f_{i}},\rho_{f_{j}}]\big{]}+\underleftrightarrow{(i,')\leftrightarrow(j,'')}\,,\nonumber \end{flalign}
and
\begin{equation}
\mymathfont{W}_{ij}^{\partial\pztr}  = \frac{\nabla f_{j}}{\sqrt{g_{j}}}\,\mymathfont{w}_{ij}^{\partial\pztr}+\underleftrightarrow{(i,')\leftrightarrow(j,'')}\label{eq:Def_W_partial_pztr}
\end{equation}
where
\begin{eqnarray}
\mymathfont{w}_{ij}^{\partial\pztr} & :=&\dcw2_{f}[\partial_{'}\rho_{c_{i}}\nabla_{''}\delta\rho_{j}]-2H_{i}\,\dcw2_{f}[u'\partial_{'}\rho_{c_{i}}\nabla_{''}\delta\rho_{j}]\nonumber \\
 & &\quad+\, K_{i}\,\dcw2_{f}[(u')^{2}\partial_{'}\rho_{c_{i}}\nabla_{''}\delta\rho_{j}]\,.\label{eq:Def_w_partial_pztr}
 \end{eqnarray}
Here, we already have omitted higher order terms which can be found in Ref.\,\cite{Hiester:2005}.
Applying the curvature expansion (Eq.\,(\ref{eq:Def_curex})) for each
density and its surface we obtain
\begin{flalign}
\mymathfont{W}_{ij}^{\partial\delta} & =\,2H_{i}\,\dcw2_{f}[\partial_{''}\rho_{c_{j}}\partial_{'}\rho_{H_{i}}]+K_{i}\,\dcw2_{f}[\partial_{''}\rho_{c_{j}}\partial_{'}\rho_{K_{i}}]\nonumber \\
 & +(2H_{i})^{2}\,\Big{(}\dcw2_{f}[\partial_{''}\rho_{c_{j}}\partial_{'}\rho_{H_{i}^{2}}]-\dcw2_{f}[u'\,\partial_{'}\rho_{H_{i}}\partial_{''}\rho_{c_{j}}]\Big{)}\nonumber \\
 & +4H_{j}H_{i}\,\Big{(}\,\frac{1}{2}\,\dcw2_{f}[\partial_{'}\rho_{H_{i}}\partial_{''}\rho_{H_{j}}]-\dcw2_{f}[u'\,\partial_{'}\rho_{c_{i}}\partial_{''}\rho_{H_{j}}]\Big{)}\nonumber \\
 & \quad+\underleftrightarrow{(i,')\leftrightarrow(j,'')}+\mathcal{O}(f^{3})\label{eq:W_partial_delta_CUREX}\end{flalign}
and up to second order
\begin{equation}
\mymathfont{W}_{ij}^{\partial\pztr}\approx\frac{\nabla(2H_{j})\nabla f_{j}}{\sqrt{g_{j}}}\,\,\dcw2_{f}[\partial_{'}\rho_{c_{i}}\rho_{H_{j}}]+\underleftrightarrow{(i,')\leftrightarrow(j,'')}\,.\label{eq:W_partial_pztr_CUREX}
\end{equation}

\subsection{Hard Sphere Part}

In this subsection we use $f^{*}$ as introduced in Eq.\,(\ref{eq:f_star_approx}) and instead
of $h(\rho_{1},\rho_{2})$ we write (see the introductory remarks in  Subsec.\,\ref{sub:MeanSurfaceApprox}
and Eq.\,(\ref{eq:Def_mean_density}))\begin{equation}
h(\rho_{1},\rho_{2})\stackrel{!}{=}h^{*}(\md)\,\label{eq:hh_Ersetzung}\end{equation}
and similarly $h_{f}^{*}\equiv h^{*}(\md_{f^{*}})$ and $h_{c}^{*}\equiv h^{*}(\md_{c^{*}})$.
For the reasoning below it is not necessary to specify $h^{*}(\md)$ explicitly. Using $u\equiv u^{*}$
(distance from $f^{*}$), $\mathbf{S}\equiv\mathbf{S}^{*}$, and $\delta\md\equiv\delta\md_{f^{*}}$
we find, after an integration by parts with respect to $z$,
\begin{subequations}
\begin{eqnarray}
\hspace*{-1.2cm}& & \int_{A}d^{2}R\,\,\mathcal{H}_{h}\big{(}\mathbf{f}(\mathbf{R})\big{)}\nonumber \\
\hspace*{-1.2cm}& = & \int_{A}d^{2}R\int\lts_{-\infty}^{\infty}\! dz\,(c^{*}-z)\,\big{[}\partial h_{f}^{*}\,\partial_{z}\md_{f}-\partial h_{c}^{*}\,\partial_{z}\md_{c}\big{]}\label{eq:H_Term_intbyparts_z}
\end{eqnarray}
so that
\begin{equation}
 \int_{A}d^{2}R\,\,\mathcal{H}_{h}\big{(}\mathbf{f}(\mathbf{R})\big{)}  =  \int\lts_{A}d^{2}S\,\,\tr\md\,\Big{[}\mymathfont{H}^{\partial}+\mymathfont{H}^{\delta}+\mymathfont{H}^{\mathcal{R}}\Big{]}\,.\label{eq:h_Term_Trafo}
\end{equation}
\end{subequations}
Similar to Eq.\,(\ref{eq:Def_delta_bar}) we define the abbreviations
(using $\tr\md$ and $g^{*}$ instead of $\tr\rho_{i}$ and $g_{i}$)
\begin{equation}
\gdcb{n}{k}[\mathcal{A}^{*}]\::=\:\frac{\gdc{n}{k}[\mathcal{A}^{*}]}{\sqrt{(g^{*})^{n}}}\,\::=\:\bar{\delta}_{n}[\partial^{k}h_{c}^{*}\,\,\mathcal{A}^{*}]\label{eq:Def_gdc_bar}\end{equation}
and apply the expansion\begin{equation}
\partial h_{f}^{*}\equiv\partial h^{*}(\md_{c}+\delta\md)\approx\partial h_{c}^{*}+\partial^{2}h_{c}^{*}\,\delta\md+\frac{\partial^{3}h_{c}^{*}}{2}\,\big{[}\delta\md\big]^{2}\label{eq:partial_h_expansion}\end{equation}
to Eq.\,(\ref{eq:H_Term_intbyparts_z}). With $\fcstar:=f^{*}-c^{*}$
we find
\begin{eqnarray}
\mymathfont{H}^{\partial} & = & -\,\fcstar\,\dcop{o}+2H^{*}\sqrt{g^{*}}\,\dcopb{2}\label{eq:Def_H_partial}\\
 &  & +\,\fcstar\,2H^{*}\,\dcop{1}-K^{*}g^{*}\,\dcopb{3}\nonumber \\
 &  & +\,\big{(}1-\frac{1}{\sqrt{g^{*}}}\big{)}\dcop{1}\,,\nonumber \end{eqnarray}
\begin{equation}
\mymathfont{H}^{\delta}=\frac{\dco{o}}{\sqrt{g^{*}}}-4H^{*}\,\dcob{1}+\frac{\dct{o}}{2\sqrt{g^{*}}}\,,\label{eq:Def_H_delta}\end{equation}
and
\begin{equation}
\mymathfont{H}^{\pztr}=\frac{\,-1}{\tr\md\sqrt{g^{*}}}\,\int\lts_{-\ibfu}^{\ibfu}\!\!\! du\,\,\Big{(}\fcstar+\frac{u}{\sqrt{g^{*}}}\Big{)}\,\partial h_{f}^{*}\,\nabla f^{*}\nabla\delta\md\,.\label{eq:Def_H_pztr}\end{equation}
Now we use the curvature expansion from Eq.\,(\ref{eq:Def_curex}) for
$\delta\md$. Up to second order in $f^{*}$ one has
\begin{flalign}
\mymathfont{H}^{\delta} & =\frac{1}{\sqrt{g^{*}}}\,(2H^{*})\,\gdc{o}{1}[\md_{H}]+\frac{1}{\sqrt{g^{*}}}\, K^{*}\,\gdc{o}{1}[\md_{K}]\label{eq:H_delta_CUREX}\\
 & +(2H^{*})^{2}\,\Big{(}\frac{\gdc{o}{1}[\md_{H^{2}}]}{\sqrt{g^{*}}}\,-2\,\gdcb{1}{1}[\md_{H}]+\frac{\gdc{o}{2}[(\md_{H})^{2}]}{2\sqrt{g^{*}}}\Big{)}\,.\nonumber
\end{flalign}
The contribution $\mymathfont{H}^{\mathcal{R}}$ is treated similarly as $\mymathfont{G}_{j}^{\pztr}$
(see Eq.\,(\ref{eq:Def_G_pztr})). The curvature expansion  finally leads to
\begin{eqnarray}
\mymathfont{H}^{\pztr} & = & \frac{1}{\sqrt{g^{*}}}\,\tr f^{*}\,(2H^{*})\,\gdcb{1}{1}[\md_{H}]+\mathcal{O}((f^{*})^{3})\,.\label{eq:H_pztr_CUREX}\end{eqnarray}

\section{Second Order approximation\label{app:Second-Order-approximation}}

In this appendix we provide the explicit expressions up to  second order
which result from applying the equilibrium condition in Eq.\,(\ref{eq:eq_condition_1})
to the expressions given in Appendix\:\ref{app:Explicit-Form-of-H}.
Furthermore, the approximation $2H=g^{-3/2}\big{(}f_{xx}(1+f_{y}^{2})+f_{yy}(1+f_{x}^{2})-2f_{x}f_{y}f_{xy}\big{)}\approx\tr f$
and the Gauß-Bonnet theorem, $\int K(\mathbf{S})\, d^{2}S=2\pi\chi_{E}$,
where $\chi_{E}$ denotes the Euler characteristic of the surface
$f$, are used. As a consequence, since we consider laterally flat,
connected surfaces, we have $\chi_{E}=0$. Moreover it is convenient
to present the terms in Fourier space using Eq.\,(\ref{eq:Def_Fourier_1}).
Additional details are provided in Ref.\,\cite{Hiester:2005}.

\subsection{Gravity Part}

Together with the interaction terms we obtain from Eq.\,(\ref{eq:G_Term_Trafo})
with Eqs. (\ref{eq:Def_G_partial}), (\ref{eq:G_delta_CUREX}), and (\ref{eq:G_pztr_CUREX}) by using
the equilibrium condition in Eq.\,(\ref{eq:eq_condition_2}) several times
\setformulalength
\begin{eqnarray}
\int\! d^{2}S\,\,\mathcal{H}_{V}^{G}\big{(}\mathbf{f}(\mathbf{S})\big{)}&=&\frac{G}{4\pi}\int\! d^{2}q\,\,\sum_{j=1}^{2}\mathcal{G}_{j}(q)\,|\hat{f}_{j}^{c}(\mathbf{q})|^{2}\,,\label{eq:H_V_Gauss_Fourier}\\
\textrm{\parbox{0.34\formulalength}{where \hfill}}& &\textrm{\parbox{0.66\formulalength}{ \hfill}} \,\nonumber\\
\mathcal{G}_{j}(q) & := & m_{j}\tr\rho_{j}\,\big{(}1-2q^{2}\,\delta_{o}[\rho_{H_{j}}]\big{)}\,.\label{eq:Def_G_j_Fourier}
\end{eqnarray}
Here, some expressions arising from the special treatment of the hard
sphere part are not taken into account; we shall add them in Subsec.\,\ref{sub:Hard-Sphere-Part_GAUSS}
where the derivation of those terms is explained. The expressions in Eqs.\,(\ref{eq:H_V_Gauss_Fourier})
and (\ref{eq:Def_G_j_Fourier}) are identical to those derived  previously
for a single interface \cite{Mecke:6766(1999)}.

\subsection{Interaction Part\label{sub:Interaction-Part_GAUSS}}

$\hat{w}_{ij}^{(k)}[q,\delta c_{ij}]:=\hat{w}_{ij}^{(k)}(q,u'-u''+\delta c_{ij})$
denotes the Fourier transformed interaction potential (see Eq.\,(\ref{eq:w_Fourier_general_form})),
or the Fourier transformed integrals of $w_{ij}$ (see Eqs.\,(\ref{eq:Def_w^(1)}) and  (\ref{eq:Def_w^(2)})),
respectively. Thus, similar to Eq.\,(\ref{eq:Def_dcw_f}) for
$\mathcal{A}_{ij}\equiv\mathcal{A}_{ij}(u',u'',\ldots)$ we introduce
\begin{equation}
\dcwh{k}(q,\mathcal{A}_{ij}):=\iint\lts_{-\ibfu}^{\ibfu}\!\!\! du'du''\,\,\hat{w}_{ij}^{(k)}[q,\delta c_{ij}]\,\mathcal{A}_{ij}(u',u'',\ldots)\label{eq:Def_dcwh}
\end{equation}
and\begin{equation}
\delta\dcwh{k}(q,\cdots):=\dcwh{k}(q,\cdots)-\dcwh{k}(0,\cdots)\,,\label{eq:Def_delta_dcwh}
\end{equation}
but we suppress the  index $c$, because all quantities $\dcw k_{f}$ have been
expanded in terms of $f^{c}$ and $\nabla f$, respectively. Moreover,
we have omitted the square brackets indicating the functional
dependence of $\dcwh{k}\big{(}q,[\mathcal{A}_{ij}(u',u'',\ldots)]\big{)}$
(see also Eq.\,(\ref{eq:Def_dcw_f})) and we simplify $\dcwh{0}(\ldots)\equiv\dcws(\ldots)$
due to $w_{ij}^{(0)}(\ldots)\equiv w_{ij}(\ldots)$. \\
\\
From Eq.\,(\ref{eq:W_Term_Trafo}) with Eqs.\,(\ref{eq:Def_W_partial}), (\ref{eq:W_partial_delta_CUREX}), and
(\ref{eq:W_partial_pztr_CUREX}) and by using the equilibrium condition in Eq.\,(\ref{eq:eq_condition_2}) we arrive at
\begin{eqnarray}
 &  & \iint_{A}d^{2}S'd^{2}S''\,\,\mathcal{H}_{W}^{G}(\mathbf{f}(\mathbf{S}',\mathbf{S}''))\nonumber \\
 & = & \frac{1}{4\pi}\int d^{2}q\;\;\hat{\mathbf{f}}^{\dag}(\mathbf{q})\left(\begin{array}{cc}
\mathcal{W}_{12}(q) & \mathcal{V}_{21}(q)\\
\mathcal{V}_{12}(q) & \mathcal{W}_{21}(q)
\end{array}\right)\hat{\mathbf{f}}(\mathbf{q})\,.\label{eq:H_w_Gauss_Fourier}
\end{eqnarray}
The matrix elements $\mathcal{W}_{ij}$ and $\mathcal{V}_{ij}$
stem from different $\dcwh{k}(q,\ldots)$, which include the
planar density profile $\rho_{c}(u)$ and the first term $\rho_{H}(u)$ in the curvature expansion.
In order to obtain a transparent presentation we define
\begin{equation}
\mathcal{W}_{ij}(q):=\gamw_{ii}(q)+\gamv_{ii}(q)+\gamv_{ij}(q)-q^{4}\,\mathbb{K}_{ii}\label{eq:Def_H_w_Gauss_Fourier_part2}\end{equation}
and\begin{eqnarray}
\mathcal{V}_{ij}(q) & := & \begin{array}[t]{l}
\gamw_{ij}(q)-q^{4}\,\mathbb{K}_{ij}\,,\end{array}\label{eq:Def_H_w_Gauss_Fourier_part1}\end{eqnarray}
where (with the convention that quantities with an index $i$ depend on $u'$,
while those with an index $j$  depend on $u''$; see Subsec.\,\ref{sub:Interaction-Part_APPENDIX})
\begin{flalign}
\gamw_{ij}(q) & :=\quad\quad\;\hat{\dcws}\big{(}q,\partial_{'}\rho_{c_{i}}\partial_{''}\rho_{c_{j}}\big{)} &  & \,\nonumber \\
 & \quad+\, q^{2}\,\Big{[}\hat{\dcws}\big{(}q,\rho_{H_{i}}\partial_{''}\rho_{c_{j}}\big{)}+\hat{\dcws}\big{(}q,\rho_{H_{j}}\partial_{'}\rho_{c_{i}}\big{)}\Big{]} &  & \,\nonumber \\
 & \quad+\, q^{4}\,\Big{[}\hat{\dcws}\big{(}q,\rho_{H_{i}}\rho_{H_{j}}\big{)}+\mathbb{K}_{ij}\Big{]} &  & \label{eq:Def_gamma_wedge}
 \end{flalign}
and\renewcommand{\arraystretch}{1.35}
\begin{flalign}
\gamv_{ij}(q) & :=-\,\hat{\dcws}\big{(}0,\partial_{'}\rho_{c_{i}}\partial_{''}\rho_{c_{j}}\big{)}-2q^{2}\,\hat{\dcws}\big{(}0,\rho_{H_{i}}\partial_{''}\rho_{c_{j}}\big{)} &
\begin{array}[t]{c}
\end{array}\nonumber \\
 & \quad+2q^{4}\,\dcwh{2}(0,\rho_{H_{i}}\partial_{''}\rho_{c_{j}})\,.\label{eq:Def_gamma_vee}
 \end{flalign}
The  constants $\mathbb{K}_{ij}$ (Eqs.\,(\ref{eq:Def_H_w_Gauss_Fourier_part2}) and (\ref{eq:Def_H_w_Gauss_Fourier_part1}))
are given explicitly in Eq.\,(\ref{eq:Def_kappa_ij}) and are introduced here for convenience in order simplify the calculations 
using $\gamw_{ij}=\gamw_{ji}$,
$\gamv_{12}\neq\gamv_{21}$, $\gamw_{ii}+\gamv_{ii}=q^{2}\,\gamma_{ii}$,
and $\sum_{ij}(\gamw_{ij}+\gamv_{ij})=q^{2}\,\gamma^{+}$ (see Eqs.\,(\ref{eq:Def_gamma_plus}) and (\ref{eq:Def_gamma_ij})).
Since  in $\gamw_{ij}(q)$ the terms $q^4\,\mathbb{K}_{ij}$ are added with opposite sign,
Eqs.\,(\ref{eq:H_w_Gauss_Fourier})-(\ref{eq:Def_H_w_Gauss_Fourier_part1}) do not depend on $\mathbb{K}_{ij}$. 

\subsection{Hard Sphere Part\label{sub:Hard-Sphere-Part_GAUSS}}

Starting from Eq.\,(\ref{eq:h_Term_Trafo}), we obtain up to terms second
order in $f^{*}$ (except those which vanish identically due to the above-mentioned
Gauß-Bonnet theorem)
\renewcommand{\arraystretch}{1.}
\begin{eqnarray}
 &  & \int_{A}d^{2}R\,\,\mathcal{H}_{h}\big{(}\mathbf{f}(\mathbf{R})\big{)}\nonumber \\
 & = & \tr\md\int\lts_{A}d^{2}S\,\,\bigg{[}-(f^{*}-c^{*})\dcop{o}\label{eq:H_h_Gauss}\\
 &  & +\,2H^{*}\,\big{(}\dcop{2}+\gdc{o}{1}[\mnd_{H}]\big{)}-\frac{1}{2}\,(\nabla f^{*})^{2}\,\dcop{1}\nonumber \\
 &  & +\,(\tr f^{*})^{2}\,\Big{(}\gdc{o}{1}[\mnd_{H^{2}}]-\gdc{1}{1}[\mnd_{H}]+\frac{1}{2}\,\gdc{o}{2}[(\mnd_{H})^{2}]\Big{)}\bigg{]}\,.\nonumber
 \end{eqnarray}
 Our aim is to express the right hand side of Eq.\,(\ref{eq:H_h_Gauss})
in terms of $f_{i}-c_{i}$, $i\in\{1,2\}$. To this end, we use an expansion
of Eq.\,(\ref{eq:h_Term_Trafo}) around the equilibrium densities $\rho_{c_{i}}$
on the left hand side and an expansion around $\md_{c}$ on the right
hand side. A comparison of the terms leads to the relations
\setformulalength
\begin{eqnarray}
\sum_{j=1}^{2}\partial_{j}h(\rho_{c_{1}},\rho_{c_{2}})\,\delta\rho_{j} & = &\partial h_{c}^{*}\,\delta\md\label{eq:hh_back_1}\\
 \textrm{\parbox{0.6\formulalength}{and \hfill}}& &\textrm{\parbox{0.4\formulalength}{ \hfill}} \,\nonumber\\
\sum_{i,j=1}^{2}\partial_{ij}^{2}\, h(\rho_{c_{1}},\rho_{c_{2}})\,\delta\rho_{i}\delta\rho_{j} & = &\partial^{2}h_{c}^{*}\,(\delta\md)^{2}\,.\label{eq:hh_back_2}
\end{eqnarray}
Thus using the curvature expansion in Eq.\,(\ref{eq:Def_curex}) for
each $\delta\rho_{i}$ in Eq.\,(\ref{eq:hh_back_1}) one obtains similar
equations which, for instance, relate the curvatures $H_{j}$ and
$H^{*}$. Using $2H_{i}\approx\tr f_{i}$, $2H^{*}\approx\tr f^{*}$,
and $f^{*}-c^{*}=(\xi_{2}\, f_{1}^{c}+\xi_{1}\, f_{2}^{c})/(\xi_{1}+\xi_{2})$
from Eq.\,(\ref{eq:f_star_approx}) one can express the contributions in Eq.\,(\ref{eq:H_h_Gauss})
including $\gdc{1}{1}[\md_{H}]$ and $\gdc{o}{1}[\md_{H^{2}}]$ in terms of $\tr f_{j}$.
The same line of argument holds for Eq.\,(\ref{eq:hh_back_2})  which yields
\begin{equation}
\sum_{i,\, j=1}^{2}\mathbb{K}_{ij}\,\tr f_{i}\tr f_{j}\approx\tr\md\,\gdc{o}{2}[(\md_{H})^{2}]\,(\tr f^{*})^{2}\,\label{eq:hh_back_3}\end{equation}
with the constants\begin{equation}
\mathbb{K}_{ij}:=\int\lts_{-\ibfu}^{+\ibfu}du\,\,\partial_{ij}^{2}h\big{(}\rho_{c_{1}}(u),\rho_{c_{2}}(u)\big{)}\,\,\rho_{H_{i}}(u)\rho_{H_{j}}(u)\,,\label{eq:Def_kappa_ij}\end{equation}
which were already used in Eqs.\,(\ref{eq:Def_gamma_wedge}) and (\ref{eq:Def_gamma_vee}) and form
the matrix $\mathbb{K}:=(\mathbb{K}_{ij})_{i,j\in\{1,2\}}$.
If we take into account the additional terms arising from using the equilibrium condition
leading to Eqs.\,(\ref{eq:H_V_Gauss_Fourier}) and (\ref{eq:H_w_Gauss_Fourier}),
we obtain for  $\mathcal{H}_{h}^{G}(\mathbf{f})\equiv\mathcal{H}_{h}^{G}(\mathbf{f}(\mathbf{S}))$:
\begin{flalign}
\int_{A}\!\! d^{2}S\,\,\mathcal{H}_{h}^{G}(\mathbf{f}) & =\frac{1}{4\pi}\int_{A}\!\! d^{2}q\,\,\hat{\mathbf{f}}^{\dag}(\mathbf{q})\bigg{[}2q^{2}\left(\begin{array}{cc}
\ms H_{1}(q) & 0\\
0 & \ms H_{2}(q)\end{array}\right)\nonumber \\
 & \quad-\,2q^{2}\,\Clrd\,\ds{\sum_{j=1}^{2}}\,\ms H_{j}(q)+q^{4}\,\mathbb{K}\,\bigg{]}\hat{\mathbf{f}}(\mathbf{q})\label{eq:H_h_Gauss_Fourier_1}\end{flalign}
with (see \Eq{eq:Def_delta})
\begin{equation}
\ms H_{i}(q):=\tr\rho_{i}\,\Big{(}\delta_{1}\big{[}\partial_{i}h\,\partial_{u}\rho_{c_{i}}\big{]}+q^{2}\delta_{1}\big{[}\partial_{i}h\,\rho_{H_{i}}\big{]}\Big{)}\,,\label{eq:Def_ms_H}
\end{equation}
\setformulalength
\begin{eqnarray}
\clrd_{i} & := & \clr{i}\,,\label{eq:Def_clrd_entry}\\
 \textrm{\parbox{0.3\formulalength}{and \hfill}}& &\textrm{\parbox{0.7\formulalength}{ \hfill}} \,\nonumber\\
\Clrd & := & \left(\begin{array}{cc}
\clrd_{2}^{2} & \clrd_{1}\clrd_{2}\\
\clrd_{1}\clrd_{2} & \clrd_{1}^{2}\end{array}\right)\,.\label{eq:Def_clrd}
\end{eqnarray}
Furthermore, $\ms{H}_{i}(q)$ can be rewritten by using again the equilibrium
condition Eq.\,(\ref{eq:eq_condition_2})  in order to express
$\mathcal{H}_{h}^{G}(\mathbf{f}(\mathbf{S}))$ only in terms of the external
potential $V$ and the interaction potentials $w_{ij}$.
Only the constant matrix $\mathbb{K}$  remains as a direct  hard
sphere contribution:
\begin{equation}
-\ms H_{i}(q)=G\,\ms G_{i}(q)+\ds{\sum_{j=1}^{2}}\,\ms W_{ij}(q)\end{equation}
where\begin{equation}
\ms G_{i}(q):=m_{i}\tr\rho_{i}\,\Big{(}\delta_{2}[\partial_{u}\rho_{c_{i}}]+q^{2}\,\delta_{2}[\rho_{H_{i}}]\Big{)}\label{eq:Def_msG_i}\end{equation}
and\begin{equation}
\ms W_{ij}(q):=\dcwh{1}\big{(}0,u'\partial_{'}\rho_{c_{i}}\partial_{''}\rho_{c_{j}}\big{)}+q^{2}\,\dcwh{1}\big{(}0,u'\rho_{H_{i}}\partial_{''}\rho_{c_{j}}\big{)}\,.\label{eq:Def_msW_ij}\end{equation}
In combination with Eq.\,(\ref{eq:Def_G_j_Fourier}) this gives the
total gravity contribution  in matrix form (see Eq.\,(\ref{eq:Def_clrd}))
\begin{equation}
\mathbb{G}(q):=G\left(\begin{array}{cc}
\bar{\mathcal{G}}_{1}(q) & 0\\
0 & \bar{\mathcal{G}}_{2}(q)\end{array}\right)+2G\,\Clrd\, q^{2}\,\ds{\sum_{j=1}^{2}}\,\ms G_{j}(q)\label{eq:Def_G_Matrix_f}\end{equation}
with
\begin{eqnarray}
\bar{\mathcal{G}}_{i}(q) & := & \mathcal{G}_{i}(q)-2q^{2}\ms G_{i}(q)\,.\label{eq:Def_G_Matrix_f_help}\end{eqnarray}
Along the same lines the interaction contributions can be expressed as (see Eqs.\,(\ref{eq:Def_H_w_Gauss_Fourier_part2}) and 
(\ref{eq:Def_H_w_Gauss_Fourier_part1}))
\begin{equation}
\mathbb{W}(q):=\left(\begin{array}{cc}
\bar{\mathcal{W}}_{12}(q) & \mathcal{V}_{21}(q)\\
\mathcal{V}_{12}(q) & \bar{\mathcal{W}}_{21}(q)\end{array}\right)+\,2q^{2}\,\Clrd\sum_{i,j=1}^{2}\ms W_{ij}(q)\label{eq:Def_W_Matrix_f}\end{equation}
with
\begin{eqnarray}
\bar{\mathcal{W}}_{ij}(q) & := & \mathcal{W}_{ij}(q)-2q^{2}\sum_{k=1}^{2}\ms W_{ik}(q)\,.\label{eq:Def_W_Matrix_f_help}\end{eqnarray}

\section{Explicit form of the correlation functions\label{app:Explicit-correlations}}

With Eqs.\,(\ref{eq:Def_msW_ij}), (\ref{eq:Def_erb_1}), (\ref{eq:Def_erb_2}),
and (\ref{eq:Def_clrd_entry}) one has
\setformulalength
\begin{eqnarray}
\gammasym(q) & = & \frac{\delta\ms{W}(q)}{\gamma^{+}(q)}\,\Big{(}\,\erb_{1}(q)-\erb_{2}(q)-\delta\ms{W}(q)\,\Big{)}\label{eq:Def_gammasym}\\
 \textrm{\parbox{0.23\formulalength}{with \hfill}}& &\textrm{\parbox{0.77\formulalength}{ \hfill}} \,\nonumber\\
\ms{\delta W}(q) & := & 2\,\clrd_{1}\,\sum_{j=1}^{2}\ms W_{1j}(q)-2\,\clrd_{2}\,\sum_{j=1}^{2}\ms W_{2j}(q)\,.\label{eq:Def_gammasym_help}
\end{eqnarray}
In order to obtain compact expressions for the correlation functions
$\langle\hat{f}_{i}(\mathbf{q})\hat{f}_{j}(-\mathbf{q})\rangle$
we use several abbreviations. First, in order to simplify
the matrix elements $\mathbb{E}_{ij}^{-1}(q)$ which are related to the correlation
functions (see \Eq{eq:correlations_general}), we introduce the following
expressions using Eqs.\,(\ref{eq:epu_Formel})-(\ref{eq:Def_gamma_ij}),
(\ref{eq:Def_erb_1})-(\ref{eq:gamma_minus}), and (\ref{eq:Def_gammasym_help}) (the argument $q$ on the rhs is omitted):
\setformulalength
\begin{eqnarray}
\gamma_{11}^{\,\mathsf{eff}}(q) & := & \erb_{1}-\,\ms{\delta W}-\gamma_{12}^{\,\mathsf{eff}}\label{eq:Def_gamma_eff_11}\,,\\
\gamma_{22}^{\,\mathsf{eff}}(q) & := & \erb_{2}+\,\ms{\delta W}-\gamma_{12}^{\,\mathsf{eff}}\label{eq:Def_gamma_eff_22}\,,\\
                 \textrm{\parbox{0.3\formulalength}{and \hfill}}& &\textrm{\parbox{0.7\formulalength}{ \hfill}} \,\nonumber\\
\gamma_{12}^{\,\mathsf{eff}}(q) & := & \frac{\big{(}\erb_{1}-\delta\ms{W}\big{)}\,\big{(}\erb_{2}+\delta\ms{W}\big{)}}{\gamma^{+}}-\gamma^{-}\label{eq:Def_gamma_eff_12}\,.
\end{eqnarray}
This leads to the relations\begin{eqnarray}
\mathbb{E}_{ij}(q) & = & \mathbb{G}_{ij}(q)+(-1)^{i+j}\,\emi_{o}(0)+q^{2}\gamma_{ij}^{\,\mathsf{eff}}(q)\label{eq:E_ij_explicit}\end{eqnarray}
resulting in
\begin{equation}
\mathbb{E}_{11}^{-1}(q)=\frac{\mathbb{G}_{22}(q)+\emi_{o}(0)+q^{2}\gamma_{22}^{\,\mathsf{eff}}(q)}{\det\mathbb{E}}\label{eq:Appendix_f1f1_corr_explicit}\end{equation}
and similarly for $\mathbb{E}_{22}^{-1}(q)$
by interchanging the indices $1\leftrightarrow2$, and
\begin{equation}
\mathbb{E}_{12}^{-1}(q)=\frac{-\,\mathbb{G}_{12}(q)+\emi_{o}(0)-q^{2}\gamma_{12}^{\,\mathsf{eff}}(q)}{\det\mathbb{E}}\,.\label{eq:Appendix_f1f2_corr_explicit}
\end{equation}
Since all correlation functions share the same denominator
(Eqs.\,(\ref{eq:Def_G_Matrix_f}), (\ref{eq:epu_Formel}), and (\ref{eq:emi_canonical_form})), 
\begin{eqnarray}
\det\mathbb{E}(q) & = & \det\mathbb{G}(q)+\emi_{o}(0)\, G\,\mathcal{G}^{+}(q)\label{eq:det_E_explicit}\\
 &  & +\, q^{2}\,\big{[}\,\mathbb{G}_{11}(q)\,\gamma_{22}^{\,\mathsf{eff}}(q)+\mathbb{G}_{22}(q)\,\gamma_{11}^{\,\mathsf{eff}}(q)\nonumber \\
 &  & \quad-\,2\,\mathbb{G}_{12}(q)\,\gamma_{12}^{\,\mathsf{eff}}(q)\,\big{]}\nonumber \\
 &  & +\, q^{2}\,\gamma^{+}(q)\,\emi_{o}(q)\,,\nonumber \end{eqnarray}
one obtains for a vanishing external field strength, i.e., for $G\rightarrow0$
the following  long-wave limit $q\rightarrow0$ for all pairs $i,j\in\{1,2\}$:
\begin{equation}
\mathbb{E}_{ij}^{-1}(q)\stackrel{{{G\rightarrow0\atop q\rightarrow0}}}{=}\frac{1}{\epu(q)}+\mathcal{O}(G,q^{2})\,,\label{eq:Appendix_fifj_corr_limit}\end{equation}
which agrees with Ref.\,\cite{Tarazona:1357(1985)}.

It is important to mention, that by using the sharp kink profile (Eq.\,(\ref{eq:sk_profile}))
and by neglecting all curvature contributions at this stage, i.e.,
$\rho_{H}\equiv0$, expressions for $\gamma^{+}$ and $\gamma^{-}$
arise which would not follow from the original sharp kink approximation introduced in
Subsec.\,\ref{sub:SKAP_1}. Here, in this case one has $\ms{W}_{ij}=0$
and $\ms{G}_{i}=0$ (see Eqs. (\ref{eq:Def_gammasym_help}) and (\ref{eq:Def_msG_i})),
which implies $\gammasym=0$ and $\mathcal{G}_{i}^{\mathsf{{sk}}}(q)=m_{i}\tr\rho_{i}$.
Furthermore, by using $\delta\hat{w}_{ij}(q,\ldots)=\hat{w}_{ij}(q,\ldots)-\hat{w}_{ij}(0,\ldots)$
we obtain
\begin{eqnarray}
\gamma_{ij}^{\,\mathsf{sk}}(q) & = & \tr\rho_{i}\tr\rho_{j}\,\frac{\delta\hat{w}_{ij}(q,\delta c_{ij})}{q^{2}}\,,\label{eq:gamma_ij_sk}\end{eqnarray}
and, with $\gamma^{+,\,\mathsf{sk}}(q)=\sum_{ij}\gamma_{ij}^{\,\mathsf{sk}}(q)$,
\begin{eqnarray}
\gamma^{-,\,\mathsf{sk}}(q) & = & \frac{(\tr\rho_{1})^{2}(\tr\rho_{2})^{2}}{q^{2}\,\gamma^{+,\,\mathsf{sk}}(q)}\times\label{eq:gamma_minus_sk}\\
 &  & \times\Big{[}\,\delta\hat{w}_{11}(q,0)\delta\hat{w}_{22}(q,0)-\,\delta\hat{w}_{12}^{2}(q,\delta c_{12})\Big{]}\,.\nonumber \end{eqnarray}
If all interaction potentials $w_{ij}$ have the same form and their
amplitudes fulfill $w_{o}^{(12)}=\sqrt{w_{o}^{(11)}w_{o}^{(22)}}$,
all contributions to $\gamma^{-,\,\mathsf{sk}}$ vanish with the only exception the
non-vanishing contributions stemming from $\delta c_{12}$ and the difference in the particle
diameters. In addition, in this case with $\textrm{L}(\alpha):=\sqrt{(r_{o}^{(ij)})^{2}+\alpha^{2}}$
one finds a generalization of Eq.\,(\ref{eq:Def_gamma_skap}) or Eq.\,(\ref{eq:gamma_sk_lim_0}),
respectively, for $\delta c_{12}\neq0$:
\begin{flalign}
\gamma^{+,\,\mathsf{sk}}(q\rightarrow0) & =\frac{1}{16\,}\sum_{i,j=1}^{2}\frac{\tr\rho_{i}\tr\rho_{j}\, w_{o}^{(ij)}\,(r_{o}^{(ij)})^{6}}{\textrm{L}^{2}(\delta c_{ij})}\times\label{eq:gamma_plus_sk_lim_0}\\
 & \times\Big{(}1+\frac{q^{2}\,\textrm{L}^{2}(\delta c_{ij})}{4}\,\Big{[}\log(\frac{q\,\textrm{L}(\delta c_{ij})}{2})-C\Big{]}\Big{)}.\nonumber \end{flalign}

\end{document}